\documentclass[12pt]{article}
\usepackage{amsthm,amsmath,latexsym,multibox,amssymb,amsfonts}
\usepackage{multibox}
\usepackage{graphicx,lscape,fancyhdr,array,stmaryrd}

\newcommand{\bee}{\begin{eqnarray}}
\newcommand{\eee}{\end{eqnarray}}
\def\*{\ast}
\def\a{\alpha}
\def\b{\beta}
\def\g{\gamma}
\def\d{\delta}

\def\m{\mu}

\def\n{\nu}
\def\r{\rho}
\def\o{\omega}
\def\s{\sigma}

\def\t{\tau}

\def\pa{\partial}

\def\be{\begin{equation}}
\def\ee{\end{equation}}
\def\bqn{\begin{eqnarray}}
\def\eqn{\end{eqnarray}}
\def\nn{\nonumber}
\def\ca{{\cal A}}

\def\cc{{\cal C}}

\def\ci{{\cal I}}

\def\ck{{\cal K}}

\def\cs{{\cal S}}
\def\ct{{\cal T}}
\def\cu{{\cal U}}

\def\cw{{\cal W}}

\newcommand{\hu}{{{ hu(1|2\!\!:\!\![d-1,2]) }}}
\newcommand{\huin}{{{ hu_\infty (1|2\!\!:\!\![d-1,2]) }}}
\newcommand{\hup}{{{ hu_P (1|2\!\!:\!\![d-1,2]) }}}
\newcommand{\ho}{{{ ho(1|2\!\!:\!\![d-1,2]) }}}

\newcommand{\half}{\frac{1}{2}}

\newtheorem{theorem}{Theorem}[section]

\csname @addtoreset\endcsname{equation}{section}

\author{\large{X. Bekaert$^a$, S. Cnockaert$^b$\footnote{Research Fellow
of the National Fund for Scientific Research (Belgium)}$\;$, C.
Iazeolla$^c$, M.A.Vasiliev$^d$\footnote{Corresponding author. Email:
vasiliev@td.lpi.ru}}\\ \\
     \small{\it{$^a$IHES, Le Bois-Marie,}}
     \small{\it{35 route de Chartres, 91440 Bures-sur-Yvette, France
}}\\ 
      \small{\it{$^b$Physique th\'{e}orique et math\'{e}matique,
Universit\'{e} Libre de Bruxelles }}\\
      \small{\it{and International
Solvay Institutes}} \\
     \small{\it{Campus de la Plaine CP 231, blvd du Triomphe, 1050 Bruxelles,
Belgium}}\\ 
     \small{\it{$^c$Dipartimento di Fisica, Universit\`{a} di Roma ``Tor
Vergata"}}\\
     \small{\it{INFN, Sezione di Roma ``Tor Vergata" }}\\
      \small{\it{Via della Ricerca Scientifica 1, 00133 Roma, Italy }}\\ 
         \small{\it{$^d$ I.E.Tamm Department of Theoretical Physics,
Lebedev Physical Institute,}}\\
         \small{\it{Leninsky prospect 53, 119991 Moscow, Russia}}
}

\title{\Huge{Nonlinear Higher Spin Theories\\ in Various Dimensions}}

\begin{document}

\begin{titlepage}

\maketitle

\begin{abstract}
In this article, an introduction to the nonlinear equations for
completely symmetric bosonic higher spin gauge fields in anti de
Sitter space of any dimension is provided. To make the presentation
self-contained we explain in detail some related issues such as the
MacDowell-Mansouri-Stelle-West formulation of gravity, unfolded
formulation of dynamical systems in terms of free differential
algebras and Young tableaux symmetry properties in terms of  Howe
dual algebras.
   \vskip 10pt
\begin{center}
{\it Based on the lectures presented by M.A.Vasiliev  at the First
Solvay Workshop on Higher-Spin Gauge Theories held in Brussels on
May 12-14, 2004}
\end{center}
\end{abstract}

\vskip 5pt
\begin{flushleft}
\vspace{10pt} IHES/P/04/47\\ ULB-TH/04-26  \\ ROM2F-04/29\\
FIAN/TD/17/04\\
\end{flushleft}

\thispagestyle{empty}

\end{titlepage}

\tableofcontents

\pagebreak

\section{Introduction}

The Coleman-Mandula theorem \cite{cm} and its generalization
\cite{Haag} strongly restrict the S-matrix symmetries of a
nontrivial ({\it i.e.} interacting) relativistic field theory in
flat space-time. More precisely, the extension of the space-time
symmetry algebra is at most the (semi)direct sum of a
(super)conformal algebra and an internal symmetry algebra spanned by
elements that commute with the generators of the Poincar\'e algebra.
Ruling out higher symmetries  via these theorems, one rules out
higher spin (HS) gauge fields associated with them, allowing in
practice only gauge fields of low spins ({\it i.e.} $s\leqslant 2$).
However, as will be reviewed here, going beyond some assumptions of
these no-go theorems allows to overcome both restrictions, on higher
spins ({\it i.e.} $s>2$) and on space-time symmetry extensions.

By now, HS  gauge fields are pretty well understood at the free
field level. Therefore, the main open problem in this topic is to
find proper nonAbelian HS gauge symmetries which extend the
space-time symmetries. These symmetries can possibly mix fields of
different spins, as supersymmetry does. Even though one  may never
find HS particles in accelerators, nonAbelian HS symmetries might
lead us to a better understanding of the true symmetries of
unification models. {} From the supergravity perspective, the
theories with HS fields may have more than 32 supercharges and may
live in dimensions higher than $11$. {} From the superstring
perspective, several arguments support the conjecture \cite{9910096}
that the Stueckelberg symmetries of massive HS string excitations
result from a spontaneous breaking of some HS gauge symmetries. In
this picture, tensile string theory appears as a spontaneously
broken phase of a more symmetric phase with only massless fields. In
that case, superstrings should exhibit higher symmetries in the
high-energy limit as was argued long ago by Gross \cite{Gross}. A
more recent argument came from the $AdS/CFT$ side after it was
realized  \cite{Su,Witten,SS5,KP} that HS symmetries should be
unbroken in the Sundborg--Witten limit
$$\lambda=g_{YM}^2N=\Big(\frac{R^2_{AdS}}{\alpha^\prime}\Big)^2\longrightarrow
0\,,$$ because the boundary conformal theory becomes free. A dual
string theory in the highly curved $AdS$ space-time is therefore
expected to be a HS theory (see also \cite{AdSCFT,AdSCFT2} and refs
therein for recent developments).

One way to provide an explicit solution of the nonAbelian HS gauge
symmetry problem is by constructing a consistent nonlinear theory of
massless HS fields. For several decades, a lot of efforts has been
put into this direction although, from the very beginning, this line
of research faced several difficulties. The first explicit attempts
to introduce interactions between HS gauge fields and gravity
encountered severe problems \cite{diff}. However, some positive
results \cite{pos} were later obtained in flat space-time on the
existence of consistent vertices for HS gauge fields interacting
with each other, but not with gravity.

Seventeen years ago, the problem of consistent HS gravitational
interactions was partially solved in four dimensions \cite{FV1}. In
order to achieve this result, the following conditions of the no-go
theorems \cite{cm,Haag} were relaxed:
\begin{description}
      \item[(1)] the theory is formulated around a flat background.
      \item[(2)] the spectrum contains a finite number of HS fields.
\end{description}
The nonlinear HS theory in four dimensions was shown to be
consistent up to cubic order at the action level \cite{FV1} and,
later, at all orders at the level of equations of motion
\cite{con,prop}. The second part of these results was recently
extended to arbitrary space-time dimensions \cite{Vasiliev:2003ev}.
The nonlinear HS theory exhibits some rather unusual properties of
HS gauge fields:
\begin{description}
      \item[(1')] the theory is perturbed around an $(A)dS$ background
and does not admit a flat limit as long as HS symmetries are
unbroken.
      \item[(2')] the allowed spectra contain infinite towers of HS
fields and do not admit a consistent finite truncation
 with $s>2$ fields.
      \item[(3')] the vertices have higher-derivative terms (that is to
say, the higher derivatives appear in HS interactions - not at the
free field level).
\end{description}
The properties (2') and (3') were also observed by the authors of
\cite{pos} for HS gauge fields. Though unusual, these properties are
familiar to high-energy theorists. The property (1') is verified by
gauged supergravities with charged gravitinos \cite{FVFD,FVpre}. The
property (2') plays an important role in the consistency of string
theory. The property (3') is also shared by Witten's string field
theory \cite{Eliezer}.

An argument in favor of  an $AdS$ background is that the S-matrix
theorems \cite{cm,Haag} do not apply since there is no well-defined
S-matrix in $AdS$ space-time \cite{Witten98}. The $AdS$ geometry
plays a key role in the nonlinear theory because cubic higher
derivative terms are added to the free Lagrangian, requiring a
nonvanishing cosmological constant $\Lambda$. These cubic vertices
are schematically of the form
$${\cal L}^{int}=\sum_{n,p}
\Lambda^{-\frac{1}{2}(n+p)}D^n(\varphi_{\ldots})D^p(\varphi_{\ldots}){\cal
R}^{\ldots}\,,$$ where $\varphi_{\ldots}$ denotes some spin-$s$
gauge field, and $\cal R$ stands for the fluctuation of the Riemann
curvature tensor around  the $AdS$ background. Such vertices do not
admit a $\Lambda \rightarrow 0$ limit. The highest order of
derivatives which appear in the cubic vertex increases linearly with
the spin \cite{pos,FV1}: $n+p\sim s$. Since all spins $s>2$ must be
included in the nonAbelian HS algebra, the number of derivatives is
not bounded from above. In other words, the HS gauge theory is
\textit{nonlocal}\footnote{Nonlocal theories do not automatically
suffer from the higher-derivative problem. Indeed, in some cases
like string field theory, the problem is somehow cured
\cite{vt,Eliezer,nonloc} if the free theory is well-behaved and if
nonlocality is treated perturbatively (see \cite{higherder} for a
comprehensive review on this point).}.

 The purpose of these lecture notes is to present,
in a self-contained\footnote{For these lecture notes, the reader is
only assumed to have basic knowledge of Yang-Mills theory, general
relativity and group theory. The reader is also supposed to be
familiar with the notions of differential forms and cohomology
groups.} way, the nonlinear equations for completely symmetric
bosonic HS gauge fields in $AdS$  space of any dimension. The
structure of the present lecture notes is as follows.

\subsection{Plan of the paper}

In Section \ref{grav}, the MacDowell-Mansouri-Stelle-West
formulation of gravity is recalled. In Section \ref{young}, some
basics about Young tableaux and irreducible tensor representations
are introduced. In Section \ref{fshsgf}, the approach of Section
\ref{grav} is generalized to HS fields, {\it  i.e.} the free HS
 gauge theory is formulated as a theory of one-form
connections. In Section \ref{hsa}, a nonAbelian HS algebra is
constructed. The general definition of a free differential algebra
is given in Section \ref{fda} and a strategy is explained on how to
formulate nonlinear HS  field equations in these terms. Section
\ref{unfolding} presents  the unfolded form of the free massless
scalar field and linearized gravity field equations,
 which are
generalized to free HS field equations in Section
\ref{freemasslessequ}. In Section \ref{taudynamcontent} is explained
how the cohomologies of some operator $\s_-$ describe the dynamical
content of a theory. The relevant cohomologies are calculated in the
HS case in Section \ref{taucohomology}. Sections \ref{star} and
\ref{tw} introduce some tools (the star product and the twisted
adjoint representation) useful for writing the nonlinear equations,
which is done in Section \ref{Nonli}. The nonlinear equations are
analyzed perturbatively in Section \ref{pert} and they are further
discussed in Section \ref{disc}. A brief conclusion inviting to
further readings completes these lecture notes.  In Appendix
\ref{appA} some elementary material of lower spin gauge theories is
reviewed, while in Appendix \ref{appB} some technical points of the
nonlinear HS equations are addressed in more details.

\vspace{.3cm}

For an easier reading of the lecture notes, here is a guide to the
regions related to the main topics addressed in these lecture notes:
\begin{description}
  \item[-] Abelian HS gauge theory: In section \ref{fshsgf} is reviewed the
quadratic actions of free
(constrained) HS gauge fields in the metric-like and frame-like
approaches.
  \item[-] Non-Abelian HS algebra: The definition and some properties of
the two simplest HS algebras are given in Section \ref{hsa}.
  \item[-] Unfolded formulation of free HS fields:
The unfolding of the free HS equations is very important as a
starting point towards nonlinear HS equations at all orders. The
general unfolding procedure and its application to the HS gauge
theory is explained in many details in Sections \ref{unfolding},
\ref{freemasslessequ}, \ref{taudynamcontent}, \ref{taucohomology}
and \ref{tw}. (Sections \ref{taudynamcontent} and
\ref{taucohomology} can be skipped in a first quick reading since
the corresponding material is not necessary for understanding
Sections from \ref{star} till \ref{disc}.)
  \item[-] Non-linear HS equations: Consistent nonlinear equations, that
are invariant under the non-Abelian HS gauge transformations and
diffeomorphisms, and correctly reproduce the free HS dynamics at the
linearized level, are presented and discussed in Sections
\ref{unffda}, \ref{towards}, \ref{Nonli}, \ref{pert} and \ref{disc}.
  \item[-] Material of wider interest: Sections \ref{grav},
\ref{young}, \ref{defs}, \ref{taudynamcontent} and \ref{star}
introduce tools which prove to be very useful in HS gauge theories,
but which may also appear in a variety of different contexts.
\end{description}

\subsection{Conventions}

Our conventions are as follows:

A generic space-time is denoted by ${\cal M}^d$ and is a
(pseudo)-Riemannian  smooth manifold of dimension $d$, where the
metric is taken to be ``mostly minus".

Greek letters $\m,\n,\r,\s,\ldots$ denote curved ({\it i.e.,} base)
indices, while Latin letters $a,b,c,d,\ldots$ denote fiber indices
often referred to as tangent space indices. Both types of indices
run from $0$ to $d-1$. The tensor $\eta_{ab}$ is the mostly minus
Minkowski metric. Capital Latin letters $A,B,C,D,\ldots$ denote
ambient space indices and their range of values is
$0,1,\ldots,d-1,\hat{d}$, where the (timelike) $(d+1)$-th direction
is denoted by $\hat{d}$  (in order to distinguish the tangent space
index $d$ from the value $\hat{d}$ that it can take). The tensor
$\eta_{AB}$ is diagonal with entries $(+,-,\ldots,-,+)$.

The bracket $[ \ldots ]$ denotes complete antisymmetrization of
indices, with strength one ({\it e.g.} $A_{[a}B_{b]}= \frac{1}{2}
(A_a B_b-A_bB_a)$), while the bracket $\{\ldots \}$ denotes complete
symmetrization of the indices, with strength one ({\it e.g.}
$A_{\{a}B_{b\}}= \frac{1}{2} (A_a B_b+A_bB_a)$). Analogously, the
commutator and anticommutator are respectively denoted as $[\,\,,\,
]$ and $\{\,\, ,\, \}$.

The de Rham complex $\Omega^*({\cal M}^d)$ is the graded commutative
algebra of differential forms that is endowed with the wedge product
(the wedge symbol will always be omitted in this paper) and the
exterior differential $d$. $\Omega^p({\cal M}^d)$ is the space of
differential $p$-forms on the manifold ${\cal M}^d$, which are
sections of the $p$-th exterior power of the cotangent bundle
$T^*{\cal M}^d$. In the topologically trivial situation discussed in
this paper $\Omega^p({\cal M}^d)=C^\infty({\cal M}^d)\otimes
\Lambda^p{\mathbb R}^{d\,*}$, where the space $C^\infty({\cal M}^d)$
is the space of smooth functions from ${\cal M}^d$ to $\mathbb R$.
The generators $dx^\m$ of the exterior algebra $\Lambda{\mathbb
R}^{d\,*}$ are Grassmann odd ({\it i.e.} anticommuting). The
exterior differential is defined as $d=dx^\m\partial_\m$.

\section{Gravity \`a la MacDowell - Mansouri - Stelle - West}
\label{grav}

Einstein's theory of gravity is a nonAbelian gauge theory of a
spin-$2$ particle, in a similar way as Yang-Mills
theories\footnote{See Appendix \ref{YM} for a brief review of
Yang-Mills theories.} are nonAbelian gauge theories of spin-$1$
particles. Local symmetries of Yang-Mills theories originate from
internal global symmetries. Similarly, the gauge symmetries of
Einstein gravity in the vielbein formulation\footnote{See {\it e.g.}
\cite{Zanelli} for a pedagogical review on the gauge theory
formulation of gravity and some of its extensions, like
supergravity.} originate from global space-time symmetries of its
most symmetric vacua. The latter symmetries are manifest in the
formulation of MacDowell, Mansouri, Stelle and West \cite{MM,SW}.

This section is devoted to the presentation of the latter
formulation. In the first subsection \ref{MacDowell}, the
Einstein-Cartan formulation of gravity is reviewed and the link with
the Einstein-Hilbert action without cosmological constant is
explained.  A cosmological constant can be introduced into the
formalism, which is done in Subsection \ref{cosmo}. This subsection
also contains an elegant action for gravity,  written by MacDowell
and Mansouri. In Subsection \ref{mdmswg}, the improved version of
this action introduced by Stelle and West is presented, the
covariance under all symmetries being made manifest.

\subsection{Gravity as a Poincar\'e gauge theory}
\label{MacDowell}

In this subsection, the frame formulation of gravity with zero
cosmological constant is reviewed. We first introduce the dynamical
fields and sketch the link to the  metric formulation. Then the
action is written.

\vspace{.2cm}

The basic idea is as follows: instead of considering the metric
$g_{\m\n}$ as the dynamical field, two new dynamical fields are
introduced: the vielbein or frame field $e_{\m}^{a}$ and the Lorentz
connection $\o_\m^{L\hspace{.1cm} ab}$.

The relevant fields appear through the one-forms
   $e^a=e_{\mu}^{a}\, dx^{\mu}$ and
$\o^{{L}\,ab}=-\o^{{L}\,ba}=\o_{\mu}^{{L}\hspace{.1cm}
ab}\,dx^{\mu}\,$. The number of one-forms is equal to
$d+\frac{d(d-1)}{ 2}=\frac{(d+1)d}{ 2}$, which is the dimension of
the Poincar\'e group $ISO(d-1,1)$. So they can be collected into a
single one-form taking values in the Poincar\'e algebra as
$\o=e^aP_a+\frac{1}{2}\,\o^{L\,ab}M_{ab}$, where $P_a$ and $M_{ab}$
generate $iso(d-1,1)$ (see Appendix \ref{iso}). The corresponding
curvature is the two-form (see Appendix \ref{YM}):
$$R=d\o+\o^2\equiv T^aP_a+\frac{1}{2}\,R^{L\,ab}M_{ab}\,,$$ where $T^a$
is the torsion, given by $$ T^a = D^Le^a = d e^a + \o^{{L}\;
a}_{\hspace{.5cm}b} e^b \ ,$$ and $R^{{L\;ab}}\,$ is the Lorentz
curvature $$ R^{{L}\; ab}= D^L\o^{{L}\; ab}= d \o^{{L}\;
ab}+\o^{{L}\; a}_{\hspace{.5cm}c} \o^{{L}\; cb}\,, $$ as follows
from the Poincar\'e algebra (\ref{Lorentz})-(\ref{translations}).

To make contact with the metric formulation of gravity, one must
assume that the frame $e^a_\m$ has maximal rank $d$ so that it gives
rise to the nondegenerate metric tensor $g_{\m\n}=\eta_{ab}e_\m^a
e_\n^b$. As will be shown further in this section, one is allowed to
require the absence of torsion, $T_a=0$. Then one solves this
constraint and expresses the Lorentz connection in terms of the
frame field, $\o^L=\o^L(e,\partial e)$. It can be checked that the
tensor $R_{\rho\s,\,\m\n}=e^a_\m e^b_\n R^L_{\rho\s\;ab}$ is then
expressed solely in terms of the metric, and is the Riemann tensor.

The first order action of the frame formulation of gravity is due to
Weyl \cite{Weyl:1929}. In any dimension $d>1$ it can be written in
the form \bqn S[\,e^a_\m,\o^{L\;ab}_\m]=\frac{1}{2\kappa^2}
\int_{{\cal M}^d} R^{{L}\; bc}e^{a_1} \ldots e^{a_{d-2}}
\epsilon_{a_1 \ldots a_{d-2}bc}\,\,, \label{mdma} \eqn where
$\epsilon_{a_1 \ldots a_{d}}$ is the invariant tensor of the special
linear group $SL(d)$  and $\kappa^2$ is the gravitational constant,
so that $\kappa$ has dimension $(length)^{{d\over 2}-1}$.
   The Euler-Lagrange equations of the
Lorentz connection $$\frac{\d S}{\d \o^{{L}\,bc}}\propto
\epsilon_{a_1 \ldots a_{d-2}bc}\,e^{a_1} \ldots e^{a_{d-3}}
T^{a_{d-2}} = 0\label{deltacomega}$$ imply that the torsion
vanishes. The Lorentz connection is then an auxiliary field, which
can be removed from the action by solving its own (algebraic)
equations of motion. The action $S=S[\,e\,,\,\o^L(e,\partial e)\,]$
is now expressed only in terms of the vielbein \footnote{In the
context of supergravity, this action principle \cite{FVpre,NF} is
sometimes called the ``1.5 order formalism" \cite{Pvan} because it
combines in some sense the virtues of first and second order
formalism.}.  Actually, only combinations of vielbeins corresponding
to the metric appear and the action $S= S[\,g_{\m\n}]$ is indeed the
second order Einstein-Hilbert action.

The Minkowski space-time solves $R^{L\;ab}= 0$ and $T^a= 0$. It is
the most symmetrical solution of the Euler-Lagrange equations, whose
global symmetries form the Poincar\'e group. The gauge symmetries of
the action (\ref{mdma}) are the diffeomorphisms and the local
Lorentz transformations. Together, these gauge symmetries correspond
to the gauging of the Poincar\'e group (see Appendix \ref{sptime}
for more comments).

\subsection{Gravity as a theory of $o(d-1,2)$ gauge fields}
\label{cosmo}

In the previous section, the Einstein-Cartan formulation of gravity
with vanishing cosmological constant has been presented. We will now
show how a nonvanishing cosmological constant can be added to this
formalism. In these lectures, we will restrict ourselves to the
$AdS$ case but, for the bosonic case we focus on, everything can be
rephrased for $dS$. One is mostly interested in the $AdS$ case for
the reason that it is more suitable for supersymmetric extensions.
Furthermore, $dS$ and $AdS$ have rather different unitary
representations (for $dS$ there are unitary irreducible
representations the energy of which is not bounded from below).

It is rather natural to reinterpret $P_a$ and $M_{ab}$ as the
generators of the $AdS_d$ isometry algebra $o(d-1,2)$.
The curvature $R=d\o+\o^2$ then decomposes as
$R=T^aP_a+\frac{1}{2}\,R^{ab}M_{ab}$, where the Lorentz curvature
$R^{L\;ab}$ is deformed to \be R^{ab}\equiv R^{{L}\; ab}+ R^{cosm \;
ab}\equiv R^{{L}\; ab}+\Lambda \, e^a  e^b\,, \label{cosm}\ee since
(\ref{translations}) is deformed to (\ref{transv}).

MacDowell and Mansouri proposed an action \cite{MM}, the Lagrangian
of which is the (wedge) product of two curvatures (\ref{cosm}) in
$d=4$ \bqn S^{MM}[\,e,\o] =\frac{1}{4 \kappa^2 \Lambda}\int_{{\cal
M}^4} R^{a_1 a_2} R^{a_3 a_4} \epsilon_{a_1 a_2 a_3 a_{4}}\,.
\label{mmact} \eqn Expressing $R^{ab}$ in terms of $R^{{L}\;ab}$ and
$R^{cosm\;ab }$ by (\ref{cosm}), the Lagrangian is the sum of three
terms: a term $R^{{L}} R^{cosm}$, which is the previous Lagrangian
(\ref{mdma}) without cosmological constant, a cosmological term
$R^{cosm} \,R^{cosm}$ and a Gauss-Bonnet term $R^{{L}} R^L$. The
latter term contains higher-derivatives but it does not contribute
to the equations of motion because it is a  total derivative.

 In any dimension, the
$AdS_d$ space-time is defined as the most symmetrical solution of
the Euler-Lagrange equations of pure gravity with the cosmological
term. As explained in more detail in Section \ref{mdmswg}, it is a
solution of the system $R^{ab}= 0$, $T^a= 0$ such that
rank$(e^a_\m)=d$.

The MacDowell-Mansouri action  admits a higher dimensional
generalization \cite{5d} \bqn S^{MM}[\,e,\o] =\frac{1}{4 \kappa^2
\Lambda}\int_{{\cal M}^d} R^{a_1 a_2} R^{a_3 a_4} e^{a_5} \ldots
e^{a_{d}} \epsilon_{a_1 \ldots a_{d}}\,. \label{mmactd} \eqn Because
the Gauss-Bonnet term \bqn S^{GB}[\,e,\o] =\frac{1}{4 \kappa^2
\Lambda}\int_{{\cal M}^d} R^{L\,a_1 a_2} R^{L\,a_3 a_4} e^{a_5}
\ldots  e^{a_{d}} \epsilon_{a_1 \ldots a_{d}}\,
\nonumber 
\eqn is not topological beyond $d=4$, the field equations resulting
from the action (\ref{mmactd}) are different
 from the Einstein equations in $d$ dimensions. However
the difference is by nonlinear terms that do not contribute to the
free spin-$2$ equations \cite{5d} apart from replacing the
cosmological constant $\Lambda$ by $ \frac{2(d-2)}{d}\Lambda$ (in
such a way that no correction appears in $d=4$, as expected). One
way to see this is by considering the action \bqn S^{nonlin}[\,e,\o]
\equiv S^{GB}[\,e,\o]+ \frac{d-4}{4 \kappa^2}\int_{{\cal M}^d} \Big
(\frac{2}{d-2}\, R^{L\,a_1 a_2} e^{a_3} \ldots  e^{a_{d}}
+\frac{\Lambda}{d} \,e^{a_1} \ldots e^{a_{d}}\Big ) \epsilon_{a_1
\ldots a_{d}}\,, \label{nonl} \eqn which is the sum of the
Gauss-Bonnet term plus terms of the same type as the
Einstein-Hilbert and cosmological terms (note that the latter are
absent when $d=4$). The variation of (\ref{nonl}) is equal to \bqn
\delta S^{nonlin}[\,e,\o] = \frac{1}{4 \kappa^2 \Lambda}\int_{{\cal
M}^d} R^{a_1 a_2} R^{a_3 a_4} \delta ( e^{a_5} \ldots  e^{a_{d}} )\,
\epsilon_{a_1 \ldots a_{d}}\,,\label{variation} \eqn when the
torsion is required to be zero ({\it i.e.} applying the 1.5 order
formalism to see that the variation over the Lorentz connection does
not contribute). Indeed, the variation of the action (\ref{nonl})
vanishes when $d=4$, but when $d>4$ the variation (\ref{variation})
is bilinear in the $AdS_d$ field strength $R^{ab}$. Since the
$AdS_d$ field strength is zero in the vacuum $AdS_d$ solution,
variation of the action $S^{nonlin}$ is  nonlinear in the
fluctuations near the $AdS_d$ background. As a consequence, at the
linearized level the Gauss-Bonnet term does not affect the form of
the free spin-$2$ equations of motion, it merely redefines an
overall factor in front of the action and the cosmological constant
via $\kappa^2 \to (\frac{d}{2}-1)\kappa^2$ and $\Lambda \to
\frac{2(d-2)}{d}\Lambda$, respectively (as can be seen by
substituting $S^{GB}$ in (\ref{mmactd}) with its expression in terms
of $S^{nonlin}$ from (\ref{nonl})). Also let us note that from
(\ref{variation}) it is obvious that
 the  variation of the Gauss-Bonnet term contains second
  order derivatives of the metric, {\it i.e.} it  is of
the Lanczos-Lovelock type (see \cite{Zanelli} for a review).

Beyond the free field approximation the corrections to Einstein's
field equations resulting from the action (\ref{mmactd}) are
nontrivial for $d>4$ and nonanalytic in $\Lambda$ (as can be seen
from (\ref{variation})), having no meaningful flat limit. As will be
shown later, this is analogous to the structure of HS interactions
which also contain terms with higher derivatives and negative powers
of $\Lambda$. The important difference is that in the case of
gravity one can subtract the term (\ref{nonl}) without destroying
the symmetries of the model, while this is not possible in the HS
gauge theories.  In both cases, the flat limit $\Lambda\rightarrow
0$ is perfectly smooth at the level of the algebra ({\it e.g.}
$o(d-1,2)\rightarrow iso(d-1,1)$ for gravity, see Appendix
\ref{iso}) and at the level of the free equations of motion, but it
may be singular at the level of the action and nonlinear field
equations.

\subsection{MacDowell-Mansouri-Stelle-West gravity}
\label{mdmswg}

The action (\ref{mmactd}) is not manifestly $o(d-1,2)$ gauge
invariant. Its gauge symmetries  are the diffeomorphisms and the
local Lorentz transformations. It is possible to make the $o(d-1,2)$
gauge symmetry manifest by combining the vielbein and the Lorentz
connection into a single field $\o=dx^{\mu}\o_{\mu}^{\hspace{.2cm}
AB}M_{AB}$ and by introducing a vector $V^A$ called {\it
compensator}\footnote{This compensator field compensates additional
symmetries serving for them as a Higgs field. The terminology is
borrowed from application of conformal supersymmetry for the
analysis of Poincar\'e supermultiplets (see {\it e.g.} \cite{comp}).
It should \textit{not} be confused with the homonymous - but
unrelated - gauge field introduced in another approach to free HS
fields \cite{compensator,tenslimstr}.}. The fiber indices $A,B$  now
run from $0$ to $d$. They are raised and lowered by the invariant
mostly minus metric $\eta_{AB}$ of ${o}(d-1,2)\,$ (see Appendix
\ref{iso}).

In this subsection, the MacDowell-Mansouri-Stelle-West (MMSW) action
\cite{SW} is written and it is shown how to recover the action
presented in the previous subsection. The particular vacuum solution
which corresponds to $AdS$ space-time is also introduced. Finally
the symmetries of the MMSW action and of the vacuum solution are
analyzed.\vspace{.2cm}

To have a formulation   with manifest
 $o(d-1,2)$  gauge symmetries, a time-like vector compensator $V^A$  has
to be introduced which is constrained to have
 a constant norm $\r$, \bqn V^A V^B \eta_{AB} = \rho^2\,.
\label{norm} \eqn As one will see, the constant $\r$ is related to
the cosmological constant by \be \label{rco} \r^2=-\Lambda^{-1}\,.
\ee

The MMSW action is (\cite{SW} for $d=4$ and \cite{5d} for
arbitrary $d$) \bqn
S^{MMSW}[\,\o^{AB},V^A]=-\frac{\r}{4\kappa^2}\int_{{\cal
M}^d}\epsilon_{A_1 \ldots A_{d+1}}R^{A_1 A_2} R^{A_3 A_4} E^{A_5}
\ldots E^{A_d} V^{A_{d+1}}\,, \label{mmswaction}\eqn where the
curvature or field strength $R^{AB}$ is defined by $$ R^{AB}\equiv
d\o^{AB}+\o^{AC}\o_C^{~~B} $$ and the frame field $E^A$ by \be
\label{fram} E^A\equiv D V^A= d V^A+ \o^A_{~B} V^B \,. \nonumber
\ee Furthermore, in order to make link with Einstein gravity, two
constraints are imposed: (i) the norm of $V^A$ is fixed, and (ii)
the frame field $E^A_\mu$ is assumed to have maximal rank equal to
$d$. As the norm of $V^A$ is constant, the frame field satisfies
\be E^AV_A=0\,. \label{EV} \ee If the condition (\ref{norm}) is
relaxed, then the norm of $V^A$ corresponds to an additional
dilaton-like field \cite{SW}.

Let us now analyze the symmetries of the MMSW action. The action is
manifestly invariant under
\begin{itemize}
\item Local ${o}(d-1,2)$ transformations: \be\d \o_\n^{AB} (x)=
D_\n \epsilon^{AB}(x)\,,\qquad\d V^A (x)= -\epsilon^{AB}(x) V_B (x)
\label{localo}\,;\ee \item Diffeomorphisms: \be\d \o^{AB}_{\n}(x)=
\pa_{\n}\xi^{\m}(x)\,\o^{AB}_{\m}(x)\,+\,\xi^{\m}(x)\,\pa_{\m}\o^{AB}_{\n}(x)\,,
\qquad\d V^A(x)=\xi^{\n}(x)\,\pa_{\n}V^A(x)\,.\label{diffs}\ee
\end{itemize}
Let us define the covariantized diffeomorphism as the sum of a
diffeomorphism with parameter $\xi^\m$ and a $o(d-1,2)$ local
transformation with parameter
$\epsilon^{AB}(\xi^{\m})=-\xi^{\m}\o_{\m}^{AB}$. The action of this
transformation is thus \be \d^{cov}
\o^{AB}_{\m}=\xi^{\n}R_{\n\m}^{AB}\,,\qquad\d^{cov} V^A=\xi^{\n}
E_{\n}^A \label{covd} \ee by (\ref{localo})-(\ref{diffs}).

The compensator vector is pure gauge. Indeed, by local $O(d-1,2)$
rotations one can gauge fix $V^A (x)$ to any value with $V^A(x) V_A
(x)=\r^2$. In particular, one can reach the standard gauge \be
\label{stgau} V^A=\r\,\d^A_{\hat{d}}\,. \ee Taking into account
(\ref{EV}), one observes that the covariantized diffeomorphism also
makes it possible to gauge fix fluctuations of the compensator
$V^A(x)$ near any fixed value. Because the full list of symmetries
can be represented  as a combination of covariantized
diffeomorphisms, local Lorentz transformations and diffeomorphisms,
in the standard gauge (\ref{stgau}) the
 gauge symmetries are spontaneously
broken to the  $o(d-1,1)$  local Lorentz symmetry and
 diffeomorphisms. In the standard gauge, one
therefore recovers the field content and the gauge symmetries of the
MacDowell-Mansouri action. Let us note that covariantized
diffeomorphisms (\ref{covd}) do not affect the connection
$\o^{AB}_{\m}$ if it is flat ({\it i.e.} has zero curvature
$R_{\n\m}^{AB}$). In particular covariantized diffeomorphisms do not
affect the background $AdS$ geometry. \vspace{.2cm}

To show the equivalence of the action  (\ref{mmswaction}) with the
action (\ref{mmactd}), it is useful to define a Lorentz connection
by \be\o^{{L}\; AB}\equiv \o^{AB} -\r^{-2}
(E^AV^B-E^BV^A)\,.\label{lorentzcosm}\ee In the standard gauge, the
curvature can be expressed in terms of the vielbein $e^a\equiv E^a=
\rho \, \o^{a\hat d}$ and the nonvanishing components of the Lorentz
connection $\o^{{L}\; ab}= \o^{ab}$ as \bqn
R^{ab}&=&d\o^{ab}+\o^{aC}  \o_C^{~~b}=d \o^{{L}\; ab}+\o^{{L}\;
a}_{~~~c}  \o^{{L}\; cb}-\r^{-2} e^a  e^b= R^{{L}\; ab}+ R^{cosm\;
ab}\,,\nonumber\\
R^{a\hat{d}}&=& d\o^{a\hat{d}}+\o^{ac}  \o_c^{~~\hat{d}}=
\r^{-1}T^a\,.\nonumber \eqn Inserting these gauge fixed expressions
into the MMSW action yields the action (\ref{mmactd}), where
$\Lambda =-\rho^{-2}$. The MMSW action is thus equivalent to
(\ref{mmactd}) by partially fixing the gauge invariance.
Let us note that a version of the covariant compensator
formalism  applicable to the case with zero cosmological constant
was developed in \cite{ivned}. \vspace{.2cm}

Let us now consider the vacuum equations $R^{AB}(\o_0)= 0$. They
are equivalent to $T^a(\o_0)= 0$ and $R^{ab}(\o_0)= 0$ and, under
the condition that rank$(E_{\n}^A)=d\,$, they uniquely define the
local geometry of $AdS_d$ with parameter $\r$, in a coordinate
independent way. The solution $\o_0$ also obviously satisfies the
equations of motion of the MMSW action. To  find the symmetries of
the vacuum solution $\o_0$, one first notes that vacuum solutions
are sent onto vacuum solutions by diffeomorphisms and local $AdS$
transformations, because they transform the curvature
homogeneously. Since covariantized diffeomorphisms do not affect
$\o_0$, to find symmetries of the chosen solution $\o_0$ it is
enough to check its transformation law under local $o(d-1,2)$
transformation. Indeed, by adjusting an appropriate covariantized
diffeomorphism it is always possible to keep the compensator
invariant.

The solution $\o_0$ is invariant under those $o(d-1,2)$ gauge
transformations for which the parameter $\epsilon^{AB}(x)$ satisfies
\be 0= D_{0} \epsilon^{AB}(x)=d\epsilon^{AB}(x)+
\o_0{}^{A}{}_C(x)\,\epsilon^{CB}(x)-\o_0{}^{B}{}_C(x)\,\epsilon^{CA}(x)\,.
\label{AdSsol}\ee This equation fixes the derivatives $
\partial_\m\epsilon^{AB}(x)$ in terms of $\epsilon^{AB}(x)$ itself. In other
words, once $ \epsilon^{AB}(x_0)$ is chosen for some $x_0$,
$\epsilon^{AB}(x)$ can be reconstructed for all $x$ in a
neighborhood of $x_0$, since by consistency\footnote{The identity
$D_0^2=R_0=0$ ensures consistency of the system (\ref{AdSsol}),
which is overdetermined because it contains $\frac{d^2(d+1)}{2}$
equations for $\frac{d(d+1)}{2}$ unknowns. Consistency in turn
implies that higher space-time derivatives $\partial_{\nu_1} \ldots
\partial_{\nu_n} \epsilon^{AB}(x)$
obtained by hitting (\ref{AdSsol}) $n-1$ times with $D_{0\nu_k}$ are
guaranteed to be symmetric in the indices $\nu_1\ldots \nu_n$.}
   {\it all} derivatives of the parameter can be expressed as functions
of the parameter itself. The parameters $\epsilon^{AB}(x_0)$ remain
arbitrary, being parameters of the global symmetry $o(d-1,2)$. This
means that, as expected for $AdS_d$ space-time, the symmetry of the
vacuum solution $\o_0$ is the global $o(d-1,2)$. \vspace{.2cm}

The lesson is that, to describe a gauge model that has a global
symmetry $h$, it is useful to reformulate it in terms of the gauge
connections $\o$ and curvatures $R$ of $h$ in such a way that the
zero curvature  condition $R=0$ solves the field equations and
provides a solution with $h$ as its global symmetry. If a symmetry
$h$ is not known, this observation can be used the other way around:
by reformulating the dynamics \`a la MacDowell-Mansouri one might
guess the structure of an appropriate curvature $R$ and thereby the
nonAbelian algebra $h$.

\section{Young tableaux and Howe duality}
\label{young}

In this section, the Young tableaux are introduced. They
characterize the irreducible representations of ${gl}(M)$ and
$o(M)$. A representation of these algebras that will be useful in
the sequel is built.\vspace{.2cm}

A \textit{Young tableau} $\{n_i\}$ ($i=1, \ldots,p$) is a diagram
which consists of a finite number $p>0$ of rows of identical
squares. The lengths of the rows are  finite and do not increase:
$n_1\geqslant n_2\geqslant \ldots\geqslant n_p\geqslant 0$. The
Young tableau  $\{n_i\}$  is represented as follows:

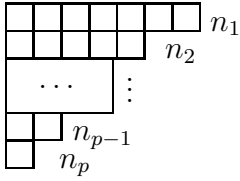
\begin{picture}(75,70)(-20,2)
\multiframe(0,0)(10.5,0){1}(10,10){}\put(20,0){$n_p$}
\multiframe(0,10.5)(10.5,0){2}(10,10){}{}\put(25,10.5){$n_{p-1}$}
\multiframe(0,21)(10.5,0){1}(40,20){\ldots}\put(45,25){$\vdots$}
\multiframe(0,41.5)(10.5,0){5}(10,10){}{}{}{}{} \put(60,41.5){$n_2$}
\multiframe(0,52)(10.5,0){7}(10,10){}{}{}{}{}{}{} \put(77,52){$n_1$}
\end{picture}

\vspace{.2cm}

Let us consider covariant tensors of ${gl}(M)$:
$A_{a,\,b,\,c,\ldots}$ where $a,b,c, \ldots=1,2,\ldots,M$. Simple
examples of these are the symmetric tensor $A^S_{a,\,b}$ such that
$A^S_{a,\,b}-A^S_{b,\,a}=0$, or the antisymmetric tensor $A^A_{a,
\,b}$ such that $A^A_{a,\,b}+A^A_{b,\,a}=0$.

A complete set of covariant tensors irreducible under ${gl}(M)$ is
given by the tensors $A_{a^1_1 \ldots a^1_{n_1},\,\ldots \,,\, a^p_1
\ldots a^p_{n_p} }$ ($n_i \geq n_{i+1}$) that are symmetric in each
set of indices $\{a^i_1 \ldots a^i_{n_{i}}\}$ with fixed $i$ and
that vanish when one symmetrizes the indices of a set $\{a^i_1
\ldots a^i_{n_{i}}\}$ with any index $a^j_l$ with $j>i$. The
properties of these irreducible tensors can be conveniently encoded
into Young tableaux. The Young tableau $\{n_i\}$ ($i=1, \ldots,p$)
is associated with the tensor $A_{a^1_1 \ldots a^1_{n_1},\,\ldots
\,,\, a^p_1 \ldots a^p_{n_p} }$. Each box of the Young tableau is
related to an index of the tensor, boxes of the same row
corresponding to symmetric indices. Finally, the symmetrization of
all the indices of a row with an index from any row below vanishes.
In this way,  the irreducible tensors $ A^S_{ab}$  and $A^A_{a,\,
b}$  are associated with the Young tableaux
\begin{picture}(65,12)(-5,0)
\multiframe(8.5,0)(8.5,0){1}(8,8){}
\multiframe(0,0)(8.5,0){1}(8,8){} \put(23,0){and}
\multiframe(50,4)(8.5,0){1}(8,8){}
\multiframe(50,-4.5)(8.5,0){1}(8,8){}
\end{picture}, respectively.
   \vspace{.2cm}

Let us introduce the polynomial algebra ${\mathbb R}[Y_i^a]$
generated by commuting generators ${Y}_{i}^a$ where $i=1, \ldots p$
; $a=1,\ldots M$. Elements of the algebra ${\mathbb R}[Y_i^a]$ are
of the form
$$A({Y})=A_{a^1_1 \ldots a^1_{n_1},\,\ldots \,,\, a^p_1 \ldots
a^p_{n_p} }{Y}_1^{a^1_1} \ldots {Y}_1^{a^1_{n_1}}
   \,\ldots\, {Y}_p^{a^p_{1}}\ldots
{Y}_p^{a^p_{n_p}}\,.
$$
The condition that $A$ is irreducible under ${gl}(M)$, {\it i.e.} is
a Young tableau $\{n_{i}\}$, can be expressed as
   \bqn
{Y}_{i}^a\frac{\pa}{\pa {Y}_{i}^a } A({Y})&= &n_{i}
A({Y})\,,\nonumber\\
{Y}_{i}^b \frac{\pa}{\pa {Y}_{j}^b } A({Y})&= &0 \,, \hspace{.5cm}i
< j \,,\label{cond1} \eqn where no sum on $i$ is to be understood in
the first equation. Let us first note that, as the generators
${Y}_{i}^{a}$ commute, the tensor $A$ is automatically symmetric in
each set of indices $\{a^i_1 \ldots a^i_{n_{i}}\}$. The operator on
the l.h.s. of the first equation of (\ref{cond1}) then counts the
number of ${Y}_{i}^{a}$'s, the index $i$ being fixed and the index
$a$ arbitrary. The first equation thus ensures that there are
$n_{i}$ indices contracted with ${Y}_{i}$, forming the set $a^i_1
\ldots a^i_{n_{i}}$. The second equation of (\ref{cond1}) is
equivalent to the vanishing of the symmetrization of a set of
indices $\{a^i_1 \ldots a^i_{n_{i}}\}$ with an index $b$ to the
right. Indeed, the operator on the l.h.s.  replaces a generator $
{Y}_{j}^b$ by a generator ${Y}_{i}^{b}$, $j> i$ , thus projecting
$A_{\ldots, \, a^i_1 \ldots a^i_{n_{i}}, \ldots b \ldots}$ on its
component symmetric in $\{a^i_1 \ldots a^i_{n_{i}} b\}$.

Two types of generators appear in the above equations: the
generators \be t^a_b={Y}_{i}^a\frac{\pa}{\pa {Y}_{i}^b } \ee of
${gl}(M)$ and the generators \be l^{j}_{i}={Y}_{i}^a\frac{\pa}{\pa
{Y}_{j}^a } \ee of $gl(p)$. These generators commute \be
[l^{j}_{i},t^a_b]=0 \ee and the algebras $gl(p)$ and $gl(M)$ are
said to be \textit{Howe dual} \cite{Howe}. The important fact is
that the irreducibility conditions (\ref{cond1}) of $gl(M)$ are the
highest weight conditions with respect to $gl(p)$.

When all the lengths $n_{i}$ have the same value $n$,  $A({Y})$ is
invariant under ${sl}(p)  \subset {gl}(p)$. Moreover, exchange of
any two rows of this  rectangular Young tableau only brings a sign
factor $(-1)^n$, as is easy to prove combinatorically. The
conditions (\ref{cond1}) are then equivalent to: \be \left({Y}_{i}^a
\frac{\pa}{\pa {Y}_{j}^a } -\frac{1}{p}\d_{i}^{j} {Y}_{k}^a
\frac{\pa}{\pa {Y}_{k}^a } \right) A({Y})= 0\;. \label{spp} \ee
Indeed,  let us first consider the equation for $i=j$. The operator
${Y}_{k}^a\frac{\pa}{\pa {Y}_{k}^a } $ (where there is a sum over
$k$ and $a$)  counts the total number $m$ of $Y$'s in $A({Y})$,
while ${Y}_{i}^a \frac{\pa}{\pa {Y}_{i}^a }$ counts the number of
$Y_i$'s for some fixed $i$. The condition can only be satisfied if
$\frac{m}{p}$ is an integer, {\it i.e.} $m=np$ for some integer $n$.
As the condition is true for all $i$'s, there are thus $n$ $Y_i$'s
for every $i$. In other words, the tensorial coefficient of $A(Y)$
has $p$ sets of $n$ indices, {\it i.e.} is rectangular. The fact
that it is a Young tableau is ensured by the condition (\ref{spp})
for $i<j$, which is simply the second condition of (\ref{cond1}).
That the condition (\ref{spp}) is true both for $i< j$ and for $i>j$
is a consequence of the simple fact that any finite-dimensional
$sl(p)$ module with zero $sl(p)$ weights (which are differences of
lengths of the rows of the Young tableau) is $sl(p)$ invariant.
Alternatively, this follows from the property that exchange of rows
leaves a rectangular Young tableau invariant.

If there are $p_1$ rows of length $n_1$, $p_2$ rows of length $n_2$,
etc. , then $A({Y})$ is invariant under ${sl}(p_1) \oplus {sl}(p_2)
\oplus sl(p_3) \oplus  \ldots$, as well as under permutations within
each set of $p_i$ rows of length $n_i$.\vspace{.2cm}

To construct irreducible representations of $o(M-N,N)$, one needs to
add the condition that $A$ is traceless\footnote{For $M=2N$ modulo
$4$, the irreducibility conditions also include the
(anti)selfduality conditions on the tensors described by Young
tableaux  with $M/2$ rows. However, these conditions are not used in
the analysis of HS dynamics in this paper.}, which can be expressed
as: \be \frac{\pa^2}{\pa {Y}_{i}^a \pa {Y}_{j}^b}\,\eta^{ab}\,
A({Y})=0 \,, \hspace{.5cm}\forall\, i,j\,,\label{trcond}\ee where
$\eta_{ab}$ is the invariant metric of $o(M-N,N)$. The generators of
$o(M-N,N)$ are given by \be
t_{ab}=\frac12(\eta_{ac}t^c_b-\eta_{bc}t^c_a)\,. \ee They commute
with the generators \be k_{ij}=\eta_{ab}{Y}_{i}^a {Y}_{j}^b \,,
\quad l^{j}_{i}={Y}_{i}^a\frac{\pa}{\pa {Y}_{j}^a }+\frac{M}{2}\delta_i^j\,, \quad
m^{ij}=\eta^{ab}\frac{\pa^2}{\pa {Y}_{i}^a \pa {Y}_{j}^b }\, \ee of
$sp(2p)$. The conditions (\ref{cond1}) and (\ref{trcond}) are
highest weight conditions for the algebra $sp(2p)$ which is Howe
dual to $o(M-N,N)$.

\vspace{.2cm}

In the notation developed here, the irreducible tensors are
manifestly symmetric in groups of indices. This is a convention: one
could as well choose to have manifestly antisymmetric groups of
indices corresponding to columns of the Young tableau. An equivalent
implementation of the conditions for a tensor to be a Young tableau
can be performed in the antisymmetric convention, by taking
fermionic generators ${Y}$.

To end up with this introduction to Young diagrams, we give two
``multiplication rules" of one box with an arbitrary Young tableau.
More precisely, the tensor product of a vector (characterized by one
box) with an irreducible tensor under $gl(M)$ characterized by a
given Young tableau decomposes as the direct sum of irreducible
tensors under $gl(M)$ corresponding to all possible Young tableaux
obtained by adding one box to the initial Young tableau,  e.g.
\begin{center}
\begin{picture}(200,20)(30,0)
\multiframe(0,10)(10.5,0){1}(10,10){}
\multiframe(10.5,10)(10.5,0){1}(10,10){}
\multiframe(0,-0.5)(10.5,0){1}(10,10){} \put(27,11){$\otimes$}
\multiframe(43,10)(10.5,0){1}(10,10){$*$} \put(65,10){$\simeq$}
\multiframe(80,10)(10.5,0){1}(10,10){}
\multiframe(90.5,10)(10.5,0){1}(10,10){}
\multiframe(101,10)(10.5,0){1}(10,10){$*$}
\multiframe(80,-0.5)(10.5,0){1}(10,10){}
\put(120,10){$\oplus$} \multiframe(140,10)(10.5,0){1}(10,10){}
\multiframe(150.5,10)(10.5,0){1}(10,10){}
\multiframe(140,-0.5)(10.5,0){1}(10,10){}
\multiframe(150.5,-0.5)(10.5,0){1}(10,10){$*$}
\put(170,10){$\oplus$} \multiframe(190,10)(10.5,0){1}(10,10){}
\multiframe(200.5,10)(10.5,0){1}(10,10){}
\multiframe(190,-.5)(10.5,0){1}(10,10){}
\multiframe(190,-11)(10.5,0){1}(10,10){$*$}
\put(230,10){.}
\end{picture}
\end{center}

For the (pseudo)orthogonal algebras $o(M-N,N)$, the tensor product
of a vector (characterized by one box) with a traceless  tensor
characterized by a given Young tableau decomposes as the direct sum
of traceless tensors under $o(M-N,N)$ corresponding to all possible
Young tableaux obtained by adding or removing one box from the
initial Young tableau (a box can be removed as a result of
contraction of indices),  e.g.
\begin{center}
\begin{picture}(200,20)(30,0)
\multiframe(0,10)(10.5,0){1}(10,10){}
\multiframe(10.5,10)(10.5,0){1}(10,10){}
\multiframe(0,-0.5)(10.5,0){1}(10,10){} \put(27,11){$\otimes$}
\multiframe(43,10)(10.5,0){1}(10,10){} \put(64,10){$\simeq$}
\multiframe(80,10)(10.5,0){1}(10,10){}
\multiframe(90.5,10)(10.5,0){1}(10,10){}
\multiframe(101,10)(10.5,0){1}(10,10){}
\multiframe(80,-0.5)(10.5,0){1}(10,10){}
\put(120,10){$\oplus$} \multiframe(140,10)(10.5,0){1}(10,10){}
\multiframe(150.5,10)(10.5,0){1}(10,10){}
\multiframe(140,-0.5)(10.5,0){1}(10,10){}
\multiframe(150.5,-0.5)(10.5,0){1}(10,10){}
\put(170,10){$\oplus$} \multiframe(190,10)(10.5,0){1}(10,10){}
\multiframe(200.5,10)(10.5,0){1}(10,10){}
\multiframe(190,-.5)(10.5,0){1}(10,10){}
\multiframe(190,-11)(10.5,0){1}(10,10){}
\put(220,10){$\oplus$} \multiframe(240,10)(10.5,0){1}(10,10){}
\multiframe(240,-.5)(10.5,0){1}(10,10){}
\put(260,10){$\oplus$} \multiframe(280,10)(10.5,0){1}(10,10){}
\multiframe(290.5,10)(10.5,0){1}(10,10){}
\put(310,10){.}
\end{picture}
\end{center}

\section{Free symmetric higher spin gauge fields as one-forms}
\label{fshsgf}

Properties of HS gauge theories are to a large extent determined by
the HS global symmetries of their most symmetric vacua. The HS
symmetry restricts interactions and fixes spectra of spins of
massless fields in HS theories as ordinary supersymmetry  does in
supergravity. To elucidate the structure of a global HS algebra $h$
it is useful to follow the approach in which fields, action and
transformation laws are formulated in terms of the connection of
$h$.

\subsection{Metric-like formulation of higher spins}

The free HS gauge theories were originally formulated in terms of
completely symmetric and double-traceless HS-fields $\varphi_{\n_1
\ldots \n_s}$ \cite{Fron,WF}, in a way analogous to the metric
formulation of gravity (see \cite{Sorokin} for recent reviews
 on the
metric-like formulation of HS theories). In Minkowski space-time
${\mathbb R}^{d-1,1}$, the spin-$s$ Fronsdal action is
\begin{eqnarray}
S^{(s)}_2[\varphi] &=& {1\over 2} \int d^dx \,
\Big(\partial_\nu\varphi_{\mu_1...\mu_s}\partial^\nu
\varphi^{\mu_1...\mu_s}\nonumber
\\
&&\,\,\,\,- {s(s-1)\over 2}\, \partial_\nu
\varphi^{\lambda}{}_{\lambda\mu_3...\mu_s}\partial^\nu
\varphi_\rho{}^{\rho\mu_3...\mu_s} + s(s-1)\,\partial_\nu
\varphi^{\lambda}{}_{\lambda\mu_3...\mu_s}\partial_\rho \varphi^
{\nu\rho\mu_3...\mu_s}
\nonumber\\
&&\,\,\,\,- s\, \partial_\nu
\varphi^{\nu}{}_{\mu_2...\mu_s}\partial_\rho
\varphi^{\rho\mu_2...\mu_s} - {s(s-1)(s-2)\over 4}\,\partial_\nu
\varphi^{\nu\rho}{}_{\rho\mu_2...\mu_s}\partial_\lambda
\varphi_{\sigma}{}^{\lambda\sigma\mu_2...\mu_s}
\Big)\,,\label{Fronsdalact}
\end{eqnarray}
where the metric-like field is double-traceless ($\eta^{\mu_1\mu_2}
\eta^{\mu_3\mu_4}\varphi_{\mu_1...\mu_s}=0$) and has the dimension
of $(length)^{1-d/2}$. This action is invariant under Abelian HS
gauge transformations \bqn \delta \varphi_{\mu_1...\mu_s}
=\partial_{\{\mu_1} \epsilon_{\mu_2...\mu_s\}} \ ,\label{Fronsdalg}
\eqn where the gauge parameter is a completely symmetric and
traceless rank-$(s-1)$ tensor,
$$\eta^{\mu_1\mu_2}\epsilon_{\mu_1...\mu_{s-1}}=0 \ .$$ For spin
$s=2$, (\ref{Fronsdalact}) is the Pauli-Fierz action that is
obtained from the linearization of the Einstein-Hilbert action via
$g_{\mu\nu}=\eta_{\mu\nu}+\kappa\varphi_{\mu\nu}$ and the gauge
transformations (\ref{Fronsdalg}) correspond to linearized
diffeomorphisms. The approach followed in this section is to
generalize the MMSW construction of gravity to the case of free HS
gauge fields in $AdS$  backgrounds. The free HS dynamics will then
be expressed in terms of one-form connections taking values in
certain representations of the $AdS_d$ isometry algebra $o(d-1,2)$.

The procedure is similar to that of Section \ref{grav} for gravity.
The HS metric-like field is replaced by a frame-like field and a set
of connections. These new fields are then united in a single
connection. The action is given in terms of the latter connection
and a compensator vector. It is constructed in such a way that it
reproduces the dynamics of the Fronsdal formulation, once auxiliary
fields are removed and part of the gauge invariance is fixed.
\vspace{.2cm}

\subsection{Frame-like formulation of higher spins}
\label{framelike}

The double-traceless metric-like HS gauge field $\varphi_{\m_1
\ldots \m_s}$ is replaced by a frame-like field $e_ {\m}^{~ a_1
\ldots a_{s-1}}$, a Lorentz-like connection $\o_{\m}^{ ~a_1 \ldots
a_{s-1}, \, b}$ \cite{V80} and a set of connections $\o_{\m}^{ ~a_1
\ldots a_{s-1}, \, b_1 \ldots b_t}$ called {\it extra fields}, where
$t=2,\ldots, s-1$, $s>2$ \cite{Fort1,LV}. All fields $e$, $\o$ are
traceless in the fiber indices $a,b$, which have the symmetry of the
Young tableaux \hspace{.1cm}
\begin{picture}(85,15)(0,2)
\multiframe(0,7.5)(13.5,0){1}(50,7){}\put(55,7.5){$s-1$}
\multiframe(0,0)(13.5,0){1}(35,7){}\put(40,0){$t \hspace{1.3cm},$}
\end{picture}
where $t=0$ for the frame-like field and $t=1$ for the Lorentz-like
connection. The metric-like field arises as the completely symmetric
part of the frame field \cite{V80}, $$ \varphi_{\m_1 \ldots
\m_s}=\r^{3-s-\frac{d}{2}} e_{\{\m_1 ,\, \m_2\ldots \m_s\}}\,,$$
where the dimensionful factor of $\r^{3-s-\frac{d}{2}}$ is
introduced for the future convenience ($\rho$ is a length scale) and
all fiber indices have been lowered using the $AdS$ or flat frame
field $e_0{}^a_\m$ defined in Section \ref{grav}. From the fiber
index tracelessness of the frame field  follows automatically that
the field $\varphi_{\m_1 \ldots \m_s}$ is double traceless.

The frame-like field and other connections are then combined
   \cite{5d} into a connection one-form
$\o^{A_1 \ldots A_{s-1},\, B_1 \ldots B_{s-1}}$ (where $A,B=
0,\ldots,d-1,\hat{d}$) taking values in the irreducible
$o(d-1,2)$-module characterized by the two-row traceless rectangular
Young tableau
\begin{picture}(60,15)(-5,2) \multiframe(0,7.5)(13.5,0){1}(50,7){}
\multiframe(0,0)(13.5,0){1}(50,7){}\put(53,0){}
\end{picture}
of length $s-1$, that is
\begin{eqnarray}
&\omega_\mu^{A_1\ldots A_{s-1},B_1\ldots
B_{s-1}}=\omega_\mu^{\{A_1\ldots A_{s-1}\},B_1\ldots
B_{s-1}}=\omega_\mu^{A_1\ldots A_{s-1},\{B_1\ldots B_{s-1}\}}\,,&\nonumber\\
&\omega_\mu^{\{A_1\ldots A_{s-1},A_s\}\,B_2\ldots B_{s-1}}=0\,,\quad
\omega_\mu^{A_1\ldots A_{s-3}\,C\,}{}_{C,}{}^{B_1\ldots
B_{s-1}}=0\,.&\label{adsy}
\end{eqnarray}

One also introduces a time-like vector $V^A$ of constant norm $\r$.
The component of the connection $\o^{A_1 \ldots A_{s-1},\, B_1
\ldots B_{s-1}}$ that is most parallel to $V^A$ is the frame-like
field
$$E^{ A_1 \ldots A_{s-1}}=\o^{A_1 \ldots A_{s-1},\, B_{1} \ldots
B_{s-1}}V_{B_{1}} \ldots V_{B_{s-1}} \,,
$$
while the less $V$-longitudinal components are the other
connections. Note that the contraction of the connection with more
than $s-1$ compensators $V^A$ is zero by virtue of (\ref{adsy}). Let
us be more explicit in a specific gauge. As in the MMSW gravity
reformulation, one can show that $V^A$ is a pure gauge field and
that one can reach the  standard gauge $V^A=\d_{\hat{d}}^A \r$ (the
argument will not be repeated here). In the standard gauge, the
frame field and the connections are given by \bqn e^{~ a_1 \ldots
a_{s-1}}&=& \r^{s-1} \o^{~a_1 \ldots a_{s-1},\, \hat{d}
\ldots \hat{d}}\,, \nonumber\\
\o^{ ~a_1 \ldots a_{s-1}, \, b_1 \ldots b_t}&=&\r^{s-1-t} \Pi
(\o^{~a_1 \ldots a_{s-1},\, b_1 \ldots b_t\, \hat{d} \ldots
\hat{d}})\,, \nonumber \eqn where the powers of $\rho$ originate
from a corresponding number of contractions with the compensator
vector $V^A$ and $\Pi$ is a projector to the Lorentz-traceless part
of a Lorentz tensor, which is needed for $t\geq 2$. These
normalization factors are consistent with the fact that the
auxiliary fields $\o_\mu^{ ~a_1 \ldots a_{s-1}, \, b_1 \ldots b_t}$
will be found to be expressed via $t$ partial derivatives of the
frame field $e_\mu^{~ a_1 \ldots a_{s-1}}$  at the linearized level.

The linearized field strength or curvature is defined as the
${o}(d-1,2)$ covariant derivative of the connection $\o^{A_1 \ldots
A_{s-1}, B_1 \ldots B_{s-1}}$, {\it {\it i.e.}} by \bqn R_1^{A_1
\ldots A_{s-1}, B_1 \ldots B_{s-1}}&= &
D_0 \o^{A_1 \ldots A_{s-1}, B_1 \ldots B_{s-1}}\nonumber \\
&=&d \o^{A_1 \ldots A_{s-1}, B_1 \ldots B_{s-1}}+ \o_{0~~C}^{~A_1}
\o^{C A_2 \ldots  A_{s-1}, \,B_1 \ldots B_{s-1}} + \ldots \nonumber \\
&& \hspace{3cm}+ \o_{0~~C}^{~B_1}  \o^{A_1 \ldots  A_{s-1}, \,C B_2
\ldots B_{s-1}} + \ldots \;, \label{R1} \eqn where the dots stand
for the terms needed to get an expression symmetric in $A_1 \ldots
A_{s-1}$ and $B_1 \ldots B_{s-1}$, and $\o_{0~B}^{~A} $ is the
$o(d-1,2)$ connection associated to the $AdS$ space-time solution,
as defined in Section \ref{grav}. The connection $\o_\mu^{A_1 \ldots
A_{s-1},\, B_{1} \ldots B_{s-1}}$ has dimension $(length)^{-1}$ in
such a way that the field strength $R_{\m\n}^{A_1 \ldots A_{s-1},\,
B_{1} \ldots B_{s-1}}$ has proper dimension $(length)^{-2}$.

As $(D_0)^2 = R_0=0$, the linearized curvature $R_1$ is invariant
under Abelian gauge transformations of the form \bqn \d_0 \o^{A_1
\ldots A_{s-1},\, B_1 \ldots B_{s-1}}&=&D_0 \epsilon^{A_1 \ldots
A_{s-1},\, B_1 \ldots B_{s-1}} \,  .\label{lingtransfo} \eqn The
gauge parameter $\epsilon^{A_1 \ldots A_{s-1},\, B_1 \ldots
B_{s-1}}$ has the symmetry
\begin{picture}(60,15)(-5,2)
\multiframe(0,7.5)(13.5,0){1}(50,7){}
\multiframe(0,0)(13.5,0){1}(50,7){}
\end{picture} and is traceless.
\vspace{.2cm}

Before writing the action, let us analyze the frame field and its
gauge transformations, in the standard gauge. According to the
multiplication rule formulated in the end of Section \ref{young},
the frame field $e_{\m}^{~a_1 \ldots a_{s-1}}$ contains three
irreducible (traceless) Lorentz components characterized by the
symmetry of their indices:
\begin{picture}(40,12)(0,2)
\multiframe(0,0)(13.5,0){1}(35,7){}\put(38,1){\tiny{$ s$}}
\end{picture}\
\,,
\begin{picture}(55,15)(-5,2)
\multiframe(0,0)(13.5,0){1}(7,7){}\put(9,1){{\tiny $ 1$}}
\multiframe(0,7.5)(13.5,0){1}(35,7){}\put(38,9){\tiny{$ s-1$}}
\end{picture}
and
\begin{picture}(40,12)(0,2)
\multiframe(0,0)(13.5,0){1}(30,7){}\put(33,1){\tiny{$ s-2$}}
\end{picture}\,\,\,\,\,\,, where the last tableau describes the trace
component of the frame field $e_{\m}^{~a_1 \ldots a_{s-1}}$. Its
gauge transformations are given by (\ref{lingtransfo}) and read
$$\d_0 e^{a_1 \ldots a_{s-1}}= D_0^L \epsilon^{a_1 \ldots a_{s-1}} -
e_{0\;c}\epsilon^{a_1 \ldots a_{s-1},\,c} \,.$$ The parameter
$\epsilon^{a_1 \ldots a_{s-1},\,c} $ is a generalized local Lorentz
parameter. It allows us to gauge away the traceless component
\begin{picture}(40,15)(0,2)
\multiframe(0,0)(13.5,0){1}(7,7){}
\multiframe(0,7.5)(13.5,0){1}(35,7){}
\end{picture}
of the frame field. The other two components of the latter just
correspond to a completely symmetric double traceless Fronsdal field
$\varphi_{\mu_1\ldots \mu_s}$. The remaining invariance is then the
Fronsdal gauge invariance (\ref{Fronsdalg}) with a traceless
completely symmetric parameter $\epsilon^{a_1 \ldots a_{s-1}}$.

\subsection{Action of higher spin gauge fields}

For a given spin $s$, the most general $o(d-1,2)$-invariant action
that is quadratic in the linearized curvatures (\ref{R1}) and, for
the rest, built only from the compensator $V^C$ and the background
frame field $E^B_0 = D_0 V^B$
   is \be S^{(s)}_2[\,\o_\mu^{A_1 \ldots
A_{s-1},\, B_1 \ldots
B_{s-1}}\,,\o_0^{AB}\,,V^C\,]\,=\,\frac{1}{2}\,
\sum_{p=0}^{s-2}\,a(s,p)\, S^{(s,p)}[\,\o_\mu^{A_1 \ldots A_{s-1},\,
B_1 \ldots B_{s-1}}\,,\o_0^{AB}\,,\,V^C\,]\,, \label{HSaction} \ee
where $a(s,p)$ is the {\it a priori} arbitrary coefficient of the
term \bqn S^{(s,p)}[\,\o,\o_0 ,V\,]&=&\epsilon_{A_1 \ldots
A_{d+1}}\int_{M^d} E_0^{A_5}\ldots E_0^{A_{d}}V^{A_{d+1}}
V_{C_1}\ldots V_{C_{2(s-2-p)}}\times
\nonumber\\
& &  \quad\quad\quad\times R_1^{A_1 B_1 \ldots B_{s-2},\, A_2 C_1
\ldots C_{s-2-p} D_1\ldots D_p} R_1^{~A_3}{}_{B_1 \ldots
B_{s-2},}{}^{ A_4 C_{s-1-p} \ldots C_{2(s-2-p)}}{}_{D_1\ldots  D_p
}\,. \nonumber \eqn This action is manifestly invariant under
diffeomorphisms, local $o(d-1,2)$ transformations (\ref{localo}) and
Abelian HS gauge transformations (\ref{lingtransfo}), which leave
invariant the linearized HS curvatures (\ref{R1}). Having fixed the
$AdS_d$ background gravitational field $\o_0^{AB}$ and compensator
$V^A$, the diffeomorphisms and the local $o(d-1,2)$ transformations
break down to the $AdS_d$ global symmetry $o(d-1,2)$.

As will be explained in Sections \ref{fda}  and
\ref{freemasslessequ}, the connections $\o_{\m}^{~a_1 \ldots
a_{s-1}, \, b_1 \ldots b_t}$ can be expressed as $t$ derivatives of
the frame-like field, via analogues of the torsion constraint.
Therefore,  to make sure that higher-derivative terms are absent
from the free theory, the coefficients $a(s,p)$ are chosen in such a
way that the Euler-Lagrange derivatives are nonvanishing only for
the frame field and the first connection ($t=1$). All extra fields,
{\it i.e.} the connections $\o_{\m}^{~a_1 \ldots a_{s-1}, \, b_1
\ldots b_t}$ with $t>1$,  should appear only through total
derivatives\footnote{The extra fields show up in the nonlinear
theory and are responsible for the higher-derivatives as well as for
the terms with negative powers of $\Lambda$ in the interaction
vertices.}. This requirement
fixes uniquely the spin-$s$ free action up to a coefficient $b(s)$
in front of the action. More precisely, the coefficient $a(s,p)$ is
essentially a relative coefficient given by \cite{5d}
$$a (s,p) = b (s) (-\Lambda)^{-(s-p-1)}
\frac{(d-5 +2 (s-p-2))!!\, (s-p-1)}{  \,(s-p-2)!}\,, $$ where $b(s)$
is the arbitrary spin-dependent factor.

The equations of motion  for $\o_{\mu}^{~a_1 \ldots a_{s-1}, \, b}$
are equivalent to the ``zero-torsion condition'' $$ R_{1\,A_1 \ldots
A_{s-1} ,\,B_1\ldots B_{s-1}} V^{B_1} \ldots V^{B_{s-1}}=0\,.$$ They
imply that $\o_{\m}^{~a_1 \ldots a_{s-1}, \, b}$ is an auxiliary
field that can be expressed in terms of the first derivative of the
frame field
 modulo a pure gauge part associated with the
symmetry parameter $\epsilon_{\m}^{~a_1 \ldots a_{s-1}, \, b_1
b_2}$. Substituting the found expression for $\o_{\m}^{~a_1 \ldots
a_{s-1}, \, b}$ into the HS action yields an action only expressed
in terms of the frame field and its first derivative, modulo total
derivatives. As gauge symmetries told us, the action actually
depends only on the completely symmetric part of the frame field,
{\it i.e.} the Fronsdal field. Moreover, the action (\ref{HSaction})
has the same gauge invariance as Fronsdal's one, thus it must be
proportional to the Fronsdal action (\ref{Fronsdalact}) because the
latter is fixed up to a front factor by the requirements of being
gauge invariant and of being of second order in the derivatives of
the field \cite{curt}.\vspace{.2cm}

\section{Simplest higher spin algebras}
\label{hsa}

In the previous section, the dynamics of free HS gauge fields has
been expressed as a theory of one-forms, the $o(d-1,2)$ fiber
indices of which have symmetries characterized by two-row
rectangular Young tableaux. This suggests that there exists a
nonAbelian HS algebra $h\supset {o}(d-1,2)$ that admits a basis
formed by a set of elements $T_{A_1\ldots A_{s-1} ,B_1\ldots
B_{s-1}}$ in irreducible representations of ${o}(d-1,2)$
characterized by such Young tableaux. More precisely, the basis
elements $T_{A_1\ldots A_{s-1} ,B_1\ldots B_{s-1}}$ satisfy the
following properties $T_{\{A_1 \ldots A_{s-1},A_s\} B_2\ldots
B_{s-1} } =0$, $T_{A_1 \ldots A_{s-3}C}{}_{C,}{}^{B_1\ldots B_{s-1}
} =0$, and the basis contains the $o(d-1,2)$ basis elements
$T_{A,B}=-T_{B,A}$ such that all generators transform as $o(d-1,2)$
tensors \be [T_{C,D}\, ,\,T_{A_1 \ldots A_{s-1},B_1\ldots B_{s-1} }]
= \eta_{DA_1 } T_{C A_2 \ldots A_{s-1}, B_1\ldots B_{s-1} }
+\ldots\,.\label{tensortransf} \ee

The question is whether a nonAbelian algebra $h$ with these
properties really exists. If yes, the Abelian curvatures $R_1$ can
be understood as resulting from the linearization of the nonAbelian
field curvatures $R=dW+W^2$ of $h$ with the $h$ gauge connection $W
= \omega_0 +\omega$, where $\omega_0$ is some fixed flat ({\it i.e.}
vanishing curvature) zero-order connection of the subalgebra
$o(d-1,2)\subset h$ and $\o$ is the first-order dynamical part which
describes massless fields of various spins.

The HS algebras with these properties were originally found for the
case of $AdS_4$ \cite{FVA,V3,FVau,Konstein:ij} in terms of spinor
algebras. Then this construction was extended to HS algebras in
$AdS_3$  \cite{bl,BBS,Aq} and to $d=4$ conformal HS algebras
\cite{FLA,d4sym} equivalent to the $AdS_5$ algebras of \cite{SS5}.
The $d=7$ HS algebras \cite{Sezgin/Sundell-7} were also built in
spinorial terms. Conformal HS conserved currents in any dimension,
generating HS symmetries with the parameters carrying
representations of the conformal algebra $o(d,2)$ described by
various rectangular two-row Young tableaux, were found in
\cite{KVZ}. The realization of the conformal HS algebra $h$ in any
dimension in terms of a quotient of the universal enveloping
algebra\footnote{A universal enveloping algebra is defined as
follows. Let $\cs$ be the associative algebra that is freely
generated by the elements of a Lie algebra $s$. Let $\ci$ be the
ideal of $\cs$ generated by elements of the form $x y-y x-[x,y]$
($x,y\in s$). The quotient $\cu (s)$= $\cs/\ci$ is called the
universal enveloping algebra of $s$.} $\cu(o(d,2))$ was given by
Eastwood in \cite{Eastwood}. Here we use the construction of the
same algebra  as given in \cite{Vasiliev:2003ev}, which is based on
vector oscillator algebra ({\it i.e.} Weyl algebra).

\subsection{Weyl algebras}

The Weyl algebra $A_{d+1}$ is {the associative algebra} generated by
the oscillators $\hat{Y}^A_i$, where $i=1,2$ and $A=0,1,\ldots d$,
satisfying the commutation
relations\begin{equation}[\hat{Y}^A_i,\hat{Y}^B_j]=\epsilon_{ij}\,
\eta^{AB}\,, \label{qmcommrel}
\end{equation} where $\epsilon_{ij}
=-\epsilon_{j\,i}$ and $\epsilon_{12}=\epsilon^{12} =1$. The
invariant metrics $\eta_{AB}=\eta_{BA}$ and symplectic form
$\epsilon^{ij}$ of $o(d-1,2)$ and $sp(2)$, respectively, are used to
raise and lower indices in the usual manner $A^A = \eta^{AB} A_B $,
$ a^i =\epsilon^{ij}a_j $, $ a_i =a^j \epsilon_{j\,i}\,.$ The Weyl
algebra $A_{d+1}$ can be realized by taking as generators
$$\hat{Y}^A_1=\eta^{AB}\frac{\partial}{\partial X^B}\,,
\quad \hat{Y}_2^A=X^A\,,$$ {\it i.e.} the Weyl algebra is realized
as the algebra of differential operators acting on formal power
series $\Phi(X)$ in the variable $X^A$. One can consider both
complex ($A_{d+1}({\mathbb C})$) and real Weyl ($A_{d+1}({\mathbb
R})$) algebras. One can also construct the (say, real) Weyl algebra
$A_{d+1}({\mathbb R})$ starting from the associative algebra
${\mathbb R}<\hat{Y}^A_i>$ freely generated by the variables
$\hat{Y}^A_i$, {\it i.e.} spanned by all (real) linear combinations
of all possible products of the variables $\hat{Y}^A_i$. The real
Weyl algebra $A_{d+1}$ is realized as the quotient of ${\mathbb R
}<\hat{Y}^A_i>$ by the ideal made of all elements proportional to
$$\hat{Y}^A_i\hat{Y}^B_j-\hat{Y}^B_j\hat{Y}^A_i-\epsilon_{ij} \eta^{AB}\,.$$
In order to pick one representative of each equivalence class, we
work with Weyl ordered operators. These are the operators completely
symmetric under the exchange of  $\hat{Y}_i^A$'s. The generic
element of $A_{d+1}$ is then of the form \be \label{genel}
f(\hat{Y}) = \sum_{p=0}^\infty     \phi^{i_1 \ldots i_p}_{A_1 \ldots
A_p}\hat{Y}_{i_1}^{A_1}\ldots \hat{Y}_{i_p}^{A_p}\,, \ee where
$\phi^{i_1 \ldots i_p}_{A_1 \ldots A_p}$ is symmetric under the
exchange $(i_k,A_k) \leftrightarrow (i_l,A_l)$. Equivalently, one
can define basis elements $S^{A_1 \ldots A_m\,,B_1 \ldots B_n}$ that
are completely symmetrized products of $m$ $\hat{Y}_1^A$'s and $n$
$\hat{Y}_2^B$'s ({\it e.g.} $S^{A ,B}=\{\hat{Y}_1^A\ ,
\hat{Y}_2^B\}$), and write the generic element as \be \label{exp}
f(\hat{Y}) = \sum_{m,n} f_{A_1 \ldots A_m\,,B_1 \ldots B_n} S^{A_1
\ldots A_m\,,B_1 \ldots B_n}\,, \ee where the coefficients $f_{A_1
\ldots A_m\,,B_1 \ldots B_n}$ are symmetric in the indices $A_i$ and
$B_j$.

The elements \be T^{AB}=
-T^{BA}=\frac14\,\{\hat{Y}^{Ai}\,,\hat{Y}^B_i\}\, \label{AdSgen} \ee
satisfy the ${o}(d-1,2)$ algebra
\begin{equation}[T^{AB},T^{CD}]=\frac{1}{2} \Big (
\eta^{BC}T^{AD}-\eta^{AC}T^{BD} -\eta^{BD}T^{AC}+\eta^{AD}T^{BC}\Big
)\, \nonumber \end{equation} because of (\ref{qmcommrel}). When the
Weyl algebra is realized as the algebra of differential operators,
then $T^{AB}=X^{[A}\partial^{B]}$ generates rotations of ${\mathbb
R}^{d-1,2}$ acting on a scalar $\Phi(X^A)$.

The operators \bqn t_{ij}=t_{ji}=\frac12 \{\hat{Y}^A_i\
,\hat{Y}^B_j\}\eta_{AB} \label{spgen} \eqn generate $sp(2)$. The
various bilinears $T^{AB}$ and $t_{ij}$ commute \be
[T^{AB},t_{ij}]=0\,, \ee thus forming a Howe dual pair
$o(d-1,2)\oplus sp(2)$.

\subsection{Definition of the higher spin algebras}\label{sp2etc}

Let us consider the subalgebra $\cs$ of elements $f(\hat{Y})$ of the
complex Weyl algebra $A_{d+1} (\mathbb C)$ that are invariant under
$sp(2)$, {\it i.e.} $[f(\hat{Y}),t_{ij}]=0$. Replacing $f(\hat{Y})$
by its Weyl symbol $f({Y})$, which is the ordinary function of
commuting variables $Y$ that has the same power series expansion as
$f(\hat{Y})$ in the Weyl ordering,  the $sp(2)$ invariance condition
takes the form (a simple proof will be given in Section \ref{star})
\be \label{sp2c} \Big (\epsilon_{k\,j} {Y}_{i}^A \frac{\pa}{\pa
{Y}_{k}^A }+ \epsilon_{k\,i} {Y}_{j}^A \frac{\pa}{\pa {Y}_{k}^A
}\Big ) f(Y^A_i )=0\,, \ee which is equivalent to (\ref{spp}) for
$p=2$. This condition implies that the coefficients $f_{A_1 \ldots
A_n,\, B_1 \ldots B_m}$ vanish except when $n=m$, and the
nonvanishing coefficients carry irreducible representations of
$gl(d+1)$ corresponding to two-row rectangular Young tableaux. The
$sp(2)$ invariance condition means in particular that (the symbol
of) any element of $\cs$ is an even function of $Y^A_i$. Let us note
that the r\^ole of $sp(2)$ in our construction is reminiscent of
that of $sp(2)$ in the conformal framework description of dynamical
models (two-time physics) \cite{Marnelius,Bars}.

However, the associative algebra $\cs$ is not simple. It contains
the ideal $\ci$ spanned by the elements of $ \cs$ of the form $g=
t_{ij}\,g^{ij}=g^{ij}\,t_{ij}$. Due to the definition of $t_{ij}$
(\ref{spgen}), all traces of two-row Young tableaux are contained in
$\ci$. As a result, the associative algebra $\ca=\cs / \ci$ contains
only all traceless two-row rectangular tableaux. Let us choose a
basis $\{T_s\}$  of $\cal A$ where the elements $T_s$ carry an
irreducible representation of $o(d-1,2)$ characterized by a two-row
Young tableau with $s-1$ columns:
\begin{picture}(80,15) \put(0,0){$T_s \sim $}
\multiframe(30,7.5)(13.5,0){1}(50,7){}\put(83,9.5){{\tiny $s-1$}}
\multiframe(30,0)(13.5,0){1}(50,7){}\put(83,1.5){{\tiny
$s-1$}}\put(99,0){$\;.$}
\end{picture}

Now consider the complex Lie algebra $h_{\mathbb C}$ obtained from
the associative algebra $\ca$ by  taking the commutator as Lie
bracket, the associativity property of $\ca$ thereby translating
into the Jacobi identity of $h_{\mathbb  C}$. It admits several
inequivalent real forms $h_{\mathbb R}$ such that $h_{\mathbb
C}\,=\,h_{\mathbb R}\,\oplus \,i\, h_{\mathbb R}$. The particular
real form that corresponds to a unitary HS theory is denoted by
$\hu$. This notation\footnote{  This notation was introduced in
\cite{0404124} instead of the more complicated one
${hu}(1/sp(2)[d-1,2])$ of \cite{Vasiliev:2003ev}. } refers to the
Howe dual pair $sp(2)\oplus o(d-1,2)$ and to the fact that the
related spin-$1$ Yang-Mills subalgebra is $u(1)$. The algebra $\hu$
is spanned by the elements satisfying the following reality
condition \be \label{rcond} (f(\hat{Y}))^\dagger=-f(\hat{Y}) \ , \ee
where $\dagger$ is an involution\footnote{This means that $\dagger$
conjugates complex numbers, reverses the order of operators and
squares to unity: $(\mu f)^\dagger = \bar{\mu} f^\dagger$, $
(fg)^\dagger = g^\dagger f^\dagger$, $((f)^\dagger)^\dagger = f$. To
be an involution of the Weyl algebra, $\dagger$ is required to leave
invariant its defining relation (\ref{qmcommrel}). Any condition of
the form (\ref{rcond}) singles out a real linear space of elements
that form a closed Lie algebra with respect to commutators.}  of the
complex Weyl algebra defined by the relation
 \be\label{invo}
(\hat{Y}_i^A)^\dagger=i\hat{Y}_i^A\,. \ee Thanks to the use of  the
Weyl ordering prescription, reversing the order of the oscillators
has no effect so that $(f(\hat{Y}))^\dagger =\bar{f}(i\hat{Y})$
where the bar means complex conjugation of the coefficients in the
expansion (\ref{genel}) \be \bar{f}(\hat{Y}) = \sum_{p=0}^\infty
\bar{\phi}^{i_1 \ldots i_p}_{A_1 \ldots
A_p}\hat{Y}_{i_1}^{A_1}\ldots \hat{Y}_{i_p}^{A_p}\ . \ee As a
result, the reality condition (\ref{rcond}) implies that the
coefficients in front of the generators $T_s$ ({\it i.e.} the basis
elements $S^{A_1 \ldots A_{s-1}\,,B_1 \ldots B_{s-1}}$) with even
and odd $s$ are, respectively, real and pure imaginary. In
particular, the spin-$2$ generator $T^{AB}$ enters with a real
coefficient.

What singles out $\hu$ as the physically relevant real form is that
it allows lowest weight unitary representations to be identified
with the spaces of single particle states in the free HS theory
\cite{0404124}. For these unitary representations (\ref{rcond})
becomes the antihermiticity property of the generators with
$\dagger$  defined via a positive definite  Hermitian form. As was
argued in \cite{V3,FVau,Konstein:ij} for the similar problem in the
case of $d=4$ HS algebras, the real HS algebras that share this
property are obtained by imposing the reality conditions based on an
involution of the underlying complex associative algebra ({\it
i.e.,} Weyl algebra).

Let us note that, as pointed out in \cite{SSS} and
 will be demonstrated below in Section \ref{compsome}, the Lie
algebra with the ideal $\ci$ included, {\it i.e.,} resulting from
the algebra $\cs$ with the commutator as a Lie product and the
reality condition (\ref{rcond}), (\ref{invo}), underlies the
off-mass-shell formulation of the HS gauge theory. We call this
off-mass-shell HS algebra $\huin$.\footnote{ There exist also
intermediate factor algebras $\hup$ with the smaller ideals $\ci_P$
factored out, where $\ci_P$ is spanned by the elements of $\cs$ of
the form $t_{ij} P(c_2) g^{ij}$, where $c_2=\half t_{ij} t^{ij}$ is
the quadratic Casimir operator of $sp(2)$ and $P(c_2 )$ is some
polynomial. The latter HS algebras are presumably of less importance
because they should correspond to HS models  with higher derivatives
in the field equations even at the free field level. }

\subsection{Properties of the higher spin algebras}

The real Lie algebra $\hu$ is infinite-dimensional. It contains the
space-time isometry algebra ${o}(d-1,2)$ as the subalgebra generated
by $T^{AB}$. The basis elements $T_s$ ($s\geqslant 1$) will be
associated with a spin-$s$ gauge field. In Section \ref{pert} we
will show that $\hu$ is indeed a global symmetry algebra of the
$AdS_d$ vacuum solution in the nonlinear HS gauge theory.

Taking two HS generators $T_{s_1}$ and $T_{s_2}$, being
homogeneous polynomials of degrees $2(s_1 -1)$ and $2(s_2-1)$ in
$\hat Y$, respectively, one obtains (modulo some coefficients) \be
\label{ee} [T_{s_1} \,,T_{s_2} ]= \sum_{m=1}^{min(s_1,s_2)-1}
T_{s_1+s_2-2m}=T_{s_1+s_2-2}+T_{s_1+s_2-4}+\ldots +T_{|s_1-s_2|+2}
\,. \ee Let us notice that the formula (\ref{ee}) is indeed
consistent with the requirement (\ref{tensortransf}). Furthermore,
once a gauge field of spin $s>2$ appears, the HS symmetry algebra
requires an infinite tower of HS gauge fields to be present,
together with gravity. Indeed, the commutator $[T_s,T_s]$ of two
spin-$s$ generators gives rise to generators $T_{2s-2}$,
corresponding to a gauge field of spin $s'=2s-2>s$, and also gives
rise to generators $T_2$ of $o(d-1,2)$, corresponding to gravity
fields. The spin-$2$ barrier separates theories with usual
finite-dimensional lower-spin symmetries from those with
infinite-dimensional HS symmetries. More precisely, the maximal
finite-dimensional subalgebra of $\hu$ is the direct sum:
$u(1)\oplus o(d-1,2)$, where $u(1)$ is the center associated with
the elements proportional to the unit. Another consequence of the
commutation relations (\ref{ee}) is that even spin generators
$T_{2p}$ ($p\geqslant 1$) span a proper subalgebra of the HS
algebra,  denoted as $\ho$ in \cite{0404124}.

The general structure of the commutation relations (\ref{ee})
follows from the simple fact that the associative Weyl algebra
possesses an antiautomorphism such that $\rho(f(\hat{Y}))=
f(i\hat{Y})$ in the Weyl ordering. (The difference between $\rho$
and $\dagger$ is that the former does not conjugate complex
numbers.) It induces an automorphism $\tau$ of the Lie algebra $\hu$
with \be\label{even} \tau(f(\hat{Y}))= -f(i\hat{Y})\,. \ee This
automorphism is involutive in the HS algebra, {\it i.e.} $\tau^2 =
identity$, because $f(-\hat{Y})$ = $f(\hat{Y})$. Therefore the
algebra decomposes into subspaces of $\tau$--odd and $\tau$--even
elements. Clearly, these are the subspaces of odd and even spins,
respectively. This determines the general structure of the
commutation relation (\ref{ee}), implying in particular that the
even spin subspace forms the proper subalgebra $\ho\subset \hu$.

Alternatively, the commutation relation (\ref{ee}) can be obtained
from the following reasoning. The oscillator commutation relation
(\ref{qmcommrel}) contracts two $\hat Y$ variables and produces a
tensor $\epsilon_{ij}$. Thus, since the commutator of two
polynomials is antisymmetric, only odd numbers of contractions can
survive. A HS generator $T_s$ is a polynomial of degree $2(s-1)$
in $\hat Y$ with the symmetries associated with the two-row Young
tableau of length $s-1$. Computing the commutator
$[T_{s_1},T_{s_2}]$, only odd numbers $2m-1$ of contractions
survive ($m\geqslant 1$) leading to polynomials of degree
$2(s_1+s_2-2m-1)$ in $\hat Y$. They correspond to two-row
rectangular Young tableaux\footnote{Note that the formal tensor
product of two two-row rectangular Young tableaux contains various
Young tableaux having up to four rows. The property that only
two-row Young tableaux appear in the commutator of HS generators
is a consequence of the $sp(2)$ invariance  condition.} of length
$s_1+s_2-2m-1$ that are associated to basis elements
$T_{s_1+s_2-2m}$. The maximal number, say $2n-1$, of possible
contractions is at most equal to the lowest polynomial degree in
$\hat Y$ of the two generators. Actually, it must be one unit
smaller since the numbers of surviving contractions are odd
numbers while the polynomial degrees of the generators are even
numbers. The lowest polynomial degree in $Y$ of the two generators
$T_{s_1}$ and $T_{s_2}$ is equal to $2\Big(min(s_1,s_2)-1\Big)$.
Hence, $n=min(s_1,s_2)-1$. Consequently, the lowest possible
polynomial degree of a basis element appearing on the
right-hand-side of (\ref{ee}) is equal to
$2(s_1+s_2-2n-1)=2(|s_1-s_2|+1)$. The corresponding generators are
$T_{|s_1-s_2|+2}$.

The gauge fields of $\hu$ are the components of the connection
one-form \be \label{gop}
\o(\hat{Y},x)=\sum_{s=1}^{\infty}dx^{\mu}\,i^{s-2}\o_{\mu~A_1 \ldots
A_{s-1}\,,B_1 \ldots B_{s-1}}(x) \hat{Y}_{1}^{A_1}\ldots
\hat{Y}_{1}^{A_{s-1}}\hat{Y}_{2}^{B_1}\ldots
\hat{Y}_{2}^{B_{s-1}}\,. \ee They take values in the traceless
two-row rectangular Young tableaux of ${o}(d-1,2)$. It is obvious
from this formula why the basis elements $T_s$ are associated to
spin-$s$ fields. The curvature and gauge transformations have the
standard Yang-Mills form \be \label{HScur} R=d\o+ \o^2\Big |_{\ci
\sim 0}\,,\qquad \d \o= D \epsilon\equiv d\epsilon +[\o
,\epsilon]\Big |_{\ci \sim 0}\, \, \ee except that the product of
two elements (\ref{gop}) with traceless coefficients is not
necessarily traceless so that the ideal $\ci$ has to be factored out
in the end. More precisely, the products  $\o^2$ and $[\o
,\epsilon]$ have to be represented as sums of elements of the
algebra with traceless coefficients and  others of the form
$g_{ij}t^{ij}$ (equivalently, taking into account the $sp(2)$
invariance condition, $t^{ij}\tilde{g}_{ij}$). The latter terms have
then to be dropped out,  and the resulting factorization is denoted
by the symbol $\Big |_{\ci \sim 0}$ (see also the discussion in
Subsections \ref{factorization} and \ref{refsi}).

The formalism here presented is equivalent \cite{0404124} to the
spinor formalism developed previously for lower dimensions, where
the HS algebra was realized in terms of commuting spinor oscillators
$\hat{y}^\alpha,\hat{\bar{y}}^{\dot{\alpha}}$ (see, for example,
\cite{9910096,Vasiliev:1995dn} for reviews) as $$
[\hat{y}^\alpha,\hat{y}^\beta]=i\epsilon^{\alpha\beta} \ ,\quad
[\hat{\bar{y}}^{\dot{\alpha}},\hat{\bar{y}}^{\dot{\beta}}]
=i\epsilon^{\dot{\alpha}\dot{\beta}} \ , \quad
[\hat{y}^\alpha,\hat{\bar{y}}^{\dot{\beta}}]=0 \ .$$

Though limited to $d=3,4$, the definition of the HS algebra with
spinorial oscillators is simpler than that with vectorial
oscillators $Y^A_i$, since the generators are automatically
traceless (because $\hat{y}^\alpha
\hat{y}_\alpha=\hat{\bar{y}}^{\dot{\alpha}}
\hat{\bar{y}}_{\dot{\alpha}}=const$), and there is no ideal to be
factored out. However, spinorial realizations of $d=4$ conformal HS
algebras \cite{FLA,d4sym} (equivalent to the $AdS_5$ algebras
\cite{SS5}) and $AdS_7$ HS algebras \cite{Sezgin/Sundell-7} require
the factorization of an ideal.

\section{Free differential algebras and unfolded dynamics}\label{fda}

Subsection \ref{defs} reviews some general definitions of the
unfolded formulation of dynamical systems, a particular case of
which are the HS field equations \cite{Vasfda,Vasfda1}. The strategy
of the unfolded formalism is presented in Subsection \ref{unffda}.
It makes use of free differential {algebras \cite{FDA} in order to
write consistent nonlinear dynamics. In more modern terms the
fundamental underlying concept is $L_\infty$ algebra \cite{stash}.}

\subsection{Definition and examples of free differential algebras}
\label{defs}

Let us consider an arbitrary set of differential  forms
$W^\alpha\in\Omega^{p_\a}({\cal M}^d)$ with degree
$p_\alpha\geqslant 0$ (zero-forms are included), where $\a$ is an
index enumerating various forms, which, generically, may range in
the infinite set $1\leqslant \a<\infty$.

Let $R^\a \in\Omega^{p_\a+1}({\cal M}^d)$ be the generalized
curvatures defined by the relations \be \label{uncur} R^\a= dW^\a
+G^\a(W)\,, \ee where \be G^\a(W^\b)=\sum_{n=1}^\infty f_{\b_1
\ldots \b_n}^\a W^{\b_1}\ldots W^{\b_n} \ee
 are some power series in
$W^\b$ built with the aid of the exterior product of differential
forms. The (anti)symmetry properties of the structure constants
$f_{\b_1 \ldots \b_n}^\a$ are such that $f_{\b_1 \ldots \b_n}^\a\neq
0$ for $p_\a+1 = \sum_{i=1}^n p_{\b_i}$ and the permutation of any
two indices $\b_i$ and $\b_j$ brings a factor of $(-1)^{p_{\b_i}
p_{\b_j} }$ (in the case of bosonic fields,  {\it i.e.} with no
extra Grassmann grading in addition to that of the exterior
algebra).

The choice of a function $G^\a(W^\b)$ satisfying the generalized
Jacobi identity \bqn \label{BI} G^\b \frac{\delta^L G^\a }{\delta
W^\b} \equiv 0\, \label{prop}\eqn (the derivative with respect to
$W^\b$ is left) defines a free differential algebra\footnote{We
remind the reader that a differential $d$ is a Grassmann odd
nilpotent derivation of degree one, {\it i.e.} it satisfies the
(graded) Leibnitz rule and $d^2=0$. A differential algebra is a
graded algebra endowed with a differential $d$. Actually, the
``free differential algebras" (in physicist terminology) are more
precisely christened ``graded commutative free differential
algebra" by mathematicians (this means that the algebra does not
obey algebraic relations apart from graded commutativity). In the
absence of zero-forms (which however play a key role in the
unfolded dynamics construction) the structure of these algebras is
classified by Sullivan \cite{Sullivan}. } \cite{FDA} introduced
originally in the field-theoretical context in \cite{AF}. We
emphasize that the property (\ref{BI}) is a condition on the
function $G^\a (W)$ to be satisfied identically for all $W^\b$. It
is equivalent to the following generalized Jacobi identity on the
structure coefficients \be \label{jid} \sum_{n=0}^{m} (n+1)
f_{[\b_1 \ldots \b_{m-n}}^\gamma  f_{\gamma\b_{m-n+1} \ldots
\b_m\}}^\a =0\,, \ee where the brackets $[\,\}$ denote an
appropriate (anti)symmetrization of all indices $\b_i$. Strictly
speaking, the generalized Jacobi identities (\ref{jid}) have to be
satisfied only at $p_{\a} < d$ for the case of a $d$-dimensional
manifold ${\cal M}^d$ where any $d+1$-form is zero. We will call a
free differential algebra {\it universal} if the generalized
Jacobi identity is true for all values of indices, {\it i.e.,}
independently of a particular value of space-time dimension. The
HS free differential algebras discussed in this paper belong to
the universal class. Note that every universal free differential
algebra defines some $L_\infty$ algebra\footnote{The minor
difference is that
 a form degree $p_\a$ of $W^\a$ is fixed in a universal
free differential algebra
while $W^\a$ in $L_\infty$  are treated as coordinates of
a graded manifold. A universal free differential algebra
can therefore be obtained from $L_\infty$ algebra by an
appropriate projection to specific form degrees.}.

The property (\ref{prop}) guarantees the generalized Bianchi
identity $$ dR^\a = R^\b \frac{\delta^L G^\a }{\delta W^\b}\,, $$
which tells us that the differential equations on $W^\b$ \be
\label{eq} R^\a(W) =0 \ee are consistent with $d^2=0$ and
supercommutativity. Conversely, the property (\ref{BI}) is necessary
for the consistency of the equation (\ref{eq}).

For universal free differential algebras
one defines the gauge transformations as \be \label{delw} \delta
W^\a = d \varepsilon^\a -\varepsilon^\b \frac{\delta^L G^\a }{\delta
W^\b}\,, \ee where $\varepsilon^\a (x) $ has form degree equal to
$p_\a -1$ (so that zero-forms $W^\a$ do not give rise to any gauge
parameter). With respect to these gauge transformations the
generalized curvatures transform as $$ \delta R^\a =-R^\g
\frac{\delta^L }{\delta W^\g} \left (\varepsilon^\b \frac{\delta^L
G^\a }{\delta W^\b} \right )\, $$ due to the property (\ref{BI}).
This implies the gauge invariance of the equations (\ref{eq}). Also,
since the equations (\ref{eq}) are formulated entirely in terms of
differential forms, they are explicitly general coordinate
invariant.

{\it Unfolding} means  reformulating the dynamics of a system into
an equivalent system of the form (\ref{eq}), which, as is
explained below, is always possible by virtue of introducing
enough auxiliary fields. Note that, according to (\ref{uncur}), in
this approach the exterior differential of all fields is expressed
in terms of the fields themselves. A nice property of the
universal free differential algebras is that they allow an
equivalent description of unfolded systems in larger (super)spaces
simply by adding additional coordinates corresponding to a larger
(super)space as was demonstrated for some particular examples in
\cite{d4sym,Sundell,s3,GV}.

Let $h$ be a Lie (super)algebra, a basis of which is the set
$\{T_\a\}$. Let $\o=\o^\a T_\a$ be a one-form taking values in $h$.
If one chooses $G (\,\o)=\o^2\equiv \frac{1}{2} \o^\a \o^\b [T_\a ,
T_\b ]$, then the equation (\ref{eq}) with $W=\o$ is the
zero-curvature equation $d\o+\o^2=0$. The relation (\ref{BI})
amounts to the usual Jacobi identity for the Lie (super)algebra $h$
as is most obvious from (\ref{jid}) (or its super version).  In the
same way, (\ref{delw}) is the usual gauge transformation of the
connection $\o$.

If the set $W^\a$ also contains some $p$-forms denoted by $\cc^i$
({\it e.g.} zero-forms) and if the functions $G^i$ are linear in
$\o$ and $\cc$, \be \label{lin}
   G^i = \o^\a(T_\a)^i {}_j \cc^j\,,
\ee
   then the relation
(\ref{BI}) implies that the coefficients $(T_\a)^i {}_j$ define some
matrices $T_{\a}$ forming a representation $T$ of $h$, acting in a
module $V$ where the $\cc^i$ take their values. The corresponding
equation (\ref{eq}) is a covariant constancy condition $D_\o \cc=0$,
where $D_\o\equiv d+\o$ is the covariant derivative in the
$h$-module $V$.

\subsection{Unfolding strategy}\label{unffda}

{}From the previous considerations, one knows that the system of
equations \be d\o_0 +\o^2_0 =0\,, \label{fdacur} \ee \be D_{\o_0}
\cc=0 \label{fdar} \ee forms a free differential algebra. The
first equation usually describes a background (for example
Minkowski or $AdS$) along with some pure gauge modes. The
connection one-form $\o_0$ takes value in some Lie algebra $h$.
The second equation may describe nontrivial dynamics if $\cc$ is a
zero-form $C$ that forms an infinite-dimensional $h$-module $T$
appropriate to describe the space of all moduli of solutions ({\it
i.e.}, the initial data). One can wonder how the set of equations
(\ref{fdacur}) and \be D_{\o_0} C=0 \label{fda0} \ee could
describe any dynamics, since it implies that (locally) the
connection $\o_0$ is pure gauge and $C$ is covariantly constant,
so that
\bqn\o_0(x)&=&g^{-1}(x)\,dg(x)\,, \label{pureg1} \\
C(x)&=&g^{-1}(x)\,\cdot\, \mathrm{C}\,,\label{pureg}\eqn where
$g(x)$ is some function of the position $x$ taking values in the
 Lie group $H$ associated with $h$ (by exponentiation),
$\mathrm{C}$ is a constant vector of the $h$-module $T$ and the dot
stands for the corresponding action of $H$ on $T$. Since the gauge
parameter $g(x)$ does not carry any physical degree of freedom, all
physical information is contained in the value
$C(x_0)=g^{-1}(x_0)\cdot\mathrm{C}$ of the zero-form $C(x)$ at a
fixed point $x_0$ of space-time. But as one will see in Section
\ref{unfolding}, if the zero-form $C(x)$ somehow parametrizes all
the derivatives of the original dynamical fields, then, supplemented
with some algebraic constraints (that, in turn, single out an
appropriate $h$-module), it can actually describe nontrivial
dynamics.  More precisely, the restrictions imposed on values of
some zero-forms at a fixed point $x_0$ of space-time can lead to
nontrivial dynamics if the set of zero-forms is rich enough to
describe all space-time derivatives of the dynamical fields at a
fixed point of space-time, provided that the constraints just single
out those values of the derivatives that are compatible with the
original dynamical equations. By knowing a solution (\ref{pureg})
one knows all the derivatives of the dynamical fields compatible
with the field equations and can therefore reconstruct these fields
by analyticity in some neighborhood of $x_0$.

The $p$-forms with $p>0$ contained in $\cc$ (if any) are still pure
gauge in these equations. As will be clear from the examples below,
the meaning of the zero-forms $C$ contained in $\cc$ is that they
describe all gauge invariant degrees of freedom  ({\it e.g.} the
spin-$0$ scalar field, the spin-$1$ Maxwell field strength, the
spin-$2$ Weyl tensor, etc., and all their on-mass-shell nontrivial
derivatives). When the gauge invariant zero-forms are identified
with derivatives of the gauge fields which are $p>0$ forms, this is
expressed by a deformation of the equation (\ref{fdar}) \be D_{\o_0}
\cc= P(\o_0 ) \cc\,, \label{fdadel} \ee where $P(\o_0 ) $ is a
linear operator (depending on $\o_0$ at least quadratically) acting
on $\cc$. The equations (\ref{fdacur}) and (\ref{fdadel}) are of
course required to be consistent, {\it i.e.} to describe some free
differential algebra, which is a deformation of (\ref{fdacur}) and
(\ref{fdar}). If the deformation is trivial, one can get rid of the
terms on the right-hand-side of (\ref{fdadel}) by a field
redefinition. The interesting case therefore is when the deformation
is nontrivial. A useful criterium of whether the deformation
(\ref{fdadel}) is trivial or not is given in terms of the $\sigma_-$
cohomology in Section \ref{taudynamcontent}.

The next step is to interpret the equations (\ref{fdacur}) and
(\ref{fdadel}) as resulting from the linearization of some nonlinear
system of equations with \be W=\o_0 +\cc \label{perturbation}\ee in
which $\o_0$ is some fixed zero-order background field chosen to
satisfy (\ref{fdacur}) while $\cc$ describes first order
fluctuations. Consistency of this identification however requires
nonlinear corrections to the original linearized equations because
the full covariant derivatives built of $W=\o_0 +\cc$ develop
nonzero curvature due to the right hand side of (\ref{fdadel}).
Finding these nonlinear corrections is equivalent to finding
interactions.

This  suggests the following strategy for the analysis of HS gauge
theories:
\begin{description}
    \item[1.] One starts from a space-time with
some symmetry algebra $s$ ({\it e.g.} Poincar\'e or anti-de Sitter
algebra) and a vacuum gravitational gauge field $\o_0$, which is a
one-form taking values in $s$ and satisfying the zero curvature
equations (\ref{fdacur}).
    \item[2.] One reformulates the field equations of a given free
dynamical system in the ``unfolded form" (\ref{fdadel}). This can
always be done in principle (the general procedure is explained in
Section \ref{unfolding}). The only questions are: ``how simple is
the explicit formulation?" and ``what are the modules $T$ of $s$ for
which the unfolded equation (\ref{fda0}) can be interpreted as a
covariant constancy condition?".
    \item[3.] One looks for a  nonlinear free differential
algebra such that (\ref{eq}) with (\ref{perturbation}) correctly
reproduces the free field equations (\ref{fdacur}) and
(\ref{fdadel}) at the linearized level. More precisely, one looks
for some function $G(W)$ verifying (\ref{BI}) and the Taylor
expansion of which around $\o_0$ is given by
$$G(W)=\o_0^2+\Big(\o_0-P(\o_0)\Big)\cc+O(\cc^2)\,,$$
where $\o_0 \cc $ denotes the action of $\o_0$ in the $h$-module
$\cc$ and
   the terms denoted by $O(\cc^2)$ are at least quadratic in
the fluctuation.
\end{description}
It is not {\it a priori} guaranteed that some nonlinear deformation
exists at all. If not, this would mean that no consistent nonlinear
equations exist. But if the deformation is found, then the problem
is solved because the resulting equations are formally consistent,
gauge invariant and generally coordinate invariant\footnote{Note
that any fixed choice of $\o_0$ breaks down the diffeomorphism
invariance to a global symmetry of the vacuum solution. This is why
the unfolding formulation works equally well both in the theories
with fixed background field $\o_0$ and no manifest diffeomorphism
invariance and those including gravity  where $\o_0$ is a zero order
part of the dynamical gravitational field.}
   as a consequence of the general properties of free
differential algebras, and, by construction, they describe the
correct dynamics at the free field level.

To find some nonlinear deformation, one has to address two related
questions. The first one is ``what is a relevant $s$-module $T$ in
which  zero-forms that describe physical degrees of freedom in the
model can take values?" and the second is ``which
infinite-dimensional (HS) extension $h$ of $s$, in which one-form
connections take their values, can act on the $s$-module $T$?". A
natural candidate is a Lie algebra $h$ constructed via commutators
from the associative algebra $\ca$
$$ \ca=\cu(s)/Ann(T)\,, $$ where $\cu(s)$ is the universal
enveloping algebra of $s$ while $Ann(T)$ is the annihilator, {\it
i.e.} the ideal of $\cu(s)$ spanned by the elements which trivialize
on the module $T$. Of course, this strategy may be too naive in
general because not all algebras can be symmetries of a consistent
field-theoretical model and only some subalgebras of $h$ resulting
from this construction may allow a consistent nonlinear deformation.
A useful criterium is the \textit{admissibility condition}
\cite{KV0}
   which requires that there should
be a unitary $h$-module which describes a list of quantum
single-particle states corresponding to all HS gauge fields
described in terms of the connections of $h$. If no such
representation exists, there is no chance to find a nontrivial
consistent (in particular, free of ghosts) theory that admits $h$
as a symmetry of its most symmetric vacuum. In any case, $\cu(s)$
is the reasonable starting point to look for a HS
algebra\footnote{Based on somewhat different arguments, this idea
was put forward by Fradkin and Linetsky in \cite{FLU}.}. It seems
to be most appropriate, however, to search for conformal HS
algebras. Indeed, the associative algebra $\ca$ introduced in
Section \ref{hsa} is a quotient $ \ca=\cu(s)/Ann(T)\,, $ where $T$
represents its conformal realization in $d-1$ dimensions
 \cite{Eastwood,0404124}. The related real Lie
algebra $h$ is $\hu$. The space of single-particle quantum states of
free massless HS fields of Section \ref{freemasslessequ} provides a
unitary module of $\hu$ in which all massless completely symmetric
representations of $o(d-1,2)$ appear just once \cite{0404124}.

\section{Unfolding lower spins}
\label{unfolding}

The dynamics of any consistent system can in principle be rewritten
in the unfolded form (\ref{eq}) by adding enough auxiliary variables
\cite{Vasiliev:1992gr}. This technique is explained in Subsection
\ref{technique}.  Two particular examples of the general procedure
are presented: the unfolding of the Klein-Gordon equation and the
unfolding of gravity,  in Subsections \ref{scalar} and
\ref{unfgrav}, respectively.

\subsection{Unfolded dynamics}\label{technique}

Let $\o_0=e_0^a\,P_a+\frac12 \o_0^{ab}M_{ab}$ be a vacuum
gravitational gauge field taking values in some space-time symmetry
algebra $s$. Let $C^{(0)}(x)$ be a given space-time field satisfying
some dynamical equations to be unfolded. Consider for simplicity the
case where $C^{(0)}(x)$ is a zero-form. The general procedure of
unfolding free field equations goes schematically as follows:

For a start, one writes the equation \be D_0^L C^{(0)} \,=\,
e_0^a\,\,\, C_a^{(1)}\,,\label{unf1} \ee where $D_0^L$ is the
covariant Lorentz derivative and the field $C_a^{(1)}$ is auxiliary.
Next, one checks whether the original field equations for $C^{(0)}$
impose any restrictions on the first derivatives of $C^{(0)}$. More
precisely, some part of $D^L_{0\,,\,\m} C^{(0)}$ might vanish
on-mass-shell ({\it e.g.} for Dirac spinors). These restrictions in
turn impose some restrictions on the auxiliary fields $C_a^{(1)}$.
If these constraints are satisfied by $C_a^{(1)}$, then these fields
parametrize all on-mass-shell nontrivial components of first
derivatives.

Then, one writes for these first level auxiliary  fields an equation
similar to (\ref{unf1})
   \be D^L_0 C^{(1)}_a =  e_0^b\,\, C^{(2)}_{a,b}\,, \label{unf2}\ee
where the new fields $C^{(2)}_{a,b}$ parametrize the second
derivatives of $C^{(0)}$. Once again one checks (taking into account
the Bianchi identities) which components of  the second level fields
$C^{(2)}_{a,b}$ are nonvanishing provided that the original
equations of motion are satisfied.

This process continues indefinitely, leading to a chain of equations
having the form of some covariant constancy condition for the chain
of fields $C^{(m)}_{a_1,a_2,\ldots,a_m}$ ($m\in\mathbb N$)
parametrizing all  on-mass-shell nontrivial derivatives of the
original dynamical field. By construction, this leads to a
particular unfolded equation (\ref{eq}) with $G^i$ in (\ref{uncur})
given by (\ref{lin}). As explained in Section \ref{defs}, this means
that the set of fields realizes some module $T$ of the space-time
symmetry algebra $s$. In other words, the fields
$C^{(m)}_{a_1,a_2,\ldots,a_m}$ are the components of a single field
$C$ living in the infinite-dimensional $s$--module $T$. Then the
infinite chain of equations can be rewritten as a single covariant
constancy condition $D_0C=0$, where $D_0$ is the $s$-covariant
derivative in $T$.

\subsection{The example of the scalar field}\label{scalar}

 As a preliminary to the gravity example considered in the next
subsection, the simplest field-theoretical case of unfolding is
reviewed, {\it i.e.} the unfolding of a massless scalar field,
 which was first described in \cite{Vasiliev:1992gr}.

For simplicity, for the remaining of Section \ref{unfolding}, we
will consider the flat space-time background. The Minkowski solution
can be written as \be \o_0=dx^\m\d_\m^a P_a\label{omeg0}\ee {\it
i.e.} the flat frame is $(e_0)_\m^a=\d_\m^a$ and the Lorentz
connection vanishes. The equation (\ref{omeg0}) corresponds to the
``pure gauge" solution (\ref{pureg1}) with \be g(x)=\exp (x^\mu
\,\delta_\mu ^a\, P_a)\,,\label{gx} \ee where the space-time Lie
algebra $s$ is identified with the Poincar\'e algebra $iso(d-1,1)$.
Though the vacuum $\o_0$ solution has a pure gauge form (\ref{gx}),
this solution  cannot be gauged away because of the constraint
rank$(e_0)=d$ (see Section \ref{MacDowell}).

The ``unfolding" of the massless Klein-Gordon equation \be \Box
C(x)=0\label{KG} \ee is relatively easy to work out, so we give
directly the final result and we comment about how it is obtained
afterwards.

To describe the dynamics of the spin-$0$ massless field $C (x)$, let
us introduce the infinite collection of zero-forms $C_{a_1\ldots
a_n}(x)$ ($n=0,1,2,\ldots$) that are completely symmetric traceless
tensors \be \label{tr} C_{a_1\ldots a_n}=C_{\{a_1\ldots
a_n\}}\,,\quad \eta^{bc}C_{bca_3\ldots a_n}=0\,. \ee The ``unfolded"
version of the Klein-Gordon equation (\ref{KG}) has the form of the
following infinite chain of equations \be \label{un0} d C_{a_1\ldots
a_n } =e_0^b C_{a_1 \ldots a_n b}\quad (n=0,1,\ldots)\,, \ee where
we have used the opportunity to replace the Lorentz covariant
derivative $D_0^L$ by the ordinary exterior derivative $d$. It is
easy to see that this system is formally consistent because applying
$d$ on both sides of (\ref{un0}) does not lead to any new condition,
$$ d^2 C_{a_1\ldots a_n } \,=\,-\,
e_0^b\, dC_{a_1 \ldots a_n b}\,=\, -\,e_0^b\, e_0^c\, C_{a_1 \ldots
a_n bc}\,=\, 0 \quad (n=0,1,\ldots)\ ,$$ since $e_0^b e_0^c=-e_0^c
e_0^b$ because $e_0^b$ is a one-form.
   As we know from Section \ref{defs}, this property
implies that the space $T$ of zero-forms $C_{a_1 \ldots a_n}(x)$
spans some representation of the Poincar\'e algebra $iso(d-1,1)$. In
other words, $T$ is an infinite-dimensional
$iso(d-1,1)$-module\footnote{ Strictly speaking, to apply the
general argument of Section \ref{defs} one has to check that the
equation remains consistent for any flat connection in $iso(d-1,1)$.
It is not hard to see  that this is true indeed.}.

To show that this system of equations is indeed equivalent to the
free massless field equation (\ref{KG}), let us identify the scalar
field $C (x)$ with the member of the family of zero-forms $C_{a_1
\ldots a_n}(x)$ at $n=0$. Then the first two equations of the system
(\ref{un0}) read
$$\partial_\nu C =C_\nu \,,$$
$$\partial_\nu C_\mu= C_{\mu\nu}\,,$$
where we have identified the world and tangent indices via
$(e_0)_\mu^a=\delta_\mu^a$. The first of these equations just tells
us that $C_\nu$ is the first derivative of $C$. The second one tells
us that $C_{\nu\mu}$ is the second derivative of $C$. However,
because of the tracelessness condition (\ref{tr}) it imposes the
Klein-Gordon equation (\ref{KG}). It is easy to see that all other
equations in (\ref{un0}) express highest tensors in terms of the
higher-order derivatives \be \label{hder} C_{\nu_1 \ldots \nu_n}=
\partial_{\nu_1}\ldots\partial_{\nu_n}C
\ee and impose no new conditions on $C$. The tracelessness
conditions (\ref{tr}) are all satisfied once the Klein-Gordon
equation is true. {}From this formula it is clear that the meaning
of the zero-forms $C_{\nu_1 \ldots \nu_n}$ is that they form a basis
in the space of all on-mass-shell nontrivial derivatives of the
dynamical field $C(x)$ (including the derivative of order zero which
is the field $C(x)$ itself).

Let us note that the system (\ref{un0}) without the constraints
(\ref{tr}), which was originally considered in \cite{Shaynk},
remains formally consistent but is dynamically empty just expressing
all highest tensors in terms of derivatives of $C$ according to
(\ref{hder}). This simple example illustrates how algebraic
constraints like tracelessness of a tensor can be equivalent to
dynamical equations.

The above consideration can be simplified further by means of
introducing the auxiliary coordinate $u^a$ and the generating
function $$ C (x,u)=\sum_{n=0}^\infty \frac{1}{n\,!}C_{a_1 \ldots
a_n}(x) u^{a_1} \ldots u^{a_n} $$ with the convention that
$$ C (x,0)=C(x)\,. $$ This generating function accounts
for all tensors $C_{a_1 \ldots a_n}$ provided that the tracelessness
condition is imposed, which in these terms implies that \be
\label{ubox} \Box_u C (x,u)\equiv \frac{\partial}{\partial u^a}
\frac{\partial}{\partial u_a} C =0\,.\ee In other words, the
$iso(d-1,1)$-module $T$ is realized as the space of formal harmonic
power series in $u^a$. The equations (\ref{un0}) then acquire the
simple form \be \label{xu} \frac{\partial}{\partial x^\mu} C
(x,u)=\delta_\mu^a\,\frac{\partial}{\partial u^a} C (x,u)\,. \ee
{}From this realization one concludes that the translation
generators in the infinite-dimensional module $T$ of the Poincar\'e
algebra are realized as translations in the $u$--space, {\it i.e.}
$$P_a=-\frac{\partial}{\partial u^a}\,,$$ for which the equation
(\ref{xu}) reads as a covariant constancy condition (\ref{fdar}) \be
dC (x,u)+e_0^a P_aC (x,u)=0\,.\label{xu2}\ee One can find a general
solution of the equation (\ref{xu2}) in the form \be C (x,u )=C
(x+u,0) =C (0,x+u )\, \label{gensol}\ee from which it follows in
particular that \be \label{tay} C (x)\equiv C
(x,0)=C(0,x)=\sum_{n=0}^\infty \frac{1}{n!} C_{\nu_1\ldots\nu_n}(0)
x^{\nu_1} \ldots x^{\nu_n}\,. \ee From (\ref{tr}) and (\ref{hder})
one can see that this is indeed the Taylor expansion for any
solution of the Klein-Gordon equation which is analytic at $x_0=0$.
Moreover the general solution (\ref{gensol}) expresses the covariant
constancy of the vector $C(x,u)$ of the module $T$,
$$C(x,u )=C(0,x+u)=\exp(-x^\mu \,\delta_\mu ^a\,
P_a)C(0,u)\,.$$ This is a particular realization of the pure gauge
solution (\ref{pureg}) with the gauge function $g(x)$ of the form
(\ref{gx}) and $\mathrm{C}=C(0,u)$.

The example of a free scalar field is so simple that one might think
that the unfolding procedure is always like a trivial mapping of the
original equation (\ref{KG}) to the equivalent one (\ref{ubox}) in
terms of additional variables. This is not true, however, for the
less trivial cases of dynamical systems in nontrivial backgrounds
and, especially, nonlinear systems. The situation here is
   analogous to that in the Fedosov quantization prescription \cite{fed}
which reduces the nontrivial problem of quantization in a curved
background to the standard problem of quantization of the flat phase
space, that, of course, becomes an identity when the ambient space
is flat itself. It is worth to mention that this parallelism is not
accidental because, as one can easily see, the Fedosov quantization
prescription provides a  particular case of the general unfolding
approach \cite{Vasfda} in the dynamically empty situation ({\it
i.e.}, with no dynamical equations imposed).

\subsection{The example of gravity}\label{unfgrav}

The set of fields in Einstein-Cartan's formulation of gravity is
composed of the frame field $e_\m^a$ and the Lorentz connection
$\o_\m^{ab}$. One supposes that the torsion constraint $T_a=0$ is
satisfied, in order to express the Lorentz connection in terms of
the frame field. The Lorentz curvature can be expressed as
$R^{ab}=e_ce_d\,R^{[cd]\,;\,[ab]}$, where $R^{ab\,;\,cd}$ is a rank
four tensor with indices in the tangent space and which is
antisymmetric both in  $ab$ and in $cd$,  having the symmetries of
the tensor product
\begin{picture}(60,16)(0,0)
\multiframe(0,5)(10.5,0){1}(10,10){$a$}
\multiframe(0,-5.5)(10.5,0){1}(10,10){$b$}\put(20,1){$\bigotimes$}
\multiframe(40,5)(10.5,0){1}(10,10){$c$}
\multiframe(40,-5.5)(10.5,0){1}(10,10){$d$}\end{picture}. The
algebraic Bianchi identity $e_b R^{ab}=0$, which follows from the
zero torsion constraint, imposes that the tensor $R^{ab\,;\,cd}$
possesses the symmetries of the Riemann tensor, {\it i.e.}
$R^{[ab\,;\,c]d}=0$. More precisely, it carries an irreducible
representation of $GL(d)$ characterized by the Young tableau
\begin{picture}(30,16)(0,0)
\multiframe(1,4)(10.5,0){2}(10,10){$a$}{$c$}
\multiframe(1,-6.5)(10.5,0){2}(10,10){$b$}{$d$}
\end{picture}
in the antisymmetric basis. The vacuum Einstein equations state that
this tensor is traceless, so that it is actually irreducible under
the pseudo-orthogonal group $O(d-1,1)$ on-mass-shell. In other
words, the Riemann tensor is equal on-mass-shell to the Weyl tensor.

For HS generalization, it is more convenient to use the symmetric
basis. In this convention, the Einstein equations can be written as
\be T^a=0\,,\qquad
R^{ab}\,=\,e_c\,e_d\,C^{ac,\,bd}\,,\label{Einst}\ee where the
zero-form $C^{ac,\,bd}$ is the Weyl tensor in the symmetric basis.
More precisely, the tensor $C^{ac,bd}$ is symmetric in the pairs
$ac$ and $bd$ and it satisfies the algebraic identities
$$C^{\{ac,\,b\}d}=0\,,\qquad \eta_{ac}C^{ac,\,bd}=0\,.$$

Let us now start the unfolding of linearized gravity around the
Minkowski background described by a frame one-form $e^a_0$ and
Lorentz covariant derivative $D^L_0$. The linearization of the
second equation of (\ref{Einst}) is \be\label{HSpin2}
R_1^{ab}\,=\,e_{0\;c}\,e_{0\;d}\,C^{ac,\,bd}\,,\ee where $R_1^{ab}$
is the linearized Riemann tensor. This equation is a particular case
of the equation (\ref{fdadel}). What is lacking at this stage is the
equations containing the differential of the Weyl zero-form
$C^{ac,\,bd}$. Since we do not want to impose any additional
dynamical restrictions on the system, the only restrictions on the
derivatives of the Weyl zero-form $C^{ac,\,bd}$ may result from the
Bianchi identities applied to (\ref{HSpin2}).

{\it A priori}, the first Lorentz covariant derivative of the Weyl
tensor is a rank five tensor in the following representation
\begin{eqnarray}\footnotesize
\begin{picture}(38,15)(0,0)
\multiframe(-50,0)(10.5,0){1}(10,10){}\put(-30,2.5){$\bigotimes$}
\multiframe(-10,0)(10.5,0){2}(10,10){}{}
\multiframe(-10,-10.5)(10.5,0){2}(10,10){}{}\put(20,0){$=$}
\multiframe(40,0)(10.5,0){2}(10,10){}{}
\multiframe(40,-10.5)(10.5,0){2}(10,10){}{}
\multiframe(40,-21)(10.5,0){1}(10,10){}\put(70,0){$\bigoplus$}
\multiframe(90,0)(10.5,0){3}(10,10){}{}{}
\multiframe(90,-10.5)(10.5,0){2}(10,10){}{}
\end{picture}\normalsize
\label{Yougdec}\\\nonumber\end{eqnarray}decomposed according to
irreducible representations of $gl(d)$. Since the Weyl tensor is
traceless, the right hand side of (\ref{Yougdec}) contains only one
nontrivial trace, that is for traceless tensors we have the
$o(d-1,1)$ Young decomposition by adding a three cell hook tableau,
{\it i.e.}
\begin{eqnarray}\footnotesize \nonumber
\begin{picture}(38,15)(0,0)
\multiframe(-50,0)(10.5,0){1}(10,10){}\put(-30,2.5){$\bigotimes$}
\multiframe(-10,0)(10.5,0){2}(10,10){}{}
\multiframe(-10,-10.5)(10.5,0){2}(10,10){}{}\put(20,0){$=$}
\multiframe(40,0)(10.5,0){2}(10,10){}{}
\multiframe(40,-10.5)(10.5,0){2}(10,10){}{}
\multiframe(40,-21)(10.5,0){1}(10,10){}\put(70,0){$\bigoplus$}
\multiframe(90,0)(10.5,0){3}(10,10){}{}{}
\multiframe(90,-10.5)(10.5,0){2}(10,10){}{} \put(135,0){$\bigoplus$}
\multiframe(155,0)(10.5,0){2}(10,10){}{}
\multiframe(155,-10.5)(10.5,0){1}(10,10){}
\end{picture}\nonumber\normalsize\label{lyd}
\\\end{eqnarray}
The linearized Bianchi identity $D_0^L R_1^{ab}=0$ leads to \be
e_{0\;c}e_{0\;d}\,D_0^L C^{ac,\,bd}=0\,.\label{linB1}\ee The
components of the left-hand-side written in the basis $dx^\m dx^\nu
dx^\rho$ have the symmetry property corresponding to the tableau
$$\begin{picture}(38,15)(0,0)
\multiframe(10,9.5)(10.5,0){2}(10,10){$\m$}{$a$}
\multiframe(10,-1)(10.5,0){2}(10,10){$\n$}{$b$}
\multiframe(10,-11.5)(10.5,0){1}(10,10){$\r$}
\end{picture}\sim D_0^L{}_{[\r}C^a{}_\m{}^{\,,b}{}_{\n]}\,,$$
which also contains the single trace part with the symmetry
properties of the three-cell hook tableau.

Therefore the consistency condition (\ref{linB1}) says that in the
decomposition  (\ref{lyd}) of the Lorentz covariant derivative of
the Weyl tensor, the first and third terms vanish and the second
term is traceless and otherwise arbitrary. Let $C^{abf,\,cd}$ be the
traceless tensor corresponding to the second term in the
decomposition (\ref{Yougdec}) of the Lorentz covariant derivative of
the Weyl tensor. This is equivalent to say that \be \label{D1} D_0^L
C^{ac,\,bd}\,=
\,{e_0}_f\,(2C^{acf,\,\,bd}+C^{acb,\,\,df}+C^{acd,\,\,bf})\,,
\ee where the right hand side is fixed
   by the Young symmetry properties of the left hand side
modulo an overall normalization coefficient. This equation looks
like the first step (\ref{unf1}) of the unfolding procedure.
$C^{acf,\,\,bd}$ is irreducible under $o(d-1,1)$.

One should now perform the second step of the general unfolding
scheme and write the analogue of (\ref{unf2}). This process goes on
indefinitely. To summarize the procedure, one can analyze the
decomposition of the $k$-th Lorentz covariant derivatives (with
respect to the  Minkowski vacuum background, so they commute) of the
Weyl tensor $C^{ac,\,bd}$. Taking into account the Bianchi identity,
the decomposition goes as follows
\begin{eqnarray}\footnotesize
\begin{picture}(38,15)(0,0)
\multiframe(-100,0)(10.5,0){1}(50,10){}{} \put(-45,1){$k$}
\put(-30,0){$\bigotimes$} \multiframe(-10,0)(10.5,0){2}(10,10){}{}
\multiframe(-10,-10.5)(10.5,0){2}(10,10){}{}\put(20,0){$\rightarrow$}
\multiframe(40,0)(10.5,0){2}(10,10){}{}{}\multiframe(51,0)(10.5,0){1}(50,10){}{}
\put(106,1){$k+2$} \multiframe(40,-10.5)(10.5,0){2}(10,10){}{}
   \end{picture}\normalsize
\label{Youngdecomp}\\\nonumber\end{eqnarray} As a result, one
obtains
\begin{eqnarray}
D_0^L C^{a_1\ldots a_{k+2},\,b_1b_2}&=&
e_{0\;c}\,\Big(\,(k+2)\,C^{a_1 \ldots a_{k+2}c,\, b_1
b_2}\,+\,C^{a_1 \ldots a_{k+2} b_1,\,b_2c}\,+\,C^{a_1 \ldots a_{k+2}
b_2,\,b_1c}\,\Big)\,,\nn\\
&&\qquad\qquad(0\leqslant k\leqslant
\infty)\,,\label{unfoldingggravity}
\end{eqnarray}
where the fields $C^{a_1\ldots a_{k+2},\,b_1b_2}$ are in the
irreducible representation of $o(d-1,1)$ characterized by the
traceless two-row Young tableau on the right hand side of
(\ref{Youngdecomp}), {\it i.e.}
$$C^{\{a_1\ldots a_{k+2},\,b_1\}b_2}=0\,,\quad
\eta_{a_1a_2}C^{a_1a_2\ldots a_{k+2},\,b_1b_2}=0\,.$$ Note that, as
expected, the system (\ref{unfoldingggravity}) is consistent with
$(D_0^L )^2 C^{a_1\ldots a_{k+2},\,b_1b_2}=0$.

Analogously to the spin-$0$ case, the meaning of the zero-forms
$C^{a_1\ldots a_{k+2},\,b_1b_2}$ is that they form a basis in the
space of all on-mass-shell nontrivial gauge invariant combinations
of the derivatives of the spin-$2$ gauge field.

In order to extend this analysis to nonlinear gravity, one replaces
the background derivative $D_0^L$ and frame field $e_0^a $ by the
full Lorentz covariant derivative $D^L$ and dynamical frame $e^a$
satisfying the zero torsion condition $ D^L e^a =0 $ and \be
\label{D2R} D^L D^L = R\,, \ee where $R$ is the Riemann tensor
taking  values in the adjoint representation of the Lorentz algebra.
The unfolding procedure goes the same way up to the equation
(\ref{D1}) but needs nonlinear corrections starting from the next
step. The reason is that Bianchi identities for (\ref{D1}) and
analogous higher equations give rise to terms nonlinear in $C$ via
(\ref{D2R}) and (\ref{Einst}). All terms of second order in $C$ in
the nonlinear deformation of (\ref{unfoldingggravity}) were obtained
in \cite{Vasfda1} for the case of four dimensions. The problem of
unfolding nonlinear gravity in all orders remains unsolved.

\section{Free massless equations for any spin}
\label{freemasslessequ}

In order to follow the strategy exposed in Subsection \ref{unffda}
and generalize the example of gravity treated along these lines in
Subsection \ref{unfgrav}, we shall start by writing unfolded HS
field equations in terms of the linearized HS curvatures (\ref{R1}).
This result is christened the ``central on-mass-shell theorem".  It
was originally obtained in \cite{Fort1,Vasfda} for the case of $d=4$
and then extended to any $d$ in \cite{LV,5d}. That these HS
equations of motion  indeed reproduce the correct physical degrees
of freedom will be shown later in Section \ref{taucohomology}, via a
cohomological approach  explained in Section \ref{taudynamcontent}.

\subsection{Connection one-form sector}

The linearized curvatures $R_1^{A_1 \ldots A_{s-1},\, B_1 \ldots
B_{s-1}}$ were defined by (\ref{R1}). They decompose  into the
linearized curvatures with Lorentz ({\it i.e.} $V^A$ transverse)
fibre indices which have the symmetry properties associated with the
two-row traceless Young tableau
\begin{picture}(71,15)(-2,2)
\multiframe(0,7.5)(13.5,0){1}(50,7){}\put(55,8.5){{\tiny $s-1$}}
\multiframe(0,0)(13.5,0){1}(35,7){}\put(40,0){{\tiny$t$}}
\end{picture}. It is convenient to use
   the standard gauge $V^A=\d^A_{\hat{d}}$ (from now on we
normalize $V$ to unity). In the Lorentz basis, the linearized HS
curvatures have the form \be R_1^{a_1 \ldots a_{s-1},\, b_1 \ldots
b_{t}}\,=\,D^L_0\o^{a_1 \ldots a_{s-1},\, b_1 \ldots
b_{t}}\,+\,e_{0\;c}\, \o^{a_1 \ldots a_{s-1},\, b_1 \ldots b_{t}c}\,
+\,O(\Lambda)\,.\label{tauminus}\ee For simplicity, in this section
we discard the complicated $\Lambda$--dependent terms which do not
affect the general analysis, {\it i.e.} we present explicitly the
flat-space-time part of the linearized HS curvatures. It is
important to note however that the $\Lambda$--dependent terms in
(\ref{tauminus}) contain only the field $\o^{a_1 \ldots a_{s-1},\,
b_1 \ldots b_{t-1}}$ which carries one index less than the
linearized HS curvatures. The explicit form of the
$\Lambda$--dependent terms is given in \cite{LV}.

For $t=0$, these curvatures generalize the torsion of gravity, while
for $t>0$ the curvature corresponds to the Riemann tensor. In
particular, as we will demonstrate in Section \ref{taucohomology},
the analogues of the Ricci tensor and scalar curvature are contained
in the curvatures with $t=1$ while the HS analog of the Weyl tensor
is contained in the curvatures with $t=s-1$. (For the case of $s=2$
they combine into the level $t=1$ traceful Riemann tensor.)

The first on-mass-shell theorem states that the following free field
equations in Minkowski or $(A)dS$ space-time \be R_1^{a_1 \ldots
a_{s-1},\, b_1 \ldots
b_{t}}\,=\,\d_{t,\,s-1}\,\,\,e_{0\;c}\,e_{0\;d}\,\,C^{a_1 \ldots
a_{s-1}c,\, b_1 \ldots b_{s-1}d}\,,\qquad (0\leqslant t\leqslant
s-1) \label{onmshell} \ee properly describe completely symmetric
gauge fields of generic spin $s\geqslant 2$. This means that they
are equivalent to the proper generalization of the $d=4$ Fronsdal
equations of motion to any dimension, supplemented with certain
algebraic constraints on the auxiliary HS connections which express
the latter via derivatives of the dynamical HS fields. The zero-form
$C^{a_1 \ldots a_s,\, b_1 \ldots b_s}$ is the spin-$s$ Weyl-like
tensor. It is irreducible under $o(d-1,1)$ and is characterized by a
rectangular two-row Young tableau
\begin{picture}(60,15)(-5,2)
\multiframe(0,7.5)(13.5,0){1}(50,7){}\put(53,7.5){$s$}
\multiframe(0,0)(13.5,0){1}(50,7){}\put(53,0){$s$}
\end{picture} . The field
equations generalize (\ref{HSpin2}) of linearized gravity. The
equations of motion put to zero all curvatures  with $t\neq s-1$ and
require $C^{a_1 \ldots a_s,\, b_1 \ldots b_s}$ to be traceless.

\subsection{Weyl zero-form sector}\label{Weylzerosect}

Note that the equations (\ref{onmshell}) result from  the first step
in the unfolding of the Fronsdal equations.\footnote{ Actually, the
action and equations of motion for totally symmetric massless HS
fields in $AdS_d$ with
   $d>4$ were originally obtained in \cite{LV} in the
frame-like formalism.} The analysis of the Bianchi identities of
(\ref{onmshell}) works for any spin $s\geqslant 2$ in a way
analogous to gravity. The final result is the following equation
\cite{5d}, which presents itself like a covariant constancy
condition \bqn 0=\widetilde{D}_0C^{a_1 \ldots a_{s+k},\, b_1 \ldots
b_s}&\equiv& D_0^LC^{a_1 \ldots a_{s+k},\, b_1 \ldots
b_s}\nn\\&&-\,e_{0\;c}\Big((k+2)C^{a_1 \ldots a_{s+k}c,\, b_1 \ldots
b_s}\,+\,s\,C^{a_1 \ldots a_{s+k} \{b_1,\,b_2 \ldots
b_s\}c}\Big)+O(\Lambda)\,,\nn\\&&(0\leqslant k\leqslant \infty
)\,,\label{unfHS} \eqn where $C^{a_1 \ldots a_{s+k},\, b_1 \ldots
b_s}$ are $o(d-1,1)$ irreducible ({\it i.e.}, traceless) tensors
    characterized by the Young tableaux
\begin{picture}(85,15)(0,2)
\multiframe(0,7.5)(13.5,0){1}(50,7){}\put(55,7.5){$s+k$}
\multiframe(0,0)(13.5,0){1}(35,7){}\put(40,0){$s$}
\end{picture}
 They describe on-mass-shell nontrivial $k$-th derivatives of the
spin-$s$ Weyl-like tensor, thus forming a basis in the space of
gauge invariant combinations of $(s+k)$-th derivatives of a spin-$s$
HS gauge field. The system (\ref{unfHS}) is the generalization of
the spin-$0$ system (\ref{un0}) and  the spin-$2$ system
(\ref{unfoldingggravity}) to arbitrary spin and to $AdS$ background
(the explicit form of the $\Lambda$--dependent terms is given in
\cite{5d}). Let us stress that for $s \geqslant 2$ the infinite
system of equations (\ref{unfHS}) is a consequence of
(\ref{onmshell}) by the Bianchi identity. For $s=0$ and $s=1$, the
system (\ref{unfHS}) contains the dynamical Klein-Gordon and Maxwell
equations, respectively. Note that (\ref{onmshell}) makes no sense
for $s=0$ because there is no spin-$0$ gauge potential while
(\ref{unfHS}) with $s=0$ reproduces the unfolded spin-$0$ equation
(\ref{un0}) and its $AdS$ generalization. For the spin-$1$ case,
(\ref{onmshell}) only gives a definition of the spin-$1$ Maxwell
field strength $C^{a,b} = - C^{b,a}$ in terms of the potential
$\o_\mu$. The dynamical equations for spin-$1$, {\it i.e.} Maxwell
equations, are contained in (\ref{unfHS}). The fields $C^{a_1 \ldots
a_{k+1},\, b}$, characterized by the Lorentz irreducible ({\it i.e.}
traceless) two-row Young tableaux with one cell in the second row,
form a basis in the space of on-mass-shell nontrivial derivatives of
the Maxwell tensor $C^{a,\,b}$.

It is clear that  the complete set of zero-forms $C^{a_1 \ldots
a_{s+k},\, b_1 \ldots b_s}$ \begin{picture}(85,15)(0,2)
\put(-2,5){$\sim$}
\multiframe(11,7.5)(13.5,0){1}(51,7){}\put(66,7.5){$s+k$}
\multiframe(11,0)(13.5,0){1}(36,7){}\put(51,0){$s$}
\end{picture}
$\,$ covers the set of all two-row Young tableaux. This suggests
that the Weyl-like zero-forms take values in the linear space of
$\hu$, which obviously forms an $o(d-1,1)$-({\it i.e.} Lorentz)
module. Following Sections \ref{defs} and \ref{unffda}, one expects
that the zero-forms belong to an $o(d-1,2)$-module $T$. But the idea
to use the adjoint representation of $\hu$ does not work because,
according to  the commutation relation (\ref{tensortransf}),  the
commutator of the background gravity connection
$\o_0=\o_0^{AB}T_{AB}$ with a generator of $\hu$ preserves the rank
of the generator, while the covariant derivative $\widetilde{D}_0$
in (\ref{unfHS}) acts on the infinite set of Lorentz tensors of
infinitely increasing ranks. Fortunately, the appropriate
representation only requires a slight modification compared to the
adjoint representation. As will be explained in Section \ref{tw},
the zero-forms $C$ belong to the so-called ``twisted adjoint
representation".

Since  $k$ goes from zero to infinity for any fixed $s$ in
(\ref{unfHS}), in agreement with the general arguments of Section
\ref{fda}, each irreducible spin-$s$ submodule of the twisted
adjoint representation is infinite-dimensional. This means that, in
the unfolded formulation, the dynamics of any fixed spin-$s$ field
is described in terms of an infinite set of fields related by the
first-order unfolded equations. Of course, to make it possible to
describe a field-theoretical dynamical system with an infinite
number of degrees of freedom, the set of auxiliary zero-forms
associated with all gauge invariant combinations of derivatives of
dynamical fields should be infinite. Let us note that the
right-hand-side of the equation (\ref{onmshell}) is a particular
realization of the deformation terms (\ref{fdadel}) in free
differential algebras. \vspace{.2cm}

The system of equations (\ref{onmshell})-(\ref{unfHS}) provides the
unfolded form of the free equations of motion for completely
symmetric massless fields of all spins in any dimension. This fact
is referred to as ``central on-mass-shell theorem" because it plays
a distinguished role in various respects. The idea of the proof will
be explained in Section \ref{taucohomology}. The proof is based on a
very general cohomological reformulation of the problem, which is
reviewed in Section \ref{taudynamcontent}.

\section{Dynamical content via $\s_-$ cohomology}
\label{taudynamcontent}

In this section, we perform a very general analysis of equations
of motion of the form \be \widehat{D}_0\cc =0\ ,
\label{sfieldequs} \ee via a cohomological reformulation
\cite{Shaynk,d4sym,s3,GV}  of the problem\footnote{Note that a
similar approach was  applied in a recent paper \cite{BGST} in the
context of $BRST$ formalism.}. It will be applied to the HS
context in the next section.

\subsection{General properties of $\s_-$}

 Let us introduce the number of Lorentz indices as a grading
$G$ of the space of tensors with fiber (tangent) Lorentz indices. In
the unfolded HS equations the background covariant derivative
$\widehat{D}_0$ decomposes as the sum \be
\widehat{D}_0=\s_-+D_0^L+\s_+\,,\label{Dzero} \ee where the operator
$\s_\pm$ modifies the rank of a Lorentz tensor by $\pm 1$ and the
background Lorentz covariant derivative $D_0^L$ does not change it.

In the present consideration we do not fix the module $V$ on which
$\widehat{D}$ acts and, in (\ref{sfieldequs}), $\cc$ denotes a set
of differential forms taking values in $V$. In the context of the HS
theory, the interesting cases are when $\widehat{D}$ acts either  on
the adjoint representation or on the twisted adjoint representation
of the HS algebra. For example, the covariant constancy condition
(\ref{unfHS}) takes the form $$
\widetilde{D}_0C=(D_0^L+\s_-+\s_+)\,C=0\,,$$
where $\s_+$ denotes the $\Lambda$--dependent terms. The
cohomological classification exposed here only assumes the following
abstract properties:
\begin{description}
  \item[(i)] The grading operator $G$ is diagonalizable in the vector space $V$
and it possesses a spectrum bounded from below.
  \item[(ii)]  The grading
properties of the Grassmann odd operators $D_0^L$ and $\s_-$ are
summarized in the commutation relations
$$[G,D_0^L]=0\,,\qquad
[G,\s_-]=-\s_-\,.$$ The operator $\s_+$ is a sum of operators of
strictly positive grade. (In HS applications $\s_+$ has grade one,
{\it i.e.} $[G,\s_+]=\s_+\,$, but this is not essential for the
general analysis.)
  \item[(iii)] The operator $\s_-$ acts
vertically in the fibre $V$,  {\it i.e.} it does not act on
space-time coordinates. (In HS models, only the operator $D_0^L$
acts nontrivially on the space-time coordinates (differentiates).)
  \item[(iv)] The background covariant derivative
$\widehat{D}_0$ defined by (\ref{Dzero}) is nilpotent.
The graded decomposition of the nilpotency equation
$(\widehat{D}_0)^2=0$ gives the following identities \be
(\s_-)^2=0\,,\quad D_0^L\s_-+\s_- D_0^L=0\,,\quad
(D_0^L)^2+\s_+\s_-+\s_-\s_+ + D_0^L \s_+ +\s_+ D_0^L
+(\s_+)^2=0\,.\quad \ee If $\s_+$ has definite grade $+1$ (as is the
case in the HS theories under consideration) the last relation is
equivalent to the three conditions $ (\s_+)^2=0\,,\quad D_0^L\s_+
+\s_+ D_0^L=0\,,\quad (D_0^L)^2+\s_+\s_-+\s_-\s_+ =0\,.$
\end{description}

\noindent An important property is the nilpotency of $\s_-$. The
point is that the analysis of Bianchi identities (as was done in
details in Section \ref{unfgrav} for gravity) is, in fact,
equivalent to the analysis of the cohomology of $\s_-$, that is
$$H(\s_-)\equiv \frac{Ker(\s_-)}{Im(\s_-)}\,.$$

\subsection{Cohomological classification of the dynamical content}

 Let $\cc$ denote a differential form of degree $p$ taking
values in $V$, that is an element of the complex $V\otimes
\Omega^p({\cal M}^d)$. The field equation (\ref{sfieldequs}) is
invariant under the gauge transformations \be\d
\cc=\widehat{D}_0\varepsilon\,,\label{sgginv}\ee since
$\widehat{D}_0$ is nilpotent by the hypothesis (iv). The
 gauge parameter $\varepsilon$
is a $(p-1)$-form. These gauge transformations contain both
differential gauge transformations (like linearized diffeomorphisms)
and Stueckelberg gauge symmetries (like linearized local Lorentz
transformations\footnote{Recall that, at the linearized level, the
metric tensor corresponds to the symmetric part $e_{\{\m\;a\}}$ of
the frame field. The antisymmetric part of the frame field
$e_{[\m\;a]}$ can be gauged away by fixing locally the Lorentz
symmetry, because it contains as many independent components as the
Lorentz gauge parameter $\varepsilon^{cd}$.}).

The following terminology will be used. By {\it dynamical field}, we
mean a field that is not expressed as derivatives of something else
by field equations ({\it e.g.} the frame field in gravity or a
frame-like HS one-form field $\o_{\m}^{ ~a_1 \ldots a_{s-1}}$). The
fields that are expressed by virtue of the field equations as
derivatives of the dynamical fields modulo Stueckelberg gauge
symmeries are referred to as {\it auxiliary fields} ({\it e.g.} the
Lorentz connection in gravity or its HS analogues $\o_{\m}^{ ~a_1
\ldots a_{s-1}, \, b_1 \ldots b_t}$ with $t>0$). A field that is
neither auxiliary nor pure gauge by Stueckelberg gauge symmeries is
said to be a {\it nontrivial dynamical field} ({\it e.g.} the metric
tensor or the metric-like gauge fields of Fronsdal's
approach).\vspace{.2cm}

Let $\cc (x)$ be an element of the complex $V\otimes \Omega^p({\cal
M}^d)$ that satisfies the dynamical equation (\ref{sfieldequs}).
Under the hypotheses (i)-(iv) one can prove the following
propositions \cite{Shaynk,d4sym} (see also \cite{s3,GV}):
\begin{itemize}
    \item[A.] Nontrivial dynamical fields $\cc$ are
     nonvanishing elements of $H^p(\s_-)$.
    \item[B.] Differential gauge symmetry parameters $\varepsilon$ are
    classified by $H^{p-1}(\s_-)$.
    \item[C.] Inequivalent differential field equations on the
     nontrivial dynamical fields contained in $\widehat{D}_0\cc=0$ are in
one-to-one
    correspondence with representatives of $H^{p+1}(\s_-)$.
\end{itemize}

\noindent \underline{Proof of A}: The first claim is almost obvious.
Indeed, let us decompose the field $\cc$ according to the grade $G$:
$$\cc=\sum_{n=0}\cc_n\,,\qquad G\cc_n=n\,\cc_n\,,
\qquad(n = 0,1,2,\ldots)\,.$$ The field equation (\ref{sfieldequs})
thus decomposes as \be \widehat{D}_0\cc|_{\,n-1}=
\s_-\cc_n+D_0^L\cc_{n-1}+    \Big ( \s_+\sum_{m\leqslant
n-2}\cc_{m}\Big)\Big|_{n-1}=0\,. \label{decompsfieldequ}\ee By a
straightforward induction on $n=1,2,\ldots$, one can convince
oneself that all fields $\cc_n$ that contribute to the first term of
the right hand side of the equation (\ref{decompsfieldequ}) are
thereby expressed in terms of derivatives of lower grade ({\it i.e.}
$<n$) fields, hence they are auxiliary\footnote{Here we use the fact
that the operator $\s_-$ acts vertically (that is, it does not
differentiate space-time coordinates) thus giving rise to algebraic
conditions which express auxiliary fields via derivatives of the
other fields.}. As a result only fields annihilated by $\s_-$ are
not auxiliary. Taking into account the gauge transformation
(\ref{sgginv}) \be \d
\cc_n=\widehat{D}_0\varepsilon|_n=\s_-\varepsilon_{n+1}+D_0^L\varepsilon_n+
\Big (\s_+\sum_{m\leqslant n-1} \varepsilon_{m} \Big )\Big |_{n}
\label{proofAB}\ee one observes that, due to the first term in this
transformation law, all components $\cc_n$ which are $\s_-$ exact,
{\it i.e.} which belong to the image of $\s_-$, are Stueckelberg and
they can be gauged away. Therefore, a nontrivial dynamical $p$-form
field in $\cc$ should belong to the quotient
$Ker(\s_-)/Im(\s_-)$.$\qed$ \vspace{.2cm}

\noindent For Einstein-Cartan's gravity, the Stueckelberg gauge
symmetry is the local Lorentz symmetry and indeed what distinguishes
the frame field  from the the metric tensor is that the latter
actually belongs to the cohomology $H^1(\sigma_-)$ while the former
contains a $\s_-$ exact part. \vspace{.3cm}

\noindent \underline{Proof of B}: The proof follows the same lines
as the proof of A. The first step has already been performed in the
sense that (\ref{proofAB}) already told us that the parameters such
that $\s_-\varepsilon\neq 0$ are Stueckelberg and can be used to
completely gauge away trivial parts of the field $\cc$. Thus
differential parameters must be $\s_-$ closed. The only subtlety is
that one should make use of the fact that the gauge transformation
$\d_\varepsilon \cc=\widehat{D}_0\varepsilon$ are reducible. More
precisely, gauge parameters obeying the reducibility identity
\be\varepsilon=\widehat{D}_0\zeta\label{reducid}\ee are trivial in
the sense that they do not perform any gauge transformation,
$\d_{\varepsilon=\widehat{D}_0\zeta} \cc=0$. The second step of the
proof is a mere decomposition of the reducibility identity
(\ref{reducid}) in order to see that $\s_-$ exact parameters
correspond to reducible gauge transformations\footnote{ Note that
factoring out the $\s_-$ exact parameters accounts for algebraic
reducibility of gauge symmetries. The gauge parameters in
$H^{p-1}(\s_-)$ may still have differential reducibility analogous
to differential gauge symmetries for nontrivial dynamical fields.
For the examples of HS systems considered below the issue of
reducibility of gauge symmetries is irrelevant however because there
are no $p$-form gauge parameters with $p>0$.}.$\qed$ \vspace{.3cm}

\noindent \underline{Proof of C}: Given a nonnegative integer number
$n_0$, let us suppose that one has already obtained and analyzed
(\ref{sfieldequs})  in grades ranging from $n=0$ up to $n=n_0-1$.
Let us analyze (\ref{sfieldequs}) in grade $G$ equal to $n_0$ by
looking at the constraints imposed by the Bianchi identities.
Applying the operator $\widehat{D}_0$ on the covariant derivative
$\widehat{D}_0{\cal C}$ gives identically
 zero, which is the Bianchi identity
$(\widehat{D}_0)^2\cc=0$. Decomposing the latter Bianchi identity
gives, in grade equal to $n_0-1$, \be\label{Bidty}
(\widehat{D}_0)^2\cc|_{n_0-1}=\s_-\Big(\widehat{D}_0\cc|_{n_0}\Big)
+D_0^L\Big(\widehat{D}_0\cc|_{n_0-1}\Big)+\Big(\s_+ \sum_{m\leqslant
n_0-2} \widehat{D}_0\cc|_m\Big)\Big|_{n_0-1}=0\,.\ee By the
induction hypothesis, the equations $\widehat{D}_0\cc|_{m}=0$ with
$m\leqslant n_0 -1$ have already been imposed and analyzed.
Therefore (\ref{Bidty}) leads to
$$\s_-\Big(\widehat{D}_0\cc|_{n_0}\Big)=0\,.$$ In other words,
$\widehat{D}_0\cc|_{n_0}$ belongs to $ Ker(\s_-)$. Thus it can
contain a $\s_-$ exact part and a nontrivial cohomology part:
$$\widehat{D}_0\cc|_{n_0}=\s_-(E_{n_0+1})+F_{n_0}\,,\qquad F_{n_0}\in
H^{p+1}(\s_-)\,.$$ The exact part can be compensated by a field
redefinition of the component $\cc_{n_0+1}$ which was not treated
before (by the induction hypothesis). More precisely, if one
performs
$$\cc_{n_0+1}\rightarrow \cc^\prime_{n_0+1}:=\cc_{n_0+1}-E_{n_0+1}\,,$$
then one is left with $\widehat{D}_0\cc^\prime|_{n_0}=F_{n_0}$.
 The field equation (\ref{sfieldequs}) in grade $n_0$ is
$\widehat{D}_0\cc^\prime|_{n_0}=0$. This not only expresses the
auxiliary $p$-forms $\cc^\prime_{n_0+1}$ (that are not annihilated
by $\s_-$) in terms of derivatives of lower grade $p$-forms
$\cc_{k}$ ($k\leqslant n_0$), but also sets $F_{n_0}$ to zero. This
imposes some $\cc_{n_0+1}$-independent conditions on the derivatives
of the fields $\cc_{k}$ with $k\leqslant n_0$, thus leading to
differential restrictions on the nontrivial dynamical fields.
Therefore, to each representative of $H^{p+1}(\s_-)$ corresponds a
differential field equation. $\qed$\vspace{.3cm}

Note that if $H^{p+1}(\s_-) =0$, the equation (\ref{sfieldequs})
contains only constraints which express auxilary fields via
derivatives of the dynamical fields, imposing no restrictions on the
latter. If $D^L_0$ is a first order differential operator and if
$\s_+$ is at most a second order differential operator (which is
true in HS applications) then, if $H^{p+1}(\s_-)$ is nonzero in the
grade $k$ sector, the associated differential equations on a grade
$\ell$ dynamical field are of order $k+1-\ell$.  In the next
section, two concrete examples of operator $\s_-$ will be considered
in many details, together with the physical interpretation of their
cohomologies.

\section{ $\s_-$ cohomology in higher spin gauge theories}
\label{taucohomology}

As was shown in the previous section, the analysis of generic
unfolded dynamical equations amounts to the computation of the
cohomology of $\s_-$.  In this section we apply this technique to
the analysis of the specific case we focus on: the free unfolded HS
gauge field equations of Section \ref{freemasslessequ}. The
computation of the $\s_-$ cohomology groups relevant for the zero
and one-form sectors of the theory is sketched in the subsection
\ref{compsome}. The physical interpretation of these cohomological
results is discussed in the subsection \ref{freunfhsequ}.

\subsection{Computation of some $\s_-$ cohomology groups}
\label{compsome}

 As explained in Section \ref{freemasslessequ}, the fields
entering in the unfolded formulation of the HS dynamics are either
zero- or one-forms both taking values in various two-row traceless
({\it i.e.} Lorentz irreducible) Young tableaux. These two sectors
of the theory have distinct $\s_-$ operators and thus require
separate investigations.

\vspace{.1cm}

Following Section \ref{young}, two-row Young tableaux in the
symmetric basis can be conveniently described as a subspace of the
polynomial algebra ${\mathbb R}[Y^a,Z^b]$ generated by the $2d$
commuting generators $Y^a$ and $Z^b$.  (One makes contact with the
HS algebra convention via the identification of variables $(Y,Z)$
with $(Y_1,Y_2)$. Also let us note that the variable $Z^a$ in this
section has no relation with the variables $Z^A_i$ of sections
\ref{Nonli} and  \ref{pert}.) Vectors of the space $\Omega^p({\cal
M}^d)\otimes {\mathbb R}[Y,Z]$ are $p$-forms taking values in
${\mathbb R}[Y,Z]$. A generic element reads
$$\a=\a_{a_1 \ldots a_{s-1},\, b_1 \ldots b_t}(x,dx){Y}^{a_1} \ldots
{Y}^{a_{s-1}} Z^{b_1} \ldots Z^{b_t} \,,$$ where $\a_{a_1 \ldots
a_{s-1},\, b_1 \ldots b_t}(x,dx)$ are differential forms. The
Lorentz irreducibility conditions of the HS fields are two-fold.
Firstly, there is the Young tableau condition (\ref{cond1}), {\it
i.e.} in this case \be Y^a\frac{\partial}{\partial
Z^a}\,\a=0\,.\label{Cond1} \ee The condition singles out  an
irreducible $gl(d)$-module $W \subset\Omega^p({\cal M}^d)\otimes
{\mathbb R}[Y,Z]$. Secondly, the HS fields are furthermore
irreducible under $o(d-1,1)$, which is equivalent to the
tracelessness condition (\ref{trcond}), {\it i.e.}  in our case it
is sufficient to impose \be \eta^{ab}\frac{\partial^2}{\partial
Y^a\partial Y^b}\,\a=0\,.\label{Cond2} \ee This condition further
restricts to the  irreducible $o(d-1,1)$-module $\hat{W}\subset W$.
Note that from (\ref{Cond1}) and (\ref{Cond2}) follows that all
traces are zero
$$
\eta^{ab}\frac{\partial^2}{\partial Z^a\partial Y^b}\,\a=0\,,\qquad
\eta^{ab}\frac{\partial^2}{\partial Z^a\partial Z^b}\,\a=0\,.
$$

\subsubsection{Connection one-form sector}

By looking at the definition (\ref{tauminus}) of the linearized
curvatures, and taking into account that the $\Lambda$-dependent
terms in this formula denote some operator $\s_+$ that increases the
number of Lorentz indices, it should be clear that the  $\s_-$
operator of the unfolded field equation (\ref{onmshell}) acts as \be
\label{sigmm} \s_-\,\o(Y,Z)\,\propto\, e_0^a\,
\frac{\partial}{\partial Z^a}\,\o(Y,Z)\,. \ee In other words, $\s_-$
is the ``de Rham differential" of the ``manifold" parametrized by
the $Z$-variables where the generators $dZ^a$ of the exterior
algebra are identified with the background vielbein one-forms
$e_0^a$. This remark is very helpful because it already tells us
that the cohomology of $\s_-$ is zero in the space $\Omega^p({\cal
M}^d)\otimes {\mathbb R}[Y,Z]$ with $p>0$ because its topology is
trivial in the $Z$-variable sector. The actual physical situation is
less trivial because one has to take into account the Lorentz
irreducibility properties of the HS fields. Both conditions,
(\ref{Cond1}) and (\ref{Cond2}), do commute with $\s_-$, so one can
restrict the cohomology to the corresponding subspaces.  Since the
topology in $Z$ space is not trivial any more, the same is true for
the cohomology groups.

As explained in Section \ref{taudynamcontent}, for HS equations
formulated in terms of the connection one-forms ($p=1$), the
cohomology groups of dynamical relevance are $H^q (\sigma_-)$ with
$q=0,1$ and 2. The computation of the cohomology groups obviously
increases in complexity as the form degree increases. In our
analysis we consider simultaneously the cohomology $H^p (\sigma_-
,W)$ of traceful two-row Young tableaux ({\it i.e.} relaxing the
tracelessness condition (\ref{Cond2})) and the cohomology $H^p
(\sigma_- ,\hat{W})$ of traceless two-row Young tableaux.
\vspace{.20cm}

\noindent\underline{Form degree zero}: This case corresponds to the
gauge parameters $\varepsilon$. The cocycle condition
$\s_-\varepsilon=0$ states that the gauge parameters do not depend
on $Z$. In addition, they cannot be $\s_-$-exact since they are  at
the bottom of the form degree  ladder. Therefore, the elements of
$H^0(\s_-,W)$ are the completely symmetric tensors
   which correspond to the
unconstrained zero-form gauge parameters
   $\varepsilon (Y)$ in the traceful case
(like in \cite{Francia:2002aa}), while they are furthermore
traceless in $H^0(\s_-,\hat{W})$ and correspond to Fronsdal's gauge
parameters \cite{Fron} $\eta^{ab}\frac{\partial^2} {\partial
Y^a\partial Y^b}\varepsilon (Y)=0\, $ in the traceless case.
\vspace{.20cm}

\noindent\underline{Form degree one}: Because of the Poincar\'e
Lemma, any $\sigma_-$ closed one-form $\alpha (Y,Z)$ admits a
representation \be \alpha (Y,Z)= e_0^a \frac{\partial}{\partial Z^a}
\phi (Y,Z)\,. \label{po1} \ee The right hand side of this relation
should satisfy the Young condition (\ref{Cond1}), {\it i.e.}, taking
into account that it commutes with $\sigma_-$, $$ e_0^a
\frac{\partial}{\partial Z^a} \Big(Y^b\frac{\partial}{\partial
Z^b}\phi(Y,Z)\Big)=0\,. $$ From here it follows that  $\phi (Y,Z)$
is either linear in $Z^a$ (a $Z$-independent $\phi$ does not
contribute to (\ref{po1})) or satisfies the Young property itself.
In the latter case the $\alpha (Y,Z)$ given by (\ref{po1}) is
$\sigma_-$ exact. Therefore, nontrivial cohomology can only appear
in the sector of elements of the form $\alpha (Y,Z)=e_0^a \beta_a
(Y)$, where $\beta_a(Y)$ are arbitrary in the traceful case of $W$
and harmonic in $Y$ in the traceless case of $\hat{W}$. Decomposing
$\beta_a (Y)$ into irreps of $gl(d)$
\begin{eqnarray}\footnotesize \nonumber
\begin{picture}(38,15)(0,0)
\multiframe(-140,0)(10.5,0){1}(10,10){}\put(-120,2){$\bigotimes$}\multiframe(-100,0)(10.5,0){1}(50,10){}{}
\put(-45,2){$s-1$} \put(-15,0){$\cong$}
\multiframe(5,0)(10.5,0){1}(10,10){}{}\multiframe(15,0)(10.5,0){1}(50,10){}{}\put(70,2){$s-1$}\put(100,2){$\bigoplus$}
\multiframe(5,-10.5)(10.5,0){1}(10,10){}{}
\multiframe(120,0)(10.5,0){1}(50,10){}{}\put(180,2){$s$}
\end{picture}\normalsize
\\\nonumber\end{eqnarray}
one observes that the hook ({\it i.e.}, the two-row tableau) is the
$\sigma_-$ exact part, while the one-row part describes $H^1 (\s_- ,
W)$. These are the rank $s$ totally symmetric traceful dynamical
fields which appear in the unconstrained approach
\cite{WF,Francia:2002aa}.

In the traceless case, decomposing $\beta_a (Y)$ into irreps of
$o(d-1,1)$ one obtains
\begin{eqnarray}\footnotesize
\begin{picture}(38,15)(0,0)
\multiframe(-200,0)(10.5,0){1}(10,10){}\put(-180,2){$\bigotimes$}\multiframe(-160,0)(10.5,0){1}(50,10){}{}
\put(-105,2){$s-1$} \put(-75,0){$\cong$}
\multiframe(-55,0)(10.5,0){1}(10,10){}{}\multiframe(-45,0)(10.5,0){1}(50,10){}{}\put(10,2){$s-1$}\put(40,2){$\bigoplus$}
\multiframe(-55,-10.5)(10.5,0){1}(10,10){}{}
\multiframe(60,0)(10.5,0){1}(50,10){}{}\put(120,2){$s$}\put(130,2){$\bigoplus$}
\multiframe(150,0)(10.5,0){1}(40,10){}{}\put(200,2){$s-2$}
\end{picture}\normalsize
\label{Youngdecompao}\\\nonumber\end{eqnarray} where all tensors
associated with the various Young tableaux are traceless. Again, the
hook ({\it i.e.} two-row tableau) is the $\sigma_-$ exact part,
while the one-row traceless tensors in (\ref{Youngdecompao})
describe $H^1 (\s_- ,\hat{W})$ which just matches the Fronsdal
fields \cite{Fron} because a rank $s$ double traceless symmetric
tensor is equivalent to a pair of rank $s$ and rank $s-2$ traceless
symmetric tensors. \vspace{.20cm}

\noindent\underline{Form degree two}: The analysis of $H^2(\s_-,W)$
and $H^2(\s_-,\hat{W})$ is still elementary, but a little bit more
complicated than that of $H^0(\s_-)$ and $H^1(\s_-)$. Skipping
technical details we therefore give the final results.

By following a reasoning similar to the one in the previous proof,
one can show that, in the traceful case, the cohomology group
$H^2(\s_-,W)$ is spanned by two-forms of the form \be F\,=\,e_0^a
\,e_0^b\, \frac{\partial^2}{\partial Y^a\partial Z^b}\,C(Y,Z)\,,
\label{h2} \ee where the zero-form $C(Y,Z)$ satisfies the Howe dual
$sp(2)$ invariance conditions \be Z^b\frac{\partial}{\partial Y^b}
C(Y,Z)=0\,,\qquad Y^b\frac{\partial}{\partial Z^b} C(Y,Z)=0\,,
\label{sp2invcond}\ee and, therefore, \be
(Z^b\frac{\partial}{\partial Z^b} - Y^b\frac{\partial}{\partial
Y^b}) C(Y,Z)=0\,. \label{eqdeg} \ee In accordance with the analysis
of Section \ref{young}, this means that \be C(Y,Z)= C_{a_1\ldots
a_s,\,b_1\ldots b_s}Y^{a_1}\ldots Y^{a_s}Z^{b_1}\ldots Z^{b_s}
\,,\label{oftheform} \ee where the
 zero-form components  $ C_{a_1\ldots a_s,\,b_1\ldots b_s} $ have
the symmetry properties corresponding to  the rectangular two-row
Young tableau of length $s$ \be \label{wt} C_{a_1\ldots
a_s,\,b_1\ldots b_s}\sim
\begin{picture}(70,15)(-5,2)
\multiframe(0,7.5)(13.5,0){1}(50,7){}\put(55,7.5){$s$}
\multiframe(0,0)(13.5,0){1}(50,7){}\put(55,0){$s$}
\end{picture} \ .\ee
{}From (\ref{h2}) it is clear that $F$ is $\s_- $ closed and $F\in
W$. It is also clear that it is not $\s_-$ exact in the space $W$.
Indeed, suppose that $F=\s_- G$, $G\in W$. For any polynomial $G\in
W$ its power in $Z$ cannot be higher than the power in $Y$ (because
of the Young property, the second row of a Young tableau is not
longer than the first row).  Since $\s_-$ decreases the power in
$Z$, the degree in $Z$ of exact elements $\s_- G$ is strictly less
than the degree in $Y$. This is not true for the elements (\ref{h2})
because of the condition (\ref{eqdeg}). The tensors
(\ref{oftheform}) correspond to the linearized curvature tensors
introduced by de Wit and Freedman \cite{WF}.

Let us now consider  the traceless case of $H^2(\s_-,\hat{W})$. The
formula (\ref{h2}) still gives cohomology but now $C(Y,Z)$ must be
traceless,
$$ \eta^{ab}\frac{\partial^2}
{\partial Y^a\partial Y^b}C (Y,Z)=0\,,\qquad
\eta^{ab}\frac{\partial^2} {\partial Z^a\partial Z^b}C
(Y,Z)=0\,,\qquad \eta^{ab}\frac{\partial^2} {\partial Y^a\partial
Z^b}C (Y,Z)=0\,.$$ In this case the tensors (\ref{oftheform})
correspond to Weyl-like tensors, {\it i.e.} on-shell curvatures.
They form the so-called ``Weyl cohomology". But this is not the end
of the story because there are other elements in
$H^2(\s_-,\hat{W})$.  They span the ``Einstein cohomology" and
contain two different types of elements: \be r_1=e_0^ae_0^b\, \Big
(\,(Z_b\,Y^c-Y_b\,Z^c)\,\frac{\partial^2}{
\partial Y^a\partial Y^c}\,\rho_1(Y)\Big )
\label{rtr}\,, \ee \bee r_2&=&e_0^a e_0^b \Big ( (N_Y+d-3)(d-4+2N_Y) Y_a Z_b +(N_Y+d-2) Y_cY^c Z_a \frac{\partial}{\partial Y^b}\nonumber\\
&-&(d-4+2N_Y)Y_c Z^c Y_a
\frac{\partial}{\partial Y^b}+ Y_eY^e  Y_a Z^c
\frac{\partial}{\partial Y^c} \frac{\partial}{\partial Y^b}\Big )
\,\rho_2(Y)\,, \label{rsc} \eee where $N_Y := Y^a \frac{\partial}{\partial Y^a}$ and $\rho_{1,2}(Y)$ are arbitrary
harmonic polynomials $$ \eta^{ab}\frac{\partial^2}{\partial
Y^a\partial Y^b}\rho_{1,2} (Y)=0\ , $$ thus describing completely
symmetric traceless tensors.

One can directly see that $r_1$ and $r_2$ belong to $\hat{W}$ ({\it
i.e.}, satisfy (\ref{Cond1}) and (\ref{Cond2})) and are $\sigma_-$
closed, $\sigma_- r_{1,2}=0$ (the check is particularly simple for
$r_1$). It is also easy to see that $r_{1,2}$ are in the nontrivial
cohomology class. Indeed, the appropriate trivial class is described
in tensor notations by the two-form $e_0^c \omega_{a_1\ldots
a_{s-1},bc}$, where \be \omega_{a_1\ldots
a_{s-1},bc}=e_0^f\omega_{f;\,a_1\ldots a_{s-1},bc} \label{triv2} \ee
   is a one-form that has the properties of traceless two-row Young tableau
with $s-1$ cells in the first row and two cells in the second row.
 The trivial cohomology class neither contains a
rank $s-2$ tensor like $\rho_2$, that needs a double contraction in
$\omega_{f;a_1\ldots a_{s-1},bc}$ in (\ref{triv2}), nor a rank $s$
symmetric tensor  like $\rho_1$, because symmetrization of a
contraction of the tensor $\omega_{f;\,a_1\ldots a_{s-1},bc}$ over
any $s$ indices gives zero. It can be shown that the Einstein
cohomology (\ref{rtr}) and (\ref{rsc}) together with the Weyl
cohomology (\ref{h2}) span $H^2 (\sigma_-,\hat{W})$.

\subsubsection{Weyl zero-form sector}

The $\Lambda$-dependent terms in the formula (\ref{unfHS}) denote
some operator $\s_+$ that increases the number of Lorentz indices,
therefore the operator $\sigma_-$ of the unfolded field equation
(\ref{unfHS}) is given by \be \label{stw} (\s_- {C})^{a_1 \ldots
a_{s+k},\, b_1 \ldots b_s}= -e_{0\;c}\Big((2+k)C^{a_1 \ldots
a_{s+k}c,\, b_1 \ldots b_s}\,+\,s\, C^{a_1 \ldots a_{s+k}
\{b_1,\,b_2 \ldots b_s\}c}\Big) \ee both in the traceless and in the
traceful twisted adjoint representations. One can analogously
compute the cohomology groups $H^q (\s_- , W)$ and $H^q (\s_- ,
\hat{W})$ with $q=0$ and $1$, which are dynamically relevant in the
zero-from sector ($p=0$).

\vspace{.20cm}

\noindent\underline{Form degree zero}: It is not hard to see that
$H^0 (\s_- , W)$ and $H^0 (\s_- , \hat{W})$ consist, respectively,
of the generalized Riemann ({\it i.e.} traceful) and Weyl ({\it
i.e.} traceless) tensors $C^{a_1 \ldots a_{s},\, b_1 \ldots b_s}$.
\vspace{.20cm}

 \noindent\underline{Form degree one}: The cohomology groups
$H^1 (\s_- , W)$ and $H^1 (\s_- , \hat{W})$ consist of one-forms of
the form \be \label{h1c} w^{a_1 \ldots a_{s}, b_1, \ldots b_s}=
e_{0\,c} C^{a_1 \ldots a_{s}\,, b_1 \ldots b_s\,, c}\,, \ee where
$C^{a_1 \ldots a_{s}\,, b_1 \ldots b_s\,, c}$ is, respectively, a
traceful and traceless zero-form described by the three row Young
tableaux with $s$ cells in the first and second rows and one cell in
the third row.  Indeed, a one-form $w^{a_1 \ldots a_{s}, b_1, \ldots
b_s}$ given by (\ref{h1c}) is obviously $\s_-$ closed (by
definition, $\s_-$ gives zero when applied to a rectangular Young
tableau) and not $\s_-$ exact, thus belonging to nontrivial
cohomology.

In addition, in the traceless case, $H^1 (\s_- , \hat{W})$ includes
the ``Klein-Gordon cohomology"\be \label{0ch} w^a = e_0^a k \ee and
the ``Maxwell cohomology"
 \be
\label{1ch} w^{a,\,b} = e_0^a m^b - e_0^b m^a\,, \ee where $k$ and
$m^a$ are arbitrary scalar and vector, respectively. It is obvious
that the one-forms (\ref{0ch}) and (\ref{1ch}) are $\s_-$ closed for
$\s_-$ (\ref{stw}). They are in fact exact in the traceful case with
$w=\s_- (\varphi)$, where \be \label{var} \varphi^{a_1 a_2} \propto
k\eta^{a_1 a_2}\,,\qquad \varphi^{a_1 a_2\,,b}\propto 2 m^b
\eta^{a_1 a_2} -m^{a_2}\eta^{a_1 b}-m^{a_1}\eta^{a_2 b} \ee but are
not exact in the traceless case since the tensors $\varphi$ in
(\ref{var}) are not traceless although the resulting $w=\s_-
(\varphi)$ belong to the space of traceless tensors.

\subsubsection{Unfolded system}

 One may combine the connection one-forms $\o$ and Weyl
zero-forms $C$ into the set $\cc=(\o ,C)$ and redefine
$\s_-\rightarrow \hat{\s}_-$ in such a way that $\hat{\s}_- =\s_-
+\Delta\s_- $ where $\s_-$ acts on $\o$ and $C$ following
(\ref{sigmm}) and (\ref{stw}), respectively, and  $\Delta\s_- $ maps
the rectangular zero-form Weyl tensors to the two-form sector via
\be\label{glu} \Delta{\s}_-\,C(Y,Z)\,\propto\,e_0^a \,e_0^b\,
\frac{\partial^2}{\partial Y^a\partial Z^b}\,C(Y,Z)\,, \ee thus
adding the term with the  Weyl zero-form to the linearized curvature
of one-forms.

The dynamical content of the unfolded system of equations
(\ref{onmshell})-(\ref{unfHS}) is encoded in the cohomology groups
of $\hat{\s}_-$. The gauge parameters are those of $H^0(\hat{\s}_-)$
in the adjoint module ({\it i.e.} the connection one-form sector).
The dynamical fields are symmetric tensors of $H^1(\hat{\s}_-)$ in
the adjoint module, along with the scalar field in the zero-form
sector. There are no nontrivial field equations in the traceful
case. As explained in more detail in Subsection \ref{freunfhsequ},
in the traceless case, the field equations are associated with the
Einstein cohomology (\ref{rtr})-(\ref{rsc}), Maxwell cohomology
(\ref{1ch}) and Klein-Gordon cohomology (\ref{0ch}). Note that the
cohomology (\ref{h1c}) disappears as a result of gluing the
one-form adjoint and zero-form twisted adjoint modules by
(\ref{glu}).

\subsection
{Physical interpretation of some $\s_-$ cohomology
groups}\label{freunfhsequ}

These cohomological results tell us that there are several possible
choices for gauge invariant differential equations on HS fields.


The form of $r_{1,2}$ (\ref{rtr}) and (\ref{rsc}) indicates that
the Einstein cohomology is responsible for the Lagrangian field
equations of completely symmetric {double traceless} fields.
Indeed, carrying one power of $Z^a$, they are parts of the HS
curvatures $R_{a_1\ldots a_{s-1},b}$ with one cell in the second
row of the corresponding Young tableau. For spin $s$, $\rho_1(Y)$
is a harmonic polynomial of  homogeneity degree $s$, while
$\rho_2(Y)$ is a harmonic polynomial of  homogeneity degree $s-2$.
As a result, the field equations which follow from $r_1=0$ and
$r_2=0$ are of second order\footnote{The  cohomological analysis
outlined here can be extended to the space $\hat{W}_n$ of tensors
required to have  their $n$-th trace equal zero, {\it  i.e.} with
(\ref{Cond2}) replaced by $( \eta^{ab}\frac{\partial^2}{\partial
Y^a\partial Y^b})^n \a(Y,Z)=0\,$. It is tempting to conjecture
that the resulting gauge invariant field equations will contain
$2n$ derivatives.} in derivatives of the dynamical Fronsdal fields
taking values in $H^1(\sigma_-,\hat{W})$ and, as expected for
Lagrangian equations in general, there are as many equations as
dynamical fields.

For example,  spin-$2$ equations on the trace and traceless parts of
the metric tensor associated with the elements  of
$H^1(\sigma_-,\hat{W})$ result from the conditions $r_1=0$ and
$r_2=0$ with $\rho_{1,2}$ in (\ref{rtr}) and (\ref{rsc}) of the form
$\rho_1 (Y) = r_{ab}Y^a Y^b$, $\rho_2 (Y) =r $, with arbitrary $r$
and traceless $r_{ab}$. These are, respectively, the traceless and
trace parts of the linearized Einstein equations. Analogously, the
equations $r_1=0$ and $r_2=0$  of higher orders in $Y$ correspond to
the traceless and first trace parts of the Fronsdal spin $s>1$ HS
equations (which are, of course, double traceless).

Thus, in the traceless case, the proper choice to reproduce
dynamical field equations equivalent to the equations resulting from
the Fronsdal Lagrangian is to keep only the Weyl cohomology nonzero.
Setting elements of the Einstein cohomology to zero, which imposes
the second-order field equations on the dynamical fields, leads to
\be R_1\,=\,e_0^a\,e_0^b\,\epsilon_{ij}\,\frac{\partial^2}{\partial
Y^a_i\partial Y^b_j}\,C(Y^c_k)\,, \label{REF1} \ee where one makes
contact with the HS algebra convention via the identification of
variables $(Y,Z)$ with $(Y_1,Y_2)$. Here $C(Y^c_k)$ satisfies the
$sp(2)$ invariance condition \be Y^a_i \frac{\partial}{\partial
Y^a_j} C(Y) = \half \delta^i_j Y^a_k \frac{\partial}{\partial Y^a_k}
C(Y) \ee along with the tracelessness condition \be
\eta^{ab}\frac{\partial^2}{\partial Y^a_j \partial Y^b_i} C(Y)=0 \ee
to parametrize the Weyl cohomology. The equation (\ref{REF1}) is
exactly (\ref{onmshell}). Thus, the generalized Weyl tensors on the
right hand side of (\ref{onmshell}) parametrize the Weyl cohomology
in the HS curvatures  exactly so as to make the equations
(\ref{onmshell}) for $s>1$ equivalent to the HS field equations that
follow from Fronsdal's action.\footnote{  The subtle relationship
between Fronsdal's field equations and the tracelessness of the HS
curvatures was discussed in details for metric-like spin-$3$ fields
in \cite{DD}.}

Alternatively, one can set the Weyl cohomology to zero, keeping the
Einstein cohomology arbitrary. It is well known that in the spin-$2$
case of gravity the generic solution  of the condition that the Weyl
tensor is zero leads to conformally flat metrics. Analogous analysis
for free spin 3 was performed in \cite{DD}. It is tempting to
conjecture that this property is true for
 any spin $s>2$, {\it i.e.,} the  condition that Weyl cohomology
is zero singles out the ``conformally flat" single trace HS fields
of the form \be \label{cf} \varphi_{\nu_1\ldots \nu_{s}} (x) =
g_{\{\nu_1\nu_2} (x) \psi_{\nu_3 \ldots \nu_{s}\}}(x) \ee with
traceless symmetric $\psi_{\nu_3 \ldots \nu_{s}}(x)  $. Indeed,  the
conformally flat free HS fields (\ref{cf}) have zero generalized
Weyl tensor simply because it is impossible to build a traceless
tensor (\ref{wt}) from  derivatives of $\psi_{\nu_1 \ldots
\nu_{s-2}}(x)$. \vspace{.1cm}


The spin-$0$ and spin-$1$ equations are not described by the
cohomology in the one-form adjoint sector. The Klein-Gordon and
Maxwell equations result from the cohomology of the  zero-form
twisted adjoint representation, which encodes equations that can
be imposed in terms of the generalized Weyl tensors which contain
the spin-$1$ Maxwell tensor and the spin-$0$ scalar as the lower
spin particular cases. More precisely, the spin-$0$ and spin-$1$
field equations are contained in the parts of equations
(\ref{unfHS}) associated with the cohomology groups (\ref{0ch})
and (\ref{1ch}). For spin $s\geqslant 1$, the cohomology
(\ref{h1c}) encodes the Bianchi identities for the definition of
the Weyl (Riemann) tensors by (\ref{REF1}). This is equivalent to
the fact that, in the traceless case,
 the equations (\ref{REF1})
and (\ref{unfHS}) describe properly free massless equations of all
spins supplemented with an infinite set of constraints for auxiliary
fields, which  is the content of the central on-mass-shell theorem.
\vspace{.15cm}

In the traceful case, the equation (\ref{REF1}) with traceful
$C(Y^a_i)$ does not impose any differential restrictions on the
fields in $H^1 (\sigma_- , W)$ because
$e_0^a\,e_0^b\,\epsilon_{ij}\,\frac{\partial^2}{\partial
Y^a_i\partial Y^b_j}\,C(Y^c_k)$ span the full $H^2 (\sigma_- , W)$
of the one-form adjoint sector. This means that the equation
(\ref{REF1}) describes a set of constraints which express all fields
in terms of derivatives of the dynamical fields in $H^1 (\sigma_- ,
W)$. In this sense, the equation (\ref{REF1}) for a traceful field
describes off-mass-shell constraints  identifying the components of
$C(Y)$ with the deWit-Freedman curvature tensors\footnote{ However,
as pointed out in \cite{SSS}, if one imposes $C(Y^a_i)$ to be
harmonic in $Y$, then the corresponding field equations (\ref{REF1})
imposes the deWit-Freedman curvature to be traceless. In this sense,
it may be possible to remove the tracelessness requirement in the
frame-like formulation (see Section \ref{framelike}) without
changing the physical content of the free field equations.}.
Supplementing (\ref{REF1}) by the covariant constancy equation
(\ref{unfHS}) on the traceful zero-forms $C(Y)$ one obtains an
infinite set of constraints for any spin which express infinite sets
of auxiliary fields in terms of derivatives of the dynamical fields,
imposing no differential restrictions on the latter. These
constraints provide unfolded off-mass-shell description of massless
fields of all spins. We call this fact ``central off-mass-shell
theorem". \vspace{.15cm}

Let us stress that our analysis works both in the flat space-time
and in the $(A)dS_d$ case originally considered in \cite{LV}.
Indeed, although the nonzero curvature affects the explicit form of
the background frame and the Lorentz covariant derivative $D_0^L$
and also requires a nonzero operator $\s_+$ denoted by $O(\Lambda)$
in (\ref{tauminus}), all this does not affect the analysis of the
$\s_-$ cohomology because the operator $\s_-$ remains of the form
(\ref{sigmm}) with a nondegenerate frame field $e_\mu^a$.

Let us make the following comment. The analysis of the dynamical
content of the covariant constancy equations $\widehat{D}_0\cc=0$
may depend on the choice of the grading operator $G$ and related
graded decomposition (\ref{Dzero}). This may lead to different
definitions of $\s_-$ and, therefore, different interpretations of
the same system of equations. For example, one can choose a
different definition of $\s_-$ in the space $W$ of traceful tensors
simply by decomposing $W$ into a sum of irreducible Lorentz tensors
({\it i.e.}, traceless tensors) and then defining $\s_-$ within any
of these subspaces as in $\hat{W}$. In this basis, the equations
(\ref{onmshell}) will be interpreted as dynamical equations for an
infinite set of traceless dynamical fields. This phenomenon is not
so surprising, taking into account the well-known analogous fact
that, say, an off-mass-shell
   scalar can be represented as
an integral over the parameter of mass of an infinite set of
on-mass-shell scalar fields. More generally, to avoid paradoxical
conclusions one has to take into account that $\s_-$ may or may not
have a meaning in terms of the elements of the Lie algebra Lie $h$
that gives rise to the covariant derivative (\ref{Dzero}).

\subsection{Towards nonlinear equations}\label{towards}

Nonlinear equations should replace the linearized covariant
derivative $\widetilde{D}_0$ with the full one, $\widetilde{D}$,
containing the $h$-valued connection $\o$. They should also promote
the linearized curvature $R_1$ to $R$. Indeed, (\ref{unfHS}) and
(\ref{REF1}) cannot be correct at the nonlinear level because the
consistency of $\widetilde{D}C=0$ implies arbitrarily high powers of
$C$ in the r.h.s. of the modified equations, since $$
\widetilde{D}\widetilde{D}C\sim RC \sim O(C^2)+{\rm higher\; order\;
terms} \ ,$$ the last relation being motivated by (\ref{REF1}).

Apart from  dynamical field equations, the unfolded HS field
equations contain constraints on the auxiliary components of the
HS connections, expressing the latter via  derivatives of the
nontrivial dynamical variables ({\it i.e.} Fronsdal fields),
modulo pure gauge ambiguity. Originally, all HS gauge connections
$\o_{\mu}^{~A_1 \ldots A_{s-1}, \, B_1 \ldots B_{s-1}}$ have
dimension $length^{-1}$ so that the HS field strength
(\ref{HScur}) needs no dimensionful parameter to have  dimension
$length^{-2}$. However, this means that when some of the gauge
connections are expressed via derivatives of the others, these
expressions must involve space-time derivatives in the
dimensionless combination $\rho \frac{\partial}{\partial x^\nu}$,
where $\rho$ is some parameter of dimension $length$. The only
dimensionful parameter available in the analysis of the free
dynamics is the radius $\rho$ of the $AdS$ space-time related to
the cosmological constant by (\ref{rco}). Recall that it appears
through the definition of the frame field (\ref{fram}) with
$V^A\sim \rho$ adapted to make the frame $E^A_{\mu}$ (and,
therefore, the metric tensor) dimensionless. As a result, the HS
gauge connections are expressed by the unfolded field equations
through the derivatives of the dynamical fields as \be
\label{hder1} \o^{~a_1 \ldots a_{s-1}, \, b_1 \ldots
b_t\hat{d}\ldots\hat{d}} =\Pi \left ( \rho^t
\frac{\partial}{\partial x^{b_1}}\ldots \frac{\partial}{\partial
x^{b_t}} \o^{ ~a_1 \ldots a_{s-1}, \,\hat{d}\ldots\hat{d}}\right)
+\mbox{lower derivative terms} \,, \ee where $\Pi$ is some
projector that permutes indices (including the indices of the
forms) and projects out traces. Plugging these expressions back
into the HS field strength (\ref{HScur}) one finds that HS
connections with $t>1$ ({\it i.e.} extra fields that appear for
$s>2$) contribute to the terms with higher derivatives which blow
up in the flat limit $\rho \to \infty$. This mechanism brings
higher derivatives and negative powers of the cosmological
constant into HS interactions (but not into the free field
dynamics because the free action is required to be independent of
the extra fields). Note that a similar phenomenon takes place in
the sector of the generalized Weyl zero-forms $C(Y)$ in the
twisted adjoint representation.

\section{Star product}
\label{star}

We shall formulate consistent nonlinear equations using the star
product. In other words we shall deal with ordinary commuting
variables $Y^A_i$ instead of operators $\hat{Y}^A_i$. In order to
avoid ordering ambiguities,  we use the Weyl prescription. An
operator is said to be Weyl ordered if it is completely symmetric
under the exchange of operators $\hat{Y}^A_i$. One establishes a one
to one correspondence between each Weyl ordered polynomial
$f(\hat{Y})$ (\ref{exp}) and its symbol $f(Y)$, defined by
substituting each operator $\hat{Y}^A_i$ with the commuting variable
$Y^A_i$. Thus $f(Y)$ admits a formal expansion in power series of
$Y^A_i$ identical to that of $f(\hat{Y})$, {\it i.e.} with the same
coefficients, \be\label{exp2} f({Y}) = \sum_{m,n} f_{A_1 \ldots
A_m\,,B_1 \ldots B_n}Y_1^{A_1}\ldots Y_1^{A_m}Y_2^{B_1}\ldots
Y_2^{B_n}\,.\ee

To reproduce the algebra $A_{d+1}$, one defines the \emph{star
product} in such a way that, given any couple of functions
$f_1,f_2$, which are symbols of operators $\hat{f}_1,\hat{f}_2$
respectively, $f_1\* f_2$ is the symbol of the operator
$\hat{f}_1\hat{f}_2$. The result is nontrivial because the operator
$\hat{f}_1\hat{f}_2$ should be Weyl ordered. It can be shown that
this leads to the well-known Weyl-Moyal formula \be \label{diffdef}
(f_1\*f_2)(Y)=f_1(Y)\,
e^{\frac{1}{2}\overleftarrow{\partial}^j_A\overrightarrow{\partial}_{B}^i
\eta^{AB}\epsilon_{ji}} \,f_2(Y) \ , \ee where
$\partial^j_A\equiv\frac{\partial}{\partial Y^A_j}$ and
$\overleftarrow{\partial}$, as usual, means that the partial
derivative acts to the left while $\overrightarrow{\partial}$ acts
to the right. The rationale behind this definition is simply that
higher and higher powers of the differential operator in the
exponent produce more and more contractions. One can show that the
star product is an associative product law, and that it is regular,
which means that the star product of two polynomials in $Y$ is still
a polynomial. From (\ref{diffdef}) it follows that the star product
reproduces the proper commutation relation of oscillators,
$$[Y^A_i,Y^B_j]_{\*}\equiv
Y^A_i\*Y^B_j-Y^B_j\*Y^A_i=\epsilon_{ij}\eta^{AB} \ .$$
   The star product has
also an integral definition, equivalent to the differential one
given by (\ref{diffdef}), which is \be \label{trig} (f_1\*
f_2)(Y)=\frac{1}{\pi^{2(d+1)}}\int dS dT\,
f_1(Y+S)\,f_2(Y+T)\exp(-2S_i^A T^i_A) \ . \ee
   The whole discussion
of Section \ref{hsa} can be repeated here, with the prescription of
substituting operators with their symbols and operator products with
star products. For example, the $o(d-1,2)$ generators (\ref{AdSgen})
and the $sp(2)$ generators (\ref{spgen}) are realized as
\be\label{gen2} T^{AB}=-T^{BA}=\frac{1}{2}Y^{iA}Y^B_i \ , \quad
t_{ij}=t_{ji}=Y^A_iY_{jA} \ , \ee respectively. Note that $$ Y^A_i
\* = Y^A_i +\frac{1}{2}\frac{\overrightarrow{\partial}}{\partial
Y^i_A} $$ and $$ \* Y^A_i= Y^A_i -
\frac{1}{2}\frac{\overleftarrow{\partial}}{\partial Y^i_A} \ .$$
{}From here it follows that \be \label{scom} [Y^A_i\,, f(Y)]_* =
\frac{\partial}{\partial Y^i_A} f(Y) \ee and \be \label{sacom}
\{Y^A_i\,, f(Y)\}_* = 2 Y^A_i f(Y)\,. \ee  {}With the help of these
relations it is easy to see
   that the $sp(2)$ invariance condition $[t_{ij},f(Y)]_\*=0$ indeed
has the form (\ref{sp2c}) and singles out two-row rectangular Young
tableaux, {\it i.e.} it implies that the coefficients $f_{A_1 \ldots
A_m\,,B_1 \ldots B_n}$ are nonzero only if $n=m$, and
symmetrization of any $m+1$ indices of $f_{A_1 \ldots A_m\,,B_1
\ldots B_m}$ gives zero. Let us also note that if
$[t_{ij},f(Y)]_\*=0$  then \be \label{tpr} t_{ij} * f=f*t_{ij}= \Big
( t_{ij} +\frac{1}{4} \frac{\partial^2}{\partial Y^i_A \partial
Y^{jA}}\Big ) f\,. \ee

One can then introduce the gauge fields taking values in this star
algebra as functions $\o(Y|x)$ of  oscillators, \be\label{exp2gf}
\o(Y|x) = \sum_{s\geq 1} i^{s-2}\o^{A_1 \ldots A_{s-1}\,,B_1 \ldots
B_{s-1}}(x) Y_{1 A_1}...Y_{1 A_{s-1}}Y_{2 B_1}...Y_{2 B_{s-1}} \
,\ee with their field strength defined by \be
R(Y)=d\o(Y)+(\o\*\o)(Y)\, \ee and gauge transformations \be \delta
\o(Y)=d\epsilon(Y)+[\o\,,\epsilon ]_*(Y)\, \ee (where the dependence
on the space-time coordinates $x$ is implicit). For the subalgebra
of $sp(2)$ singlets we have \be \label{sp2co} D (t_{ij})=0\,,\qquad
[t_{ij}\,,\epsilon ]_* = 0\,,\qquad [t_{ij}\,,R ]_* = 0\,. \ee Note
that $d (t_{ij} )=0$ and, therefore from the first of these
relations it follows that $[t_{ij}\,,\o ]_* = 0$, which is the
$sp(2)$ invariance relation.

Furthermore, one can get rid of traces by factoring out the ideal
$\ci$ spanned by the elements of the form $t_{ij}\* g^{ij}$. For
factoring out the ideal $\ci$ it is convenient to consider
\cite{0404124} elements of the form $\Delta *g$ where $\Delta$ is an
element satisfying the conditions $\Delta *t_{ij}=t_{ij}*\Delta =0$.
The explicit form of $\Delta $ is
   \cite{0404124,SSS}
\be \label{Delta} \Delta = \int_{-1}^1 ds (1-s^2 )^{\half(d-3)}
\exp{(s\sqrt{z}\,)} =2 \int_{0}^1 ds (1-s^2 )^{\half(d-3)}
ch{(s\sqrt{z}\,)}\,, \ee where \be
   z= \frac{1}{4} Y^A_i Y_{Aj} Y^{Bi}Y_B^j \,.
\ee

Indeed, one can see \cite{0404124} that $\Delta *f = f*\Delta$ and
that all elements of the form $f=u^{ij} *t_{ij}$ or
$f=t_{ij}*u^{ij}$ disappear in $\Delta *f$, {\it i.e.} the
factorization of $\ci$ is automatic.  The operator $\Delta$, which
we call quasiprojector, is not a projector because  $\Delta*\Delta$
does not exist (diverges) \cite{0404124,SSS}. One therefore cannot
define a product of two elements $\Delta *f$ and $\Delta *g$ in the
quotient algebra as the usual star product. A consistent definition
for the appropriately modified product law $\circ$ is \be (\Delta
*f) \circ (\Delta* g) = \Delta *f*g\, \ee (for more detail see
section 3 of  \cite{0404124}). Note that from this consideration it
follows that the star product $g_1 *g_2$ of any two elements
satisfying the strong $sp(2)$ invariance condition $t_{ij}* g_{1,2}
=0$ of \cite{SSS} is ill-defined because such elements admit a form
$g_{1,2}= \Delta *{g}^\prime_{1,2}.$

\section{Twisted adjoint representation}
\label{tw}

As announced in Section \ref{freemasslessequ}, we give here a
precise definition of the module in which the Weyl-like zero-forms
take values, in such a way that (\ref{unfHS}) is reproduced at the
linearized level.  Taking the quotient by the ideal $\cal I$ is a
subtle step the procedure of which is explained in Subsection
\ref{factorization}.

\subsection{Definition of the twisted adjoint module}

To warm up, let us start with the adjoint representation. Let $\ca$
be an associative algebra endowed with a product denoted by $\*$.
The $\*$-commutator is defined as $[a,b]_{\*}=a\* b-b\* a \ , \
a,b\in\ca$. As usual for an associative algebra, one constructs a
Lie algebra $g$ from $\ca$, the Lie bracket of which is the
$\*$-commutator. Then $g$ has an adjoint representation the module
of which coincides with the algebra itself and such that the action
of an element $a\in g$ is given by $$ [a,X]_{\*} \ , \forall X\in
\ca \ .$$

Let $\t$ be an automorphism of the algebra $\ca$, that is to say
$$ \t(a\* b) =\t(a)\*\t(b) \,,\qquad \t(\lambda a +\mu b )
= \lambda \t (a) +\mu \t(b)\\, \qquad \forall a,b\in \ca\,,$$ where
$\lambda$ and $\mu$ are any elements of the ground field ${\mathbb
R}$ or ${\mathbb C}$. The $\tau$-twisted adjoint representation of
$g$ has the same definition as the adjoint representation, except
that the action of $g$ on its elements is modified by the
automorphism $\t$:
$$ a( X) \ \to \ a\* X-X\* \t(a) \ .$$ It is easy to see that
   this gives a representation of $g$.

The appropriate choice of $\t$, giving rise to the infinite bunch of
fields contained in the zero-form $C$ (matter fields, generalized
Weyl tensors and their derivatives), is the following:
\be\label{tau} \t(f(Y))=\widetilde{f}({Y}) \ ,\ee where \be
\widetilde{f}({Y})=f(\widetilde{Y})\,,\qquad
\widetilde{Y}^A_i=Y^A_i-2V^A V^B Y_{Bi} \,, \ee {\it i.e.,}
$\widetilde{Y}^A_i$ is the oscillator $Y^A_i$ reflected with respect
to the compensator (recall that we use the normalization $V_A
V^A=1$). So one can say that the automorphism $\t$ is some sort of
parity transformation in the $V$-direction, leaving unaltered the
Lorentz components of the oscillators. More explicitly, in terms of
the transverse and longitudinal components $$^\perp Y^A_i=Y^A_i-V^A
V_B Y^B_i\,, \qquad ^\parallel Y^A_i=V^A V^B Y_{Bi}\,,$$ the
automorphism $\t$ is the transformation $$^\perp Y^A_i\rightarrow
^\perp Y^A_i\,,\qquad ^\parallel Y^A_i\rightarrow -\,^\parallel
Y^A_i\,,$$ or, in the standard gauge, $Y^a_i\rightarrow Y^a_i$,
$Y^{\hat{d}}_i\rightarrow -Y^{\hat{d}}_i$. From (\ref{diffdef}) it
is obvious that $\t$ is indeed an automorphism of the  star product
algebra.

The automorphism $\tau$ (\ref{tau}) leaves invariant the $sp(2)$
generators
$$
\tau (t_{ij}) = t_{ij}\,.
$$
This allows us to require the zero-form $C(Y|x)$ in the twisted
adjoint module to satisfy the $sp(2)$ invariance condition \be
t_{ij} \* C = C\*t_{ij} \ee and to define the covariant
derivative in the twisted adjoint module of $\hu$ as \be
\label{twcun} \widetilde{D} C=dC+\o\* C-C\*\widetilde{\o} \ . \ee

At the linearized level one obtains \be\label{twc} \widetilde{D}_0
C=dC+\o_0\* C-C\*\widetilde{\o}_0 \ . \ee One decomposes $\o_0$ into
its Lorentz and translational part, $\o_0=\o_0^{L}+\o_0^{transl}$,
via (\ref{lorentzcosm}), which gives \bqn \o_0^{L} & \equiv &
\frac{1}{2}\,\o_0^{AB}\,^\perp Y^i_A\, ^\perp Y_{Bi}=\frac{1}{2}
\,\o_0^{ab}\, Y^i_a\, Y_{bi}\ , \nonumber\\ \o_0^{transl} & \equiv &
\,\o_0^{AB}\,^\perp Y^i_A\, ^\parallel Y_{Bi}=\, e_0^a Y_a^i
Y_{Bi}V^B  \,.\nonumber \eqn Taking into account the definition
(\ref{tau}), it is clear that $\t$ changes the sign of
$\o_0^{transl}$ while leaving $\o_0^L$ untouched. This is tantamount
to say that $\widetilde{D}_0$ contains an anticommutator with the
translational part of the connection instead of a commutator, $$
\widetilde{D}_0 C=D_0^L C+ \{\o_0^{transl},C\}_{\*} \ ,$$ where
$D_0^L$ is the usual Lorentz covariant derivative, acting on Lorentz
indices. Expanding the star products, we have \be\label{expD0}
\widetilde{D}_0=D_0^L+2E_0^A V^B(^\perp Y^i_A \,^\parallel
Y_{Bi}-\frac{1}{4}\epsilon^{ij}\frac{\partial^2}{\partial \,^\perp
Y^{Ai}\partial\, ^\parallel Y^{Bj}}) \ ,\ee  the last term being due
to the noncommutative structure of the star algebra.

The equation  (\ref{expD0}) suggests that there exists a grading
operator \be N^{tw}=N_\perp -N_\parallel=^\perp Y^A_i
\frac{\partial}{\partial ^\perp Y^A_i}-\,^\parallel Y^A_i
\frac{\partial}{\partial ^\parallel Y^A_i} \label{twgr} \ee
commuting with $\widetilde{D}_0$, and whose eigenvalues $N_\perp
-N_\parallel=2s$, where $s$ is the spin, classify the various
irreducible submodules into which the twisted adjoint module
decomposes as $o(d-1,2)$-module. In other words, the system of
equations $\widetilde{D}_0 C=0$ decomposes into an infinite number
of independent subsystems,  the fields of each subset satisfying
$N^{tw} C=2s C$, for some nonnegative integer $s$. Let us give some
more detail about this fact. Recall that requiring $sp(2)$
invariance  restricts us to the rectangular two row $AdS_d$ Young
tableaux
\begin{picture}(45,13)(-3,2)
\multiframe(0,6.5)(13.5,0){1}(35,6){}\put(40,8){}
\multiframe(0,0)(13.5,0){1}(35,6){}\put(40,0){}
\end{picture}.
By means of the compensator $V^A$ we then distinguish between
transverse (Lorentz) and longitudinal indices. Clearly
$N^{tw}\geqslant 0$, since having more than  half of vector indices
in the extra direction $V$ would imply symmetrization over more than
half of all indices, thus giving zero because of the symmetry
properties of Young tableaux. Then, each independent sector
$N^{tw}=2s$ of the twisted adjoint module starts from the
rectangular Lorentz-Young tableau corresponding to the (generalized)
Weyl tensor
\begin{picture}(30,13)(-3,2)
\multiframe(0,6.5)(13.5,0){1}(20,6){}\put(23,8){\tiny $s$}
\multiframe(0,0)(13.5,0){1}(20,6){}\put(23,0){\tiny $s$}
\end{picture}
, and admits as further components all its ``descendants''
\begin{picture}(56,13)(-3,2)
\multiframe(0,6.5)(13.5,0){1}(35,6){}\put(40,8){\tiny $s+k$}
\multiframe(0,0)(13.5,0){1}(20,6){}\put(23,0){\tiny $s$}
\end{picture}
, which the equations themselves set equal to $k$ Lorentz covariant
derivatives of \begin{picture}(30,13)(-3,2)
\multiframe(0,6.5)(13.5,0){1}(20,6){}\put(23,8){\tiny $s$}
\multiframe(0,0)(13.5,0){1}(20,6){}\put(23,0){\tiny $s$}
\end{picture}. From the $AdS_d$-Young tableaux
point of view, the set of fields forming an irreducible submodule of
the twisted adjoint module with some fixed $s$ is nothing but the
components of the fields $C^{A_1...A_u, B_1...B_u}$ ($u=s, \ldots
\infty $) that have $k=u-s$ indices parallel to $V^A$:
\begin{picture}(74,13)(-3,2)
\multiframe(0,6.5)(13.5,0){1}(35,6){}\put(40,8){\tiny $s+k=u$}
\multiframe(0,0)(13.5,0){1}(20,6){}\put(23,0){\tiny $s$}
\end{picture}$\sim$
\begin{picture}(54,13)(-5,2)
\multiframe(0,6.5)(13.5,0){1}(45,6){}\put(50,8){\tiny $u$}
\multiframe(0,0)(13.5,0){1}(20,6){}\put(-6,2){}
\multiframe(20.5,0)(13.5,0){1}(6,6){\tiny $\hat{d}$}
\multiframe(39,0)(13.5,0){1}(6,6){\tiny $\hat{d}$}\put(50,0){\tiny
$u$} \multiframe(27,0)(13.5,0){1}(11.5,6){\tiny \ldots}
\end{picture}\,\,\,\,.

\subsection{Factorization procedure}\label{factorization}

The twisted adjoint module as defined in the previous section is
off-mass-shell because the zero-form $C(Y|x)$ is traceful in the
oscillators $Y^A_i$. To put it on-mass-shell one has to factor out
those elements of the twisted adjoint module $T$ that also belong to
the ideal $\ci$ of the associative algebra $\cal S$ of $sp(2)$
singlets (see Subsection \ref{sp2etc}) and form a submodule $T\cap
\cal \,I$ of the HS algebra. By a slight abuse of terminology, we
will also refer to this submodule as ``ideal $\ci$" in the sequel.
Therefore, ``quotienting by the ideal $\ci$" means dropping terms
$g\in T\cap \cal \,I$, that is \be \label{ideal} g\in \ci\quad
\Longleftrightarrow \quad [t_{ij}, g]_* =0\,, \quad g=t_{ij}
*g^{ij}=g^{ij }* t_{ij}\,. \ee

The factorization of $\cal I$ admits an infinite number of possible
ways of choosing representatives of the equivalence classes (recall
that $f_1$ and $f_2$ belong to the same equivalence class iff $f_1 -
f_2 \in \ci$).
The tracelessness condition (\ref{trcond}), which in the case of HS
gauge one-forms
amounts to $$
\frac{\partial^2}{\partial Y^A_i\partial Y_{Aj}}\o(Y)=0 \ ,
$$
is not convenient  for the twisted adjoint representation because it
does not preserve the grading (\ref{twgr}), {\it i.e.} it does not
commute with $N^{tw}$. A version of the factorization condition that
commutes with $N^{tw}$ and is more appropriate for the twisted
adjoint case
is $$
\Big ( \frac{\partial^2}{\partial {}^\perp Y^A_i\partial {}^\perp
Y_{Aj}} -4 {}^\parallel Y^{Ai} {}^\parallel
Y^j_A \Big ) C(Y)=0 \ . $$

For the computations, both in the adjoint and the twisted adjoint
representation, it may be convenient to require Lorentz
tracelessness, {\it i.e.} the tracelessness with respect to
transversal indices. Recall that the final result is insensitive to
a particular choice of the factorization condition, that is,
choosing one or another condition is a matter of convenience. In
practice the factorization procedure is implemented
   as follows.
Let $A^{tr}$ denote either a one-form HS connection or a HS Weyl
zero-form satisfying a chosen tracelessness ({\it i.e.}
factorization) condition. The left hand side of the field equations
contains the covariant derivative $D$ in the adjoint, $D=D_0$, or
twisted adjoint, $D=\widetilde{D}_0$, representation. $D(A^{tr} )$
does not necessarily satisfy the chosen tracelessness condition,
but it can be represented in the form \be \label{da} D(A^{tr}
)=(D(A^{tr} ))^{tr}+t_{ij}X_1^{ij}\,, \ee where the ordinary product
of $t_{ij}$ with some $X_1^{ij}$ parametrizes the traceful part,
while $(D(A^{tr} ))^{tr}$ satisfies the chosen tracelessness
condition. Note that $X_1^{ij}$ contains less traces than $D(A^{tr}
)$. One rewrites (\ref{da}) in the  form \be D(A^{tr} )=(D(A^{tr}
))^{tr}+t_{ij}*X_1^{ij} +B\,, \ee where ${B}= t_{ij} X_1^{ij}
-t_{ij}*X_1^{ij}$. Taking into account that \be {B}= t_{ij} X_1^{ij}
-t_{ij}*X_1^{ij} =-\frac{1}{4}\frac{\partial^2}{\partial Y^i_A
\partial Y^{jA}} X_1^{ij} \ee by virtue of (\ref{tpr}), one observes
that $B$ contains less traces than $D(A^{tr} )$. The factorization
is performed by dropping out the term $t_{ij}*X_1^{ij}$. The
resulting expression $(D(A^{tr} ))^{tr} +B$ contains less traces. If
necessary, the procedure has to be repeated to get rid of lower
traces. At the linearized level it terminates in a finite number of
steps (two steps is enough for the Lorentz tracelessness condition).
The resulting expression gives the covariant derivative in the
quotient representation which is the on-mass-shell twisted adjoint
representation.

Equivalently, one can use the factorization procedure with the
quasiprojector $\Delta$ as explained in the end of Section
\ref{star}.
 Upon application of one or another procedure for factorizing out
the traces, the spin-$s$ submodule of the twisted adjoint module
forms an irreducible $o(d-1,2)$-module. The subset of the fields $C$
in the twisted adjoint module with some fixed $s$ matches the set of
spin-$s$ generalized Weyl zero-forms of Section
\ref{freemasslessequ}. Not surprisingly, they form equivalent
$o(d-1,2)$-modules. In particular, it can be checked that, upon an
appropriate rescaling of the fields, the covariant constancy
condition in the twisted adjoint representation \be \label{tweq}
\widetilde{D}_0 C=0\, \ee  reproduces (\ref{unfHS}) in the standard
gauge. The precise form of the $\Lambda$--dependent terms in
(\ref{unfHS}) follows from this construction. It is straightforward
to compute the value of the Casimir operator in this irreducible
$o(d-1,2)$-module (see also \cite{SSS})
$$ 2 T^{AB} T_{AB} = (s-1)(s+d-3)\,,$$
where $2s$ is the eigenvalue of $N^{tw}$. This value coincides with
that for the unitary massless representations of any spin in $AdS_d$
\cite{Metsaev}. This fact is in agreement with the general
observation \cite{ShV,d4sym} that the representations carried by the
zero-form sector in the unfolded dynamics are dual by a nonunitary
Bogolyubov transform to the Hilbert space of single-particle states
in the quantized  theory.

\section{Nonlinear field equations}
\label{Nonli}

We are now ready to search for nonlinear corrections to the free
field dynamics. We will see that it is indeed possible to find a
unique form for interactions, modulo field redefinitions, if one
demands that the $sp(2)$ invariance of Section \ref{sp2etc} is
maintained at the nonlinear level. This condition is of crucial
importance because, if the $sp(2)$ invariance was broken, then the
resulting nonlinear equations might involve new tensor fields,
different from the two-row rectangular Young tableaux one started
with, and this  might have no sense (for example, the new fields may
contain ghosts). Thus, to have only usual HS fields as independent
degrees of freedom, one has to require that $sp(2)$ invariance
survives at the nonlinear level, or, in other words, that there
should be a modified $sp(2)$ generator, $$ t_{ij}^{int}=t_{ij}+O(C)
\ ,$$
   that still satisfies $D(t^{int}_{ij})=0$, which is a deformation of
the free field condition (\ref{sp2co}).

The construction of nonlinear corrections to the free field dynamics
and the check of consistency order by order is quite cumbersome.
   These have been performed explicitly up to second order in
the Weyl zero-forms \cite{Vasfda,Vasfda1,Vasiliev:1989yr} in terms
of the spinorial formulation of the $d=4$ HS theory. More refined
methods have been developed to formulate the full dynamics of HS
gauge fields in a closed form first in four dimensions
\cite{con,prop} and more recently in any dimension
\cite{Vasiliev:2003ev}. The latter is presented now.

\subsection{Doubling of oscillators}
\label{doubling}

A trick that simplifies the formulation is to introduce additional
noncommutative variables $Z$. This allows one to describe complicate
nonlinear corrections as solutions of some simple differential
equations with respect to such variables.  The form of these
equations is fixed by formal consistency and by the existence of
nonlinear $sp(2)$ generators that guarantee the correct spectrum of
fields and the gauge invariance of all nonlinear terms they encode.

More precisely, this step amounts to the doubling of the oscillators
$Y^A_i \to (Z^A_i,Y^A_i)$, and correspondingly one needs to enlarge
the star product law. It turns out that a sensible definition is the
following, \be\label{enlstar} (f\*
g)(Z,Y)=\frac{1}{\pi^{2(d+1)}}\int dS dT \, e^{-2S^A_i T^i_A}
f(Z+S,Y+S)g(Z-T,Y+T)\ ,\ee which is an associative and regular
product law in the space of polynomial functions $f(Z,Y)$, and gives
rise to the commutation relations
$$ [Z^A_i,Z^B_j]_{\*}=-\epsilon_{ij}\eta^{AB}\,, \qquad
[Y^A_i,Y^B_j]_{\*}=\epsilon_{ij}\eta^{AB}\,, \qquad
[Y^A_i,Z^B_j]_{\*}=0\ .$$ The definition (\ref{enlstar}) has the
meaning of a normal ordering with respect to the ``creation'' and
``annihilation'' operators $Z-Y$ and $Z+Y$, respectively. Actually,
from (\ref{enlstar}) follows that the left star multiplication by
$Z-Y$ and the right star multiplication by $Z+Y$ are equivalent to
usual multiplications by $Z-Y$ and $Z+Y$, respectively. Note that
$Z$ independent functions $f(Y)$ form a proper subalgebra of the
star product algebra (\ref{enlstar}) with the Moyal star product
(\ref{trig}).

One can also check that the following formulae are true:
\be\label{comy} Y^A_i\* =
Y^A_i+\frac{1}{2}\left(\frac{\overrightarrow{\partial}}{\partial
Y^i_A}-\frac{\overrightarrow{\partial}}{\partial Z^i_A}\right) \ ,
\qquad
\*Y^A_i=Y^A_i-\frac{1}{2}\left(\frac{\overleftarrow{\partial}}{\partial
Y^i_A}+\frac{\overleftarrow{\partial}}{\partial Z^i_A}\right) \ ,
\quad \ee \be \label{comz} Z^A_i\*
=Z^A_i+\frac{1}{2}\left(\frac{\overrightarrow{\partial}}{\partial
Y^i_A}-\frac{\overrightarrow{\partial}}{\partial Z^i_A}\right)\
,\qquad \*
Z^A_i=Z^A_i+\frac{1}{2}\left(\frac{\overleftarrow{\partial}}{\partial
Y^i_A}+\frac{\overleftarrow{\partial}}{\partial Z^i_A}\right) \ .
\ee Furthermore, the appropriate reality conditions for the Lie
algebra built from this associative star product algebra via
commutators are \be \label{rcond2} \bar{f}(Z,Y)=-f(-iZ,iY)\,, \ee
where the bar denotes complex conjugation of the coefficients of the
expansion of ${f}(Z,Y)$ in powers of $Z$ and $Y$. This condition
results from (\ref{rcond}) with the involution $\dagger$ defined by
the relations \be \label{reality} (Y^A_i)^\dagger=iY^A_i \ , \qquad
(Z^A_i)^\dagger=-iZ^A_i \ . \ee

\subsection{Klein operator}

The distinguishing property of the extended definition
(\ref{enlstar}) of the star product is that it admits the inner
Klein operator \be\label{defklein} \ck = \exp(-2 z_i y^i)\ , \ee
where
$$ y_i\equiv V_A Y^A_i \ ,\qquad z_i\equiv V_A Z^A_i $$ are the
projections of the oscillators along $V^A$. Using the definitions
(\ref{enlstar})-(\ref{defklein}), one can show that $\ck$ (i)
generates the automorphism $\t$ as an inner automorphism of the
extended star algebra, \be\label{prop1}
\ck\*f(Z,Y)=f(\widetilde{Z},\widetilde{Y})\*\ck \ ,\ee and (ii) is
involutive, \be\label{prop2} \ck\*\ck=1 \ee (see Appendix
\ref{appklein} for a proof of these properties).

Let us note that {\it a priori} the star product (\ref{enlstar}) is
well-defined for the algebra of polynomials (which means that the
star product of two polynomials is still a polynomial). Thus the
star product admits an ordinary interpretation in terms of
oscillators as long as we deal with polynomial functions. But $\ck$
is not a polynomial because it contains an infinite number of terms
with higher and higher powers of $z_i y^i$. So, {\it a priori} the
star product with $\ck$ might give rise to divergencies arising from
the contraction of an infinite number of terms (for example, an
infinite contribution might appear in the zeroth order like a sort
of vacuum energy). What singles out the particular star product
(\ref{enlstar}) is that this does not happen for the class of
functions which extends the space of polynomials to include $\ck$
and similar functions.

Indeed, the evaluation of  the star product of two exponentials
 like ${\cal A}=\exp( A^{ij}_{AB} W_1{}^A_i W_2{}^B_j  )$,
where $A^{ij}_{AB}$ are constant coefficients and the $W$'s are some
linear combinations of $Y{}^A_i$ and $Z{}^A_i$, amounts to
evaluating the Gaussian integral resulting from (\ref{enlstar}). The
potential problem is that the bilinear form $B$ of the integration
variables in the Gaussian integral in ${\cal A}_1 *{\cal A}_2$ may
be degenerate for some exponentials ${\cal A}_1$ and ${\cal A}_2$,
which leads to an infinite result because the Gaussian evaluates
$det^{-1/2} |B|$. As was shown originally in \cite{prop} (see also
\cite{prok}) for the analogous spinorial star product in four
dimensions, the star product (\ref{enlstar}) is well-defined for the
class of functions, which we call {\it regular}, that can be
expanded into a finite sum of functions $f$ of the  form \be
\label{class} f(Z,Y) = P(Z,Y) \int_{M^n}d^n\t \ \rho (\t)  \exp
\Big(\phi(\t) z_i y^i\Big)\,, \ee where the integration is over some
compact domain $M^n \subset {\mathbb R}^n$ parametrized by the
coordinates $\t_i$ $ (i=1,\ldots ,n )$, the functions $P(Z,Y)$ and
$\phi(\t)$ are arbitrary polynomials of $(Z,Y)$ and $\t_i$,
respectively, while $\rho(\t)$ is integrable in $M^n$. The key point
of the proof is that the star product (\ref{enlstar}) is such that
the exponential in the Ansatz (\ref{class}) never contributes to the
quadratic form in the integration variables simply because $s_i s^i
= t_i t^i \equiv 0$ (where $s_i=S^A_i V_A$ and $t_i=T^A_i V_A$). As
a result, the star product of two elements (\ref{class}) never
develops an infinity and the class (\ref{class}) turns out to be
closed under star multiplication pretty much as usual polynomials.
The complete proof is given in Appendix \ref{reg}.

The Klein operator $\ck$ obviously belongs to the regular class, as
can be seen by putting $n=1,$ $\rho (\t) = \delta (\t+2)$, $\phi(\t)
= \t$ and $P(Z,Y)=1$ in (\ref{class}). Hence our manipulations with
$\ck$ are safe. This property can be lost however if one either goes
beyond the class of regular functions (in particular, this happens
when the quasiprojector $\Delta$ (\ref{Delta}) is involved) or uses
a different star product realization of the same oscillator algebra.
For example, usual Weyl ordering prescription is not helpful in that
respect.

\subsection{Field content}
\label{eqns}

The nonlinear equations are formulated in terms of the fields
$W(Z,Y|x)$, $S(Z,Y|x)$ and $B(Z,Y|x)$, where $B$ is a zero-form,
while $$W(Z,Y|x)=dx^\mu W_\mu(Z,Y|x)\,,\qquad S(Z,Y|x)=dZ^A_i
S^i_A(Z,Y|x)$$ are connection one-forms, in space-time and auxiliary
$Z^A_i$ directions, respectively. They satisfy the reality
conditions analogous to (\ref{rcond2})
$$ \bar{W}(Z,Y|x)=-W(-iZ,iY|x)\ , \quad
\bar{S}(Z,Y|x)=-S(-iZ,iY|x)\ , $$
$$\bar{B}(Z,Y|x)=-\widetilde{B}(-iZ,iY|x)\ .
$$

The fields $\o$ and $C$ are identified with the ``initial data'' for
the evolution in $Z$ variables as follows:
$$\o(Y|x)=W(0,Y|x)\,,\qquad C(Y|x)=B(0,Y|x)\,.$$ The differentials
satisfy the standard anticommutation relations $$dx^\mu
dx^\nu=-dx^\nu dx^\mu\,, \quad dZ^A_i dZ^B_j=-dZ^B_j dZ^A_i\,,\quad
dx^\mu dZ^A_i=-dZ^A_i dx^\mu\,,$$ and commute with all other
variables. The dependence on $Z$ variables will be reconstructed by
the imposed equations (modulo pure gauge ambiguities).

We require that all $sp(2)$ indices are contracted covariantly. This
is achieved by imposing the conditions \be \label{totsp2}
[t_{ij}^{tot} , W]_* =0\,,\qquad [t_{ij}^{tot}, B]_* =0\,,\qquad
[t_{ij}^{tot} , S_k^A]_* =\epsilon_{jk} S_i^A +\epsilon_{ik} S_j^A
\,, \ee where the diagonal $sp(2)$ generator \be \label{tot}
   t_{ij}^{tot}\equiv Y_i^A Y_{Aj}-Z_i^A Z_{Aj}
\ee generates inner $sp(2)$ rotations of the star product algebra
\be [t_{ij}^{tot} , Y_k^A]_* =\epsilon_{jk} Y_i^A +\epsilon_{ik}
Y_j^A\,,\qquad [t_{ij}^{tot} , Z_k^A]_* =\epsilon_{jk} Z_i^A
+\epsilon_{ik} Z_j^A\,. \ee Note that the first of the relations
(\ref{totsp2}) can be written covariantly as $D (t^{tot}_{ij}) =0$,
by taking into account that $d (t^{tot}_{ij}) =0$.

\subsection{Nonlinear system of equations}

The full nonlinear system of equations for completely symmetric HS
fields is \be\label{vas1} dW + W\* W=0 \ ,\ee \be\label{vas2} dB +
W\* B - B\* \widetilde{W}=0 \ ,\ee \be\label{vas3} dS + W \* S + S
\* W=0 \ ,\ee \be\label{vas4} S\* B- B\* \widetilde{S}=0 \ ,\ee
\be\label{vas5} S\* S = -\frac{1}{2}(dZ^i_A
dZ_i^A+4\Lambda^{-1}dZ_{Ai}dZ^i_B V^A V^B B\* \ck) \ ,\ee where we
define \be\label{tauonS}
\widetilde{S}(dZ,Z,Y)=S(\widetilde{dZ},\widetilde{Z},\widetilde{Y})
\ .\ee  One should stress that the twisting of the basis elements
$dZ_i^A$ of the exterior algebra in the auxiliary directions is {\it
not} implemented via the Klein operator as in (\ref{prop1}).
Solutions of the system (\ref{vas1})-(\ref{vas5}) admit
factorization over the ideal generated by the nonlinear $sp(2)$
generators (\ref{tint}) defined in Section \ref{sp2} as nonlinear
deformations of the generators (\ref{gen2}) used in the free field
analysis. The system resulting from this factorization gives the
nonlinear HS interactions to all orders.

The first three equations are the only ones containing space-time
derivatives, via the de Rham differential $d=dx^\mu
\frac{\partial}{\partial x^\mu}$. They have the form of
zero-curvature equations for the space-time connection $W$
(\ref{vas1}) and the $Z$-space connection $S$ (\ref{vas3}) together
with a covariant constancy condition for the zero-form $B$
(\ref{vas2}). These equations alone do not allow any nontrivial
dynamics, so the contribution coming from (\ref{vas4}) and
(\ref{vas5}) is essential. Note that the last two equations are
constraints from the space-time point of view, not containing
derivatives with respect to the $x$-variables, and that the
nontrivial part only appears with the ``source'' term $B\* \ck$ in
the $V^A$ longitudinal sector of (\ref{vas5}) (the first term on the
right hand side of (\ref{vas5}) is a constant). The inverse power of
the cosmological constant $\Lambda$ is present in (\ref{vas5}) to
obtain a Weyl tensor with $\Lambda$ independent coefficients in the
linearized equations in such a way that their flat limit also makes
sense. In the following, however, we will again keep $V$ normalized
to $1$, which  means $\Lambda=-1$ (see Section \ref{mdmswg}).

\subsection{Formal consistency}

The system is formally consistent, {\it i.e.} compatible with
$d^2=0$ and with associativity. A detailed proof of this statement
can be found in the appendix \ref{conscheck}. Let us however point
out here the only tricky step. To prove the consistency, one has to
show that the associativity relation $S\*(S\* S)=(S\* S)\* S$ is
compatible with the equations. This is in fact the form of the
Bianchi identity with respect to the $Z$ variables, because $S$
actually acts as a sort of exterior derivative in the noncommutative
space (as will be shown in the next section). Associativity seems
then to be broken by the source term $B\*\ck$ which anticommutes
with $dz_i$ as a consequence of (\ref{vas4}) and the definition
(\ref{tauonS}), and brings in a term proportional to $dz_idz^idz_j$
(where $dz_i=V_A dZ^A_i$). This is not a problem however: since $i$
is an $sp(2)$ index, it can take only two values and, as a result,
the antisymmetrized product of three indices vanishes identically:
$dz_idz^idz_j=0$.

In a more compact way, one can prove consistency by introducing the
noncommutative extended covariant derivative $\cw=d+W+S$ and
assembling eqs. (\ref{vas1})-(\ref{vas5}) into $$
\cw\*\cw\,=\,-\frac{1}{2}\,dZ^i_A dZ_i^A\,+\,2\,dZ_{Ai}dZ^i_B V^A
V^B B\* \ck \ ,$$ $$ \cw\*B=B\*\widetilde{\cw} \ . $$ In other
words, $S\* S$ is nothing but the $ZZ$ component of an $(x,Z)$-space
curvature, and it is actually the only component of the curvature
allowed to be nonvanishing, $xx$ and $xZ$ being trivial according to
(\ref{vas1}) and (\ref{vas3}), respectively. Consistency then
amounts to the fact that the associativity relations
$\cw\*(\cw\*\cw)=(\cw\*\cw)\*\cw$ and $(\cw\*\cw)\*B=\cw\*(\cw\*B)$
are respected by the nonlinear equations. Recall, however, that it
was crucial for the consistency that the symplectic indices take
only two values.

According to the general scheme of free differential algebras, the
consistency of the nonlinear equations implies gauge symmetry under
the local transformations \be\label{gaugetr}
\delta\cw=[\epsilon,\cw]_{\*}\ , \qquad \delta B= \epsilon\*
B-B\*\widetilde{\epsilon} \ ,\ee where $\epsilon=\epsilon(Z,Y|x)$ is
$sp(2)$ invariant, {\it i.e.} $[t_{ij}^{tot} ,\epsilon ]_* =0$, and
otherwise arbitrary.

The consistency of the system, which means compatibility of the
equations with the Bianchi identities, both in the $x$ and in the
$Z$ sector (the latter case being verified by associativity of $S$),
guarantees that the perturbative analysis works systematically at
all orders.

\subsection{$Sp(2)$ invariance}
\label{sp2}

The $sp(2)$ symmetry is crucial for the consistency of the free
system, to avoid unwanted ghost degrees of freedom. But as  said
before, survival of the $sp(2)$ invariance at the full nonlinear
level  is also very important, in the sense that it fixes the form
of the nonlinear equations and prevents a mixture of unwanted
degrees of freedom at the nonlinear level.

The rationale behind this is as follows. The conditions
(\ref{totsp2}) guarantee the $sp(2)$ covariance of the whole
framework. But this is not enough because one has to remove traces
by factoring out terms which are themselves proportional to the
$sp(2)$ generators. The third commutation relation in (\ref{totsp2})
makes this difficult. Indeed, it means that the operators $S^A_i$
transform elements of the algebra proportional to $t_{ij}^{tot}$
into $t_{ij}^{tot}$ independent elements, {\it i.e.} the equations
(\ref{vas1})-(\ref{vas5}) do not allow a factorization with respect
to the ideal generated by $t_{ij}^{tot}$.

To avoid this problem at the full nonlinear level one has to build
proper generators
$$ t_{ij}^{int}=t_{ij}+t^1_{ij}+\ldots\,,$$
where $t^1_{ij}$ and higher terms denote the field-dependent
corrections to the original $sp(2)$ generators (\ref{spgen}), such
that they satisfy the $sp(2)$ commutation relations
$$[t^{int}_{ij},t^{int}_{kl}]=\epsilon_{ik}t^{int}_{j\,l}+\epsilon_{jk}t^{int}_{il}+\epsilon_{il}t^{int}_{jk}+\epsilon_{jl}t^{int}_{ik}
$$
and \be \label{invcond} Dt_{ij}^{int}=0 \ , \qquad
[S,t_{ij}^{int}]_{\*}=0 \ , \qquad [
B\*\ck,t_{ij}^{int}]_{\*}=0\,. \ee 

What fixes the form of the nontrivial equations (\ref{vas4}) and
(\ref{vas5}) is just the requirement that such
   nonlinearly deformed $sp(2)$ generators $t_{ij}^{int}$ do exist.
Actually, getting rid of the $dZ$'s in (\ref{vas4}) and (\ref{vas5})
in the longitudinal sector, these equations read as
$$ [s^i,s^j]_{\*}=-\epsilon^{ij}(1-4
B\*\ck) \ , \qquad \{s^i\,,\,B\*\ck\}_{\*}\,=\,0 $$ (where $s^i
\equiv V^A S_A^i$). This is just a realization \cite{Aq} of the so
called deformed oscillator algebra found originally by Wigner
\cite{wig} and discussed by many authors \cite{des} \be \label{def}
[\hat{y}^i,\hat{y}^j]_{\*}=\epsilon^{ij}(1+\nu \hat{k}) \ , \qquad
\{\hat{y}^i,\hat{k}\}_{\*}=0 \ , \ee $\nu$ being a central element.
The main property of this algebra is that, for any $\nu$, the
elements $\tau_{ij}=-\frac{1}{2}\{s_i,s_{j}\}_{\*}$ form the $sp(2)$
algebra  that rotates properly $s_i$
$$[\tau_{ij},s_k]_{\*}=\epsilon_{ik}s_j+\epsilon_{jk}s_i\,.$$

As a consequence, there exists an $sp(2)$ generator
$$\ct_{ij}=-\frac{1}{2}\{S^A_i,S_{Aj}\}_{\*}\,,$$
which acts on $S^A_i$ as
$$[\ct_{ij},S^A_k]_{\*}=\epsilon_{ik}S^A_j+\epsilon_{jk}S^A_i\,.$$
As a result, the difference \be\label{tint}
   t_{ij}^{int}\equiv
t_{ij}^{tot}-\ct_{ij} \ee
   satisfies the $sp(2)$ commutation relation
and the conditions (\ref{invcond}), taking into account
(\ref{totsp2}) and (\ref{vas1}) - (\ref{vas5}). Moreover, at the
linearized level, where $S^A_i=Z^A_i$ as will be shown in the next
section, $t_{ij}^{int}$ reduces to $t_{ij}$. This means that, if
nonlinear equations have the form (\ref{vas1}) - (\ref{vas5}),
interaction terms coming from the evolution along noncommutative
directions do not spoil the $sp(2)$ invariance and allow the
factorization of the elements proportional to $t^{int}_{ij}$. This,
in turn, implies that the nonlinear equations admit an
interpretation in terms of the tensor fields we started with in the
free field analysis. Let us also note that by virtue of (\ref{tint})
and (\ref{vas1})-(\ref{vas5}) the conditions (\ref{invcond}) are
equivalent to (\ref{totsp2}).

Finally, let us mention that an interesting interpretation of the
deformed oscillator algebra (\ref{def}) is \cite{Aq} that it
describes a two-dimensional fuzzy sphere of a $\nu$-dependent
radius. Comparing this with the equations (\ref{vas4}) and
(\ref{vas5}) we conclude that the nontrivial HS equations describe a
two-dimensional fuzzy sphere embedded into a noncommutative space of
variables $Z^A$ and $Y^A$. Its radius varies from point to point of
the usual (commutative) space-time with coordinates $x$, depending
on the value of the HS curvatures collectively described by the Weyl
zero-form $B(Z,Y|x)$.

\subsection{Factoring out the ideal}

The factorization procedure is performed analogously to the
linearized analysis of Subsection \ref{factorization} by choosing
one or another tracelessness condition for representatives of the
equivalence classes and then dropping the terms of the form
$f=t^{int}_{ij} *g^{ij}$, $[f,t^{int}_{ij}]=0$ as explained in more
detail in Subsection  (\ref{refsi}).

Equivalently, one can use the quasiprojector approach of
\cite{0404124} exposed in Section \ref{star}. To this end one
defines a nonlinear quasiprojector $\Delta^{int}=\Delta+\ldots$  as
a nonlinear extension of the operator $\Delta$ (\ref{Delta}),
satisfying \be [S\,,\Delta^{int}]_* =0\,,\qquad D(\Delta^{int})=0
\ee and \be \Delta^{int} *t_{ij}^{int}=t_{ij}^{int}*\Delta^{int}
=0\,. \ee The equations with factored out traces then take the
form\footnote{This form of the HS field equations was also
considered by E. Sezgin and P. Sundell (unpublished)
 as one of us (MV)
learned from a private discussion during the Solvay workshop.}
\be\label{vas1d}\Delta^{int}*( dW + W\* W)=0 \ , \ee
\be\label{vas2d} \Delta^{int}*(dB +W\* B - B\* \widetilde{W})=0 \ ,
\ee \be\label{vas3d} \Delta^{int}*(dS + W \* S + S\* W)=0 \ , \ee
\be\label{vas4d} \Delta^{int}*(S\* B- B\* \widetilde{S})=0 \ ,\ee
\be\label{vas5d} \Delta^{int}*\left (S\* S +\frac{1}{2}(dZ^i_A
dZ_i^A+4\Lambda^{-1}dZ_{Ai}dZ^i_B V^A V^B B\* \ck)\right )=0 \ . \ee
The factors of $\Delta^{int}$ here ensure that all terms
proportional to $t_{ij}^{int}$ drop out. Note that $\Delta$ and,
therefore, $\Delta^{int}$ do not belong to the regular class, and
their star products with themselves and similar operators are
ill-defined. However, as pointed out in Appendix E, since $\Delta$
and $\Delta^{int}$ admit expansions in power series in $Z^A_i$ and
$Y^A_j$, their products with regular functions are well-defined
(free of infinities), so that the equations
(\ref{vas1d})-(\ref{vas5d}) make sense  at all orders. Note that, in
practice, to derive manifest component equations on the physical HS
modes within this approach  it is anyway necessary to choose a
representative of the quotient algebra in one or another form of the
tracelessness conditions as discussed in Sections \ref{framelike}
and \ref{tw}.

Let us note that the idea to use the strong $sp(2)$ condition
suggested in \cite{SSS} \be \label{str} t_{ij}^{int}*B=B*
t_{ij}^{int}=0\, \ee is likely to lead to a problem beyond the
linearized approximation. The reason is that the elements satisfying
(\ref{str}) are themselves of the form $\Delta^{int}*B'$
\cite{5d,0404124}, which is beyond the class of regular functions
even in the linearized approximation with $\Delta$ in place of
$\Delta^{int}$. {}From (\ref{vas5}) it follows that the
corresponding field $S$ is also beyond the regular class. As a
result, star products of the corresponding functions that appear in
the perturbative analysis of the field equations may be ill-defined
(infinite), {\it i.e.} imposing this condition may cause infinities
in the analysis of HS equations \cite{SSS}. Let us note that this
unlucky situation is not accidental. A closely related point is that
the elements satisfying (\ref{str})  would form a subalgebra of the
off-mass-shell HS algebra if their product existed. This is not
true, however: the on-mass-shell HS algebra is a quotient algebra
over the ideal ${\cal I}$ but not a subalgebra. This fact manifests
itself in the nonexistence of products of elements satisfying
(\ref{str}) (see also \cite{5d,0404124,SSS}), having nothing to do
with any inconsistency of the HS field equations. The factorization
of the ideal $\ci$ in (\ref{vas1d})-(\ref{vas5d}) is both sufficient
and free of infinities.

\section{Perturbative analysis}
\label{pert}

Let us now expand the equations around a vacuum solution, checking
that the full system of HS equations reproduces the free field
dynamics at the linearized level.

\subsection{Vacuum solution}

The vacuum solution $(W_0,S_0,B_0)$ around which we will expand is
defined by $B_0=0$, which is clearly a trivial solution of
(\ref{vas2}) and (\ref{vas4}). Furthermore, it cancels the source
term in (\ref{vas5}), which is then solved  by \be\label{S0}
S_0=dZ^A_i \,Z^i_A\ . \ee 
Let us point out that $dS_0=0$ and $\widetilde{S}_0=S_0$ by the
definition (\ref{tauonS}). From (\ref{S0}) and \be\label{Zcomm}
[Z_i^A,f]_{\*} (Z,Y) =-\frac{\partial}{\partial Z_A^i}f (Z,Y) \, \ee
(see (\ref{comz})) follows that the (twisted or not) adjoint action
of $S_0$ is equivalent to the action of the exterior differential
$d_Z=dZ_i^A\frac{\partial}{\partial Z^A_i}$ in the auxiliary space.
The equation (\ref{vas3}) at the zeroth order then becomes
$\{W_0\,,S_0\}_{\*}=d_ZW_0=0$ and
 one concludes that $W_0$
can only depend on $Y$ and not on $Z$. One solution of (\ref{vas1})
is the $AdS$ connection bilinear in $Y$ \be \label{W0}
W_0=\omega_0^{AB}(x)T_{AB}(Y) \ , \ee which thus appears as a
natural vacuum solution of HS nonlinear equations. The vacuum
solution (\ref{S0}), (\ref{W0}) satisfies also the $sp(2)$
invariance condition (\ref{totsp2}).

The symmetry of the chosen vacuum solution is $\hu$. Indeed, the
vacuum symmetry parameters $\epsilon^{gl}(Z,Y|x)$ must satisfy \be
[S_0 \,, \epsilon^{gl} ]_* =0 \,,\qquad D_0 (\epsilon^{gl}) =0\,.
\ee The first of these conditions implies that
$\epsilon^{gl}(Z,Y|x)$ is $Z$--independent, {\it i.e.,}
$\epsilon^{gl}(Z,Y|x)=\epsilon^{gl}(Y|x)$
   while the second condition reconstructs the dependence
of $ \epsilon^{gl}(Y|x)$ on space-time coordinates $x$ in terms of
values of $\epsilon^{gl}(Y|x_0)$ at any fixed point $x_0$ of
space-time. Since the parameters are required to be $sp(2)$
invariant, one concludes that, upon factorization of the ideal
$\ci$, the global symmetry algebra is $\hu$. \vspace{.2cm}

Our goal is now to see whether free HS equations emerge from the
full system as first order correction to the vacuum solution. We
thus  set $$ W=W_0+W_1 \ , \qquad S=S_0+S_1 \ , \qquad B=B_0+B_1 \
,$$ and keep terms up to the first order in $W_1,S_1,B_1$ in the
nonlinear equations.

\subsection{First order correction}

 As a result of the fact that the adjoint action of $S_0$ is
equivalent to the action of $d_Z$, if treated perturbatively, the
space-time constraints (\ref{vas4}) and (\ref{vas5}) actually
correspond to differential equations with respect to the
noncommutative $Z$ variables.

We begin by looking at (\ref{vas4}). The zero-form $B=B_1$ is
already first order, so we can substitute  $S$ by $S_0$, to obtain
that $B_1$ is $Z$-independent \be\label{B1} B_1 (Z,Y)=C(Y|x) \ . \ee
Inserting this solution into (\ref{vas2}) just gives the twisted
adjoint equation  (\ref{tweq}) (with $\widetilde{D}_0$ defined by
(\ref{twc})), one of the two field equations we are looking for.

Next we attempt to find $S_1$ substituting (\ref{B1}) into
(\ref{vas5}), taking into account that
$$ f(Z,Y)\*\ck=\exp(-2 z_i y^i ) \,f(Z^A_i-V^A(z_i+y_i),\,
Y^A_i-V^A(z_i+y_i)) \,$$ (see Appendix \ref{appklein}), which one
can write as
$$ f(Z,Y)\*\ck=\exp(-2 z_i y^i ) \,f(^\perp Z-
\,^\parallel Y, ^\perp Y- \,^\parallel Z) \ .$$  This means that
$\ck$ acts on functions of $Z$ and $Y$ by interchanging their
respective longitudinal parts (taken with a minus sign) and
multiplying them by a factor of $\exp(-2 z_i y^i )$.

Looking at the $ZZ$ part of the curvature, one can see that the
$V^A$ transversal sector is trivial at first order and that the
essential $Z$-dependence is concentrated only in the longitudinal
components. One can then analyze the content of (\ref{vas5}) with
respect to the longitudinal direction only, getting \be\label{5pert}
\partial^i s_1^j-\partial^j
s_1^i=4 
\varepsilon^{ij}C (\,^\perp Y- \,^\parallel Z)\, \exp(-2 z_i y^i )
\ee with $\partial^i=\frac{\partial}{\partial z_i}$ and $s^i_1 =
S^i_{1\,A}V^A$. The general solution of the equation $\partial_i
f^i(z)=g(z)$ is $f_i(z)=\partial_i\epsilon+\int_0^1 dt \,t \,z_i
g(z)$. Applying this to (\ref{5pert}) one has
$$ s_1^i=\partial^i \epsilon_1-2 
z^i \int_0^1 dt \;t\,C(\,^\perp Y- t\,^\parallel Z)\exp (-2tz_k y^k)
\ .$$  Analogously, in the $V$ transverse sector one obtains that
${}^\perp S^A_i$ is pure gauge so that
$$ S_{1\,A}^i=\frac{\partial}{\partial Z_i^A} \epsilon_1-2
V_A z^i \int_0^1 dt \;t\,C(\,^\perp Y- t\,^\parallel Z)\exp (-2tz_k
y^k) \,,$$ where the first term on the r.h.s. is the $Z$-exact part.
This term is the pure gauge part with the gauge parameter
$\epsilon_1=\epsilon_1(Z;Y|x)$ belonging to the $Z$-extended HS
algebra. One can conveniently set $\frac{\partial}{\partial Z^A_i}
\epsilon_1=0$ by using part of the gauge symmetry (\ref{gaugetr}).
This choice fixes the $Z$-dependence of the gauge parameters to be
trivial and leaves exactly the gauge freedom one had at the free
field level, $\epsilon_1=\epsilon_1(Y|x)$. Moreover, let us stress
that with this choice one has reconstructed $S_1$ entirely in terms
of $B_1$. Note  that $s_1^i$ belongs to the regular class of
functions (\ref{class}) compatible with the star product.

We now turn our attention to the equation (\ref{vas3}), which
determines the dependence of $W$ on $z$. In the first order, it
gives
$$\partial^i W_1= ds^i_1+W_0\* s^i_1-s^i_1\* W_0 \ . $$ The general
solution of the equation $\frac{\partial}{\partial
z_i}\varphi(z)=\chi^i(z)$ is given by the line integral
$$\varphi(z)=\varphi(0)+\int_0^1 dt
\,z_i\chi^i(tz)\,,$$ provided that $ \frac{\partial}{\partial
z_i}\chi^i (z)=0 $ ($ i=1,2 $). Consequently, \be\label{W1}
W_1=\omega(Y|x)+z^j\int_0^1 dt\, (1-t)\,e^{-2tz_i y^i}E_0^B
\frac{\partial}{\partial {}^\perp Y^{jB}}\, C(\,^\perp Y-
t\,^\parallel Z) \ , \ee taking into account that the term $z_i
ds_1^i$ vanishes because $z_i z^i=0$. Again, $W_1$ is in the regular
class. Note also that in (\ref{W1}) only the frame field appears,
while the dependence on the Lorentz connection cancels out. This is
the manifestation of the local Lorentz symmetry which forbids the
presence of $\omega^{ab}$ if not inside Lorentz covariant
derivatives.

One still has to analyze (\ref{vas1}), which at first order reads
$$ dW_1+\{W_0,W_1\}_{\*}=0 \ . $$ Plugging in (\ref{W1}), one  gets
\be\label{corcur} R_1=O(C) \ , \qquad R_1 \equiv  d \o +\{\o \,, W_0
\}_*\,, \ee where corrections on the r.h.s. of the first equation in
(\ref{corcur}) come from the second term in (\ref{W1}), and prevent
(\ref{corcur}) from being trivial, that would  imply $\omega$ to be
a pure gauge solution\footnote{In retrospective, one sees that this
is the reason why it was necessary for the $Z$ coordinates to be
noncommutative, allowing nontrivial contractions, since corrections
are obtained from perturbative analysis as coefficients of an
expansion in powers of $z$ obtained by solving for the
$z$-dependence of the fields from the full system.}. The formal
consistency of the system with respect to $Z$ variables allows one
to restrict the study of (\ref{vas1}) to the physical space $Z=0$
only (with the proviso that the star products are to be evaluated
before sending $Z$ to zero). This is due to the fact that the
dependence on $Z$ is reconstructed by the equations in such a way
that if (\ref{vas1}) is satisfied for $Z=0$, it is true for all $Z$.
By elementary algebraic manipulations one obtains the final  result
$$ R_1  
=\frac{1}{2} E_0^A\,E_0^B\,\frac{\partial^2}{\partial\,^\perp
Y^{Ai}\partial\,^\perp Y_i^{B}}C(\,^\perp Y) \ , $$ which, together
with the equation for the twisted adjoint representation previously
obtained, reproduces the free field dynamics for all spins
(\ref{REF1}) and (\ref{tweq}).

\subsection{Higher order corrections and factorization of the ideal}
\label{refsi}

Following the same lines one can now reconstruct order-by-order all
nonlinear corrections to the free HS equations of motion. Note that
all expressions that appear in this analysis belong to the regular
class (\ref{class}), and therefore the computation as a whole is
free from divergencies, being well defined. At the same time, the
substitution of expressions like (\ref{hder1}) for auxiliary fields
will give rise to nonlinear corrections with higher derivatives,
which are nonanalytic in the cosmological constant.

Strictly speaking, the analysis explained so far is off-mass-shell.
To put the theory on-mass-shell one has to factor out the ideal
$\ci$. To this end, analogously to the linearized analysis of
Section \ref{tw}, one has to fix representatives of $\o (Y|x)$ and
$C(Y|x)$ in one or another way (for example, demanding $\o (Y|x)$ to
be $AdS$ traceless and $C(Y|x)$ to be Lorentz traceless). The
derived component HS equations may or may not share these
tracelessness properties.  Let us consider a resulting expression
containing some terms of the form
$$A=A_0^{tr} +t_{ij}\,A^{ij}_1 \,,$$ where $A_0^{tr}$ satisfies the
chosen tracelessness condition while the second term, with $t_{ij}$
defined by (\ref{gen2}), describes extra traces. One has to rewrite
such terms as \be A=A_0^{tr} +t^{int}_{ij}*A^{ij}_1
+A^\prime+A''\,,\label{fourterms}\ee where the third term
$$A^\prime\equiv t_{ij}\,A^{ij}_1 -t_{ij}*A^{ij}_1=-\frac{1}{4}
\frac{\partial^2}{\partial
Y^i_A \partial Y^{jA}} A^{ij}$$ contains less traces (we took into
account (\ref{tpr})), and the fourth term
$$A''\equiv(t^{int}_{ij}-t_{ij})*A^{ij}_1=O(C)\*A_1^{ij}$$
contains higher-order corrections due to the definitions (\ref{tot})
and (\ref{tint}). The factorization is performed by dropping out the
terms of the form $t^{int}_{ij}*A^{ij}_1$ in (\ref{fourterms}). The
resulting expression $A=A_0^{tr} +A^\prime+A''$ contains either less
traces or higher order nonlinearities. If necessary, the procedure
has to be repeated to get rid of lower traces at the same order or
new traces at the nonlinear order. Although this procedure is
complicated, it can be done in a finite number of well-defined steps
for any given type of  HS tensor field and given order of
nonlinearity. The process is much nicer of course within the
spinorial realization available in the lower dimension cases
\cite{con,Vasiliev:1989yr,prok} where the factorization over the
ideal $\ci$ is automatic.

\section{Discussion}
\label{disc}

A surprising issue related to the structure of the HS equations of
motion is that, within this formulation of the dynamics, one can get
rid of the space-time variables. Indeed, (\ref{vas1})-(\ref{vas3})
are zero curvature and covariant constancy conditions, admitting
pure gauge solutions. This means that, locally,
\bqn W(x) & = & g^{-1}(x)\* dg(x) \ , \nonumber \\
B(x) & = & g^{-1}(x)\* b\* \widetilde{g}(x) \ , \nonumber \\
S(x) & = & g^{-1}(x)\* s\* g(x) \ , \label{purg} \eqn the whole
dependence on space-time points being absorbed into a gauge function
$g(x)$, which is an arbitrary invertible element of the star product
algebra,  while $b = b (Z,Y)$ and $s = s (Z,Y)$ are arbitrary
$x$-independent elements of the star product algebra.  Since the
system is gauge invariant, the gauge functions disappear from the
remaining two equations (\ref{vas4})-(\ref{vas5}), which then encode
the whole dynamics though being independent of $x$. This turns out
to be possible, in the unfolded formulation, just because of the
presence of an infinite bunch of fields, supplemented by an infinite
number of appropriate constraints, determined by consistency. As
previously seen in the lower spin examples (see Section
\ref{unfolding}), the zero-form $B$ turns out to be the generating
function of all on-mass-shell nontrivial derivatives of the
dynamical fields. Thus it locally reconstructs their $x$-dependence
through their Taylor expansion which in turn is just given by the
formulas (\ref{purg}). So, within the unfolded formulation, the
dynamical problem is well posed once all zero-forms assembled  in
$b$ are given at one space-time point $x_0$, because this is
sufficient to obtain the whole evolution of fields in some
neighborhood of $x_0$  (note  that $s$ is reconstructed in terms  of
$b$ up to a pure gauge part). This way of solving the nonlinear
system, getting rid of $x$ variables, is completely equivalent to
the one used in the previous section, or, in other words, the
unfolded formulation involves a trade between space-time variables
and auxiliary noncommutative variables $(Z,Y)$. Nevertheless, the
way we see and perceive the world seems to require the definition of
local events, and it is this need for locality that makes the
reduction to the ``physical'' subspace $Z=0$ (keeping the
$x$-dependence instead of gauging it away) more appealing. On the
other hand, as mentioned in Section \ref{sp2}, the HS equations in
the auxiliary noncommutative space have the clear geometrical
meaning of describing embeddings of a two-dimensional noncommutative
sphere into the Weyl algebra.

The system of gauge invariant nonlinear equations for all spins in
$AdS_d$ here presented can be generalized \cite{Vasiliev:2003ev} to
matrix-valued fields, $W \to W_\a^{\ \b}$, $S \to S_\a^{\ \b}$ and
$B \to B_\a^{\ \b}$, $\a,\b=1,...,n$, giving rise to Yang-Mills
groups $U(n)$ in the $s=1$ sector while remaining consistent. It is
also possible to truncate to smaller inner symmetry groups $USp(n)$
and $O(n)$ by imposing further conditions based on certain
antiautomorphisms of the  star product algebra
\cite{Konstein:ij,Vasiliev:2003ev}. Apart from the possibility of
extending the symmetry group with matrix-valued fields, and modulo
field redefinitions, it seems that there is no ambiguity in the form
of nonlinear equations. As previously noted, this is due to the fact
that the $sp(2)$ invariance requires (\ref{vas4}) and (\ref{vas5})
to have the form of a deformed oscillator algebra.

HS models have just one dimensionless coupling constant $$
g^2=|\Lambda|^{{d\over 2}-1}\kappa^2 \,.$$ To introduce the coupling
constant, one has to rescale the fluctuations $\omega_1$ of the
gauge fields ({\it i.e.} additions to the vacuum field) as well as
the Weyl zero-forms by a coupling constant $g$ so that it cancels
out in the free field equations. In particular, $g$ is identified
with the Yang-Mills coupling constant in the spin-$1$ sector. Its
particular value is artificial however because it can be rescaled
away in the classical theory (although it is supposed to be a true
coupling constant in the quantum theory where it is a constant in
front of the whole action in the exponential inside the path
integral). Moreover, there is no dimensionful constant allowing us
to discuss a low-energy expansion, {\it i.e.} an expansion in powers
of a dimensionless combination of this constant and the covariant
derivative. The only dimensionful constant present here is
$\Lambda$. The dimensionless combinations $\bar{D}_\mu\equiv
\Lambda^{-1/2}D_\mu$ are not good expansion entities, since the
commutator of two of them is of order $1$, as a consequence of the
fact that the $AdS$ curvature is roughly $R_0\sim D^2 \sim \Lambda
gg$. For this reason also it would be important to find solutions of
HS field equations different from $AdS_d$, thus introducing in the
theory a massive parameter different from the cosmological constant.

Finally, let us note that a variational principle giving the
nonlinear equations (\ref{vas1})-(\ref{vas5}) is still unknown in
all orders. Indeed, at the action level, gauge invariant
interactions have been constructed only up to the cubic order
\cite{FV1,5d,cubint}.

\section{Conclusion}

 The main message of these lectures is that nonlinear dynamics
of HS gauge fields can be consistently formulated in all orders in
interactions in anti de Sitter space-time of any dimension
$d\geqslant 4$. The level of generality of the analyzes covered
has been restricted in the following points: only completely
symmetric bosonic HS gauge fields have been considered, and only at
the level of the equations of motion.

Since it was impossible to cover 
all the interesting and important directions of research in the
modern HS gauge theory, an invitation to further readings is
provided as a conclusion. For general topics in HS gauge theories,
the reader is referred to the review papers
\cite{9910096,AdSCFT,AdSCFT2,Sorokin,Vasiliev:1995dn,lectures,Sezgin:2002,Iazeolla}.
Among the specific topics that have not been addressed here, one can
mention:
\begin{description}
    \item[(i)] Spinor realizations of HS superalgebras in $d=3,4$
\cite{FVA,V3, Konstein:ij,prok}, $d=5$ \cite{FLA,SS5}, $d=7$
\cite{Sezgin/Sundell-7} and the recent developments in any dimension
\cite{0404124},
    \item[(ii)] Cubic action interactions \cite{FV1,5d,cubint},
    \item[(iii)] Spinor form of $d=4$ nonlinear HS field equations
\cite{con,9910096,SS8},
\item[(iv)] HS dynamics in larger
(super)spaces, {\it e.g.} free HS theories in tensorial superspaces
\cite{Bandos:1999qf,d4sym,GV} and HS theories in usual superspace
\cite{sup,Sundell},
    \item[(v)] Group-theoretical classification of invariant
equations via unfolded formalism \cite{STV},
    \item[(vi)] HS
gauge fields different from the completely symmetric Fronsdal
fields: {\it e.g.} mixed symmetry fields \cite{mixed}, infinite
component massless representations \cite{infs} and partially
massless fields \cite{Deser:2004},
    \item[(vii)] Light cone formulation for massless fields in
$AdS_d$ \cite{metc},
    \item[(viii)] Tensionless limit of quantized (super)string theory
    \cite{tenstring,tenslimstr}.
\end{description}

\section*{Acknowledgements}

We are grateful to the Organizers of the ``First Solvay Workshop"
for this enjoyable meeting. M.V. is grateful to I.Bandos,
M.Grigoriev and especially to A. Sagnotti and P. Sundell for very
useful discussions  and to D. Sorokin for helpful comments on the
manuscript. X.B., S.C. and C.I. are thankful to M. Tsulaia for his
careful reading of a preliminary version of the manuscript and his
comments. X.B. and S.C. also thank N. Boulanger for useful
discussions. X.B. acknowledges T. Damour for interesting
discussions. C.I. is grateful to A. Sagnotti for support and
valuable comments. We thank N. Boulanger and K. Ushakov for
drawing our attention to wrong coefficients respectively in Eqs. (3.8) and (10.12), fixed 
in the present version (v3) of this paper.  \vspace{.2cm}

The work of M.V. was supported in part by grants RFBR No.
05-02-17654, LSS No. 1578.2003-2 and INTAS No. 03-51-6346. These
lecture notes were taken while X.B. was in the university of Padova,
supported by the European Commission RTN program HPRN-CT-00131. The
work of S.C. is supported in part by the ``Actions de Recherche
Concert{\'e}es'' of the ``Direction de la Recherche Scientifique -
Communaut{\'e} Fran{\c c}aise de Belgique'', by a ``P\^ole
d'Attraction Interuniversitaire'' (Belgium), by IISN-Belgium
(convention 4.4505.86) and by the European Commission FP6 programme
MRTN-CT-2004-005104, in which she is associated to V.U.Brussel. The
work of C.I. was supported in part by INFN, by the EU contracts
HPRN-CT-2000-00122 and HPRN-CT-2000-00148, by the MIUR-COFIN
contract 2003-023852, by the INTAS contract 03-51-6346 and by the
NATO grant PST.CLG.978785.


\appendix

\section{Basic material on lower spin gauge theories}\label{appA}

\subsection{Isometry algebras}\label{iso}

By ``space-time" symmetries one means symmetries of the
corresponding space-time manifold ${\cal M}^d$ of dimension $d$,
which  may be isometries or conformal symmetries. The most
symmetrical solutions of vacuum Einstein equations, with or without
cosmological constant $\Lambda$, are (locally) Minkowski space-time
($\Lambda=0$), de Sitter ($\Lambda>0$) and Anti de Sitter
($\Lambda<0$) spaces. In this paper, we only consider ${\mathbb
R}^{d-1,1}$ and $AdS_d$ spaces though the results generalize easily
to $dS_d$ space.

The Poincar\'e group $ISO(d-1,1)= {\mathbb R}^{d-1,1}\rtimes
SO(d-1,1)$ has translation generators $P_a$ and Lorentz generators
$M_{ab}$ ($a,b=0,1,\ldots,d-1$) satisfying the algebra
\begin{equation}
[\,M_{ab}\,,\,M_{cd}\,] = \eta_{ac} \, M_{db}- \eta_{bc} \, M_{da}
-\eta_{ad} \, M_{cb} +\eta_{bd} \, M_{ca}
\,,\label{Lorentz}\end{equation}
\begin{equation}[\,P_a\,,\,M_{bc}]
=\eta_{ab}\,P_{c}-\eta_{ac}\,P_{b}\label{PM}\,,\end{equation}
\begin{equation}[P_a,P_b]=0\,.\label{translations}\end{equation}
   ``Internal" symmetries are defined as transformations that commute
with the translations generated by $P_a$ and the Lorentz
transformations generated by $M_{ab}$ \cite{cm}. The relation
(\ref{Lorentz}) defines the Lorentz algebra $o(d-1,1)$ while the
relations (\ref{PM})-(\ref{translations}) state that the Poincar\'e
algebra is a semi-direct product $iso(d-1,1)={\mathbb R}^{d} \niplus
o(d-1,1)$.

The algebra of isometries of the $AdS_d$ space-time is given by the
commutation relations (\ref{Lorentz})-(\ref{PM}) and
\begin{equation}[P_a,P_b]=-\frac{1}{\r^2}\,M_{ab}\,,
\label{transv}\end{equation} where $\r$ is proportional to the
radius of curvature of $AdS_d$ and is related to the cosmological
constant via $\Lambda=-\r^{-2}$ . The (noncommuting) transformations
generated by $P_a$ are called transvections in $AdS_d$ to
distinguish them from the (commuting) translations. By defining
$M_{\hat{d}\,a}=\r\,P_a$, it is possible to collect all generators
into the generators $M_{AB}$ where $A=0,1,\ldots,\hat{d}$. These
generators $M_{AB}$ span $o(d-1,2)$ algebra since they satisfy the
commutation relations
\begin{equation}
[M_{AB},M_{CD}]=\eta_{AC}M_{DB}-\eta_{BC}M_{DA}-\eta_{AD}M_{CB}+\eta_{BD}M_{CA}\,,\nonumber
\end{equation}
where $\eta_{AB}$ is the mostly minus invariant metric of
$o(d-1,2)$. This is easily understood from the geometrical
construction of $AdS_d$ as the hyperboloid defined by
$X^AX_A={(d-1)(d-2)\over 2}\r^2$ which is obviously invariant under
the  isometry group $O(d-1,2)$. Since transvections are actually
rotations in ambient space, it is normal that they do not commute.
It is possible to derive the Poincar\'e algebra from the $AdS_d$
isometry algebra by taking the infinite-radius limit
$\r\rightarrow\infty$. This limiting procedure is called
In\"{o}n\"{u}-Wigner contraction \cite{Inonu:1953sp}.

\subsection{Gauging internal symmetries}\label{YM}

In this subsection, a series of tools used in any gauge theory is
briefly introduced. One considers the most illustrative example of
Yang-Mills theory, which corresponds to gauging an internal symmetry
group.

   Let $g$ be a (finite-dimensional) Lie algebra of basis $\{T_\a\}$
and Lie bracket $[\,\,,\,]$. The structure constants are defined by
$[T_\a,T_\b]=T_\g f^\g{}_{\a\b}$. The Yang-Mills theory is
conveniently formulated by using differential forms taking values in
the Lie algebra $g$.
\begin{itemize}
      \item The connection $A=dx^\mu A^\a_\mu  T_\a$ is defined in terms
the vector gauge field
      $A_\m^\a$.
      \item The curvature $F=dA+A^2=\frac{1}{2}\, dx^\m dx^\n
F^\a_{\m\n} T_\a$ is associated with the field strength tensor
$F^\a_{\m\n}=\partial_{[\m}A^\a_{\nu]}+f^\a{}_{\b\g}A^\b_\m
A^\g_\n$.
      \item The Bianchi identity $dF+AF-FA=0$ is a consequence
      of $d^2=0$ and the Jacobi identity in the Lie algebra.
      \item The gauge parameter $\epsilon=\epsilon^\a T_\a$ is
associated with the infinitesimal gauge transformation
      $\d_{\epsilon} A=d\epsilon +A\epsilon-\epsilon A$ which transforms
      the curvature as
        $\d_{\epsilon} F=
        F\epsilon-\epsilon F$. In components, this reads as
$\d_{\epsilon^\a}
      A^\a_\m=\partial_\m\epsilon^\a+f^\a_{\b\g}A_\m^\b\epsilon^\g$ and
$\d F_{\m\n}^\a=f^\a_{\b\g}F^\b_{\m\n}\epsilon^\g$.
\end{itemize}
The algebra $\Omega({\cal M}^d)\otimes g$ is a Lie superalgebra, the
product of which is the graded Lie bracket denoted by
$[\,\,\,,\,\,]$ \footnote{Here, the grading is identified with that
in the exterior algebra so that the graded commutator is evaluated
in terms of the original Lie bracket $[\,\,\,,\,\,]$.}. The elements
of $\Omega^p({\cal M}^d)\otimes g$ are $p$-forms taking values in
$g$. The interest of the algebra $\Omega({\cal M}^d)\otimes g$ is
that it contains the gauge parameter $\epsilon\in\Omega^0({\cal
M}^d)\otimes g$, the connection $A\in\Omega^1({\cal M}^d)\otimes g$,
the curvature $F\in\Omega^2({\cal M}^d)\otimes g$, and the Bianchi
identity takes place in $\Omega^3({\cal M}^d)\otimes g$.

To summarize, the Yang-Mills theory is a fibre bundle construction
where the Lie algebra $g$ is the fiber, $A$ the connection and $F$
the curvature. The Yang-Mills action takes the form
$$S^{YM}[A^a_\m]\propto\int_{{\cal M}^d} Tr[F\,{}^* F]\;,$$in
which case $g$ is taken to be  compact and semisimple so that the
Killing form is negative definite (in order to ensure that the
Hamiltonian is bounded from below).
 Here ${}^*$ is the Hodge star producing a dual
form. Note that, via Hodge star, the Yang-Mills action contains the
metric tensor which is needed to achieve invariance under
diffeomorphisms. Furthermore, because of the cyclicity of the trace,
the Yang-Mills Lagrangian $Tr[F \,{}^* F]$ is also manifestly
invariant under the gauge transformations.

An operatorial formulation is also useful for its compactness. Let
us now consider some matter fields $\Phi$ living in some space $V$
on which acts the Lie algebra $g$, via a representation $T_\a$.
   In other words, the elements $T_\a$ are reinterpreted as
operators acting on some representation space (also called module)
$V$. The connection $A$ becomes thereby an operator. For instance,
if $T$ is the adjoint representation then the module $V$ is
identified with the Lie algebra $g$ and $A$ acts as
$A\cdot\Phi=[A,\Phi]$. The connection $A$ defines the covariant
derivative $D\equiv d+A$. For any representation of $g$, the
transformation law of the matter field is
$\d\Phi=-\,\epsilon\cdot\Phi$, where $\epsilon$ is a constant or a
function of $x$ according to whether
   $g$ is a global or a local symmetry. The gauge transformation
law of the connection can also be written  as $\d A=\d
D=[D,\epsilon]$ because of the identity $[d,A]=dA$, and is such that
$\d (D\Phi)=-\,\epsilon\cdot (D\Phi)$. The curvature is economically
defined as an operator $F=\frac12[D,D]=D^2$. In space-time
components, the latter equation reads as usual
$[D_\m,D_\n]=F_{\m\n}$. The Bianchi identity is a direct consequence
of the associativity of the differential algebra and Jacobi
identities of the Lie algebra  and reads in space-time components as
$[D_{\m},F_{\n\r}]+[D_{\n},F_{\r\m}]+[D_{\r},F_{\m\n}]=0$. The
graded Jacobi identity leads to $\d
F=\frac12\,\Big(\,[\,[D,\epsilon]\,,D\,]+[\,D,[D,\epsilon]\,]\,\Big)
=[D^2\,,\,\epsilon]=[F\,,\,\epsilon]$.

The present notes make an extension of the previous compact
notations and synthetic identities. Indeed, they generalize it
straightforwardly to other gauge theories formulated via a
nonAbelian connection, {\it e.g.} HS gauge theories\footnote{More
precisely, one can take $g$ as an infinite-dimensional Lie algebra
that arises from an associative algebra with product law $*$ and is
endowed with the (sometimes twisted) commutator as bracket. Up to
these subtleties and some changes of notation, all previous
relations hold for HS gauge theories considered here, and they might
simplify some explicit checks by the reader.}.

\subsection{Gauging space-time symmetries}\label{sptime}

The usual Einstein-Hilbert action $S[g]$ is invariant under
diffeomorphisms. The same is true for $S[e,\o]$, defined by
(\ref{mdma}), since everything is written in terms of differential
forms. The action (\ref{mdma}) is also manifestly invariant under
local Lorentz transformations $\d \o=d\epsilon+[\o,\epsilon]$ with
gauge parameter $\epsilon=\epsilon^{ab}M_{ab}$, because
$\epsilon_{a_1 \ldots a_{d}}$ is an invariant tensor of $SO(d-1,1)$.
The gauge formulation of gravity shares many features with a
Yang-Mills theory formulated in terms of a connection $\o$ taking
values in the Poincar\'e algebra.

However, gravity is actually \textit{not} a Yang-Mills theory with
Poincar\'e as (internal) gauge group. The aim of this section is to
express precisely the distinction between gauge symmetries which are
either internal or space-time.

To warm up, let us mention several obvious differences between
Einstein-Cartan's gravity and Yang-Mills theory. First of all, the
Poincar\'e algebra $iso(d-1,1)$ is not semisimple (since it is not a
\textit{direct} sum of simple Lie algebras, containing a nontrivial
Abelian ideal spanned by translations). Secondly, the action
(\ref{mdma}) cannot be written in a Yang-Mills form $\int Tr[F\,{}^*
F]$. Thirdly, the action (\ref{mdma}) is not invariant under the
gauge transformations $\d \o=d\epsilon+[\o,\epsilon]$ generated by
{\it all} Poincar\'e algebra generators, {\it i.e.} with gauge
parameter $\epsilon (x)=\epsilon^a (x) P_a+\epsilon^{ab} (x)
M_{ab}$.  For $d>3$, the action (\ref{mdma}) is invariant only when
$\epsilon^a= 0$. (For $d=3$, the action (\ref{mdma}) describes a
genuine Chern-Simons theory with local $ISO(2,1)$ symmetries.)

This latter fact is not in contradiction with the fact that one
actually gauges the Poincar\'e group in gravity. Indeed, the torsion
constraint allows one to relate the local translation parameter
$\epsilon^a$ to the infinitesimal change of coordinates parameter
$\xi^\m$. Indeed, the infinitesimal diffeomorphism $x^\m\rightarrow
x^\m+\xi^\m$ acts as the Lie derivative $$\d_\xi ={\cal L}_\xi
\equiv i_\xi d+d i_\xi\,,$$ where the inner product $i$ is defined
by
$$i_\xi\equiv \xi^\mu\frac{\partial}{\partial (dx^\mu)}\,,$$
where the derivative is understood to act from the left. Any
coordinate transformation of the frame field can be written as \bqn
\d_\xi e^a=i_\xi(de^a)+d(i_\xi e^a)=i_\xi
T^a+\underbrace{\epsilon^{a}{}_b
e^b+D^L\epsilon^a}_{=\d_{\epsilon}\,e^a}\,,\nonumber\eqn where the
Poincar\'e gauge parameter is given by $\epsilon=i_\xi \omega$.
Therefore, when $T^a$ vanishes any coordinate transformation of the
frame field can be interpreted as a local Poincar\'e transformation
of the frame field, and reciprocally.

To summarize, the Einstein-Cartan formulation of gravity is indeed a
fibre bundle construction where the Poincar\'e algebra $iso(d-1,1)$
is the fiber, $\o$ the connection and $R$ the curvature, but, unlike
for Yang-Mills theories, the equations of motion imposes some
constraints on the curvature ($T^a=0$), and some fields are
auxiliary ($\omega^{ab}$). A fully covariant formulation is achieved
in the $AdS_d$ case with the aid of compensator formalism as
explained in Section \ref{mdmswg}.

\section{Technical issues on nonlinear higher spin equations}\label{appB}

\subsection{Two properties of the inner Klein operator}
\label{appklein}

In this appendix, we shall give a proof of the properties
(\ref{prop1}) and (\ref{prop2}). One can check the second property
with the help of (\ref{enlstar}), which in this case amounts to $$
\ck\*\ck=\frac{1}{\pi^{2(d+1)}}\int dS dT \, e^{-2S^A_i T^i_A}
e^{-2(s_i+z_i)(y^i+s^i)}e^{-2(z_i-t_i)(y^i+t^i)}\ ,$$ with \be
\label{st} s_i\equiv V_A S^A_i\,,\qquad t_i\equiv V_A T^A_i. \ee
   Using the fact
that $s_i s^i=t_i t^i=0$ and rearranging the terms yields \bqn
\ck\*\ck&=& \frac{1}{\pi^{2(d+1)}}e^{-4z_i y^i}\int dS dT \,
e^{-2S^A_i T^i_A} e^{-2s^i(z_i-y_i)}e^{-2t^i(z_i+y_i)} \nonumber
\\ &=&
\frac{1}{\pi^{2(d+1)}}e^{-4z_i y^i}\int dS dT \,
e^{-2S^i_A[V^A(z_i-y_i)-T^A_i]}e^{-2t^i(z_i+y_i)} \ .\nonumber\eqn
Since \bqn\label{delta} \frac{1}{\pi^{2(d+1)}}\int dS
e^{-2S^i_A(Z^A_i-Y^A_i)}=\delta(Z^A_i-Y^A_i) \ ,\eqn one gets $$
\ck\*\ck=e^{-4z_i y^i}\int dT \delta(V^A(z_i-y_i)-T^A_i)e^{-2T^i_B
V^B(z_i+y_i)} $$ which, using $V_B V^B=1$, gives $\ck\*\ck=e^{-4z_i
y^i}e^{-2(-y^i z_i+z^i y_i)}=1$. The formula (\ref{delta}) might
seem unusual because of the absence of an $i$ in the exponent. It is
consistent however, since one can assume that all oscillator
variables, including integration variables, are genuine real
variables times (some fixed) square root of $i$, {\it i.e.} that the
integration is along appropriate lines in the complex plane. One is
allowed to do so without coming into conflict with the definition of
the star algebra, because its elements are analytic functions of the
oscillators, and can then always be continued to real values of the
variables.

The proof of (\ref{prop1}) is quite  similar. Let us note that, with
the help of (\ref{prop2}), it amounts to check that $\ck\*
f(Z,Y)\*\ck=f(\tilde{Z},\tilde{Y})$. One can prove, by going through
almost the same steps shown above, that $$ \ck\* f(Z,Y)=
\ck\,f(Z^i_A + V_A(y^i-z^i), Y^i_A- V_A(y^i-z^i)) \ .$$ Explicitly,
one has $$ \ck\* f(Z,Y)= \frac{\ck}{\pi^{2(d+1)}}\int dS dT
e^{-2S^A_i (T^i_A+V_A(y^i-z^i))}f(Z-T,Y+T)$$ $$=\ck\int dT \,
\delta(T^i_A+V_A(y^i-z^i))f(Z-T,Y+T)$$
$$ =\ck\,f(Z^i_A + V_A(y^i-z^i), Y^i_A- V_A(y^i-z^i)) \ ,$$ where
we have made use of (\ref{delta}). Another star product with $\ck$
leads to
$$\ck\*f(Z,Y)\*\ck=$$
$$ \frac{1}{\pi^{2(d+1)}}e^{-4z_i y^i}\!\!\!\!\int \! dS
dT e^{-2T^i_A[V^A(z_i+y_i)+S^A_i]}e^{-2s_i(y^i-z^i)}f(Z^i_A +
V_A(y^i-z^i)+S^i_A, Y^i_A- V_A(y^i-z^i)+S^i_A)\ ,$$  which,
performing the integral over $T$ and using (\ref{delta}), gives in
the end $$ e^{-4z_i
y^i}e^{2(z_i+y_i)(y^i-z^i)}f(Z^i_A+V_A(y^i-z^i)-V_A(z^i+y^i),
Y^i_A-V_A(y^i-z^i)-V_A(z^i+y^i))=f(\tilde{Z},\tilde{Y}) \ .$$

\subsection{Regularity} \label{reg}

   We will prove that the star product (\ref{enlstar}) is
well-defined for the regular class of functions (\ref{class}). This
extends the analogous result for the spinorial star product in three
and four dimension obtained in \cite{prop,prok}.

\vspace{0.5cm}

\begin{theorem}
Given two regular functions $f_1(Z,Y)$ and $f_2(Z,Y)$, their star
product (\ref{enlstar}) $(f_1\* f_2) (Z,Y)$ is a regular function.
\end{theorem}

\noindent \underline{Proof}:
$$ f_1\* f_2=P_1(Z,Y)\int_{M_1} d\t_1 \rho_1 (\t_1)  \exp
[2\phi_1(\t_1) z_i y^i]\,\*\,P_2(Z,Y)\int_{M_2} d\t_2 \rho_2 (\t_2)
\exp [2\phi_2(\t_2) z_i y^i] $$ $$ = \int_{M_1} d\t_1 \rho_1 (\t_1)
\exp [2\phi_1(\t_1) z_i y^i]\int_{M_2} d\t_2 \rho_2 (\t_2) \exp
[2\phi_2(\t_2) z_i y^i]\frac{1}{\pi^{2(d+1)}}\int dS dT\times
$$
$$\times \exp \{-2S^A_i
T^i_A+2\phi_1(\t_1)[s^i(z-y)_i]+2\phi_2(\t_2)[t^i(z+y)_i]\}
P_1(Z+S,Y+S)P_2(Z-T,Y+T)\ , $$ with $s_i$ and $t_i$ defined  in
(\ref{st}). Inserting \be\label{translosc}
P(Z+U,Y+U)=\exp\left[U^A_i\left(\frac{\partial}{\partial Z_{1
\,i}^A}+\frac{\partial}{\partial Y_{1 \,i}^A}\right)\right]
P(Z_1,Y_1)\bigg|_{{Z_1=Z} \atop {Y_1=Y}} \ , \ee one gets $$ f_1\*
f_2=\int_{M_1\times M_2}d\t_1 d\t_2 \rho_1 (\t_1)\rho_2
(\t_2)\exp\{2[\phi_1(\t_1)+\phi_2(\t_2)]z_i y^i\}\times $$
$$ \times \int dS dT
\exp\left\{-2S^i_A\left[-\phi_1(\t_1)V^A(z-y)_i+\frac{1}{2}\left(\frac{\partial}{\partial
Z_{1 \,A}^i}+\frac{\partial}{\partial Y_{1
\,A}^i}\right)\right]\right\}P_1(Z_1,Y_1)\bigg|_{{Z_1=Z} \atop
{Y_1=Y}}\times $$
$$ \times
\exp\left\{-2T^i_A\left[S^A_i-\phi_2(\t_2)V^A(z+y)_i+\frac{1}{2}\left(-\frac{\partial}{\partial
Z_{2 \,A}^i}+\frac{\partial}{\partial Y_{2
\,A}^i}\right)\right]\right\}P_2(Z_2,Y_2)\bigg|_{{Z_2=Z} \atop
{Y_2=Y}}$$
$$ =\int_{M_1\times M_2}d\t_1
d\t_2 \rho_1 (\t_1)\rho_2
(\t_2)\exp\{2[\phi_1(\t_1)+\phi_2(\t_2)+2\phi_1(\t_1)\phi_2(\t_2)]z_i
y^i\}\times $$
$$ \times \exp\left\{-\frac{1}{2}\left(\frac{\partial}{\partial Z_{1
\,A}^i}+\frac{\partial}{\partial Y_{1
\,A}^i}\right)\left(\frac{\partial}{\partial Z_{2
\,A}^i}-\frac{\partial}{\partial Y_{2
\,A}^i}\right)\right\}P_1[Z_1-\phi_2(\t_2)
\,^\parallel(Z+Y),Z_2-\phi_2(\t_2)\,^\parallel(Z+Y)]\times $$
$$ P_2(Z_2+\phi_1(\t_1)\,^\parallel(Z-Y),
Y_2-\phi_1(\t_1)\,^\parallel(Z-Y))\bigg|_{{Z_1=Z_2=Z} \atop
{Y_1=Y_2=Y}} \ , $$ 
as one can check by using (\ref{delta}) (or, equivalently, Gaussian
integration, the rationale for the equivalence being given in
Appendix \ref{appklein}). The product of two compact domains
$M_1\subset R^n$ and $M_2\subset R^m$ is a compact domain in
$R^{n+m}$ and $P_1, P_2$ are polynomials, so one concludes that the
latter expression is a finite sum of regular functions of the form
(\ref{class}). $\qed$\vspace{0.5cm}

As explained in Section \ref{doubling}, it is  important for this
proof that the exponential in the Ansatz (\ref{class}) never
contributes to the quadratic form in the integration variables,
which is thus independent of the particular choice of the functions
$f_1$ and $f_2$. This property makes sure that the class of
functions (\ref{class}) is closed under star multiplication, and it
is a crucial consequence of the definition (\ref{enlstar}).

Analogously one can easily prove that the star products $f*g$ and
$g*f$ of $f(Z,Y)$ being a power series in $Z^A_i$ and $Y^A_i$ with a
regular function $g(Z,Y)$ are again some power series. In other
words, the following theorem is true:
\begin{theorem}
The space of power series $f(Z,Y)$  forms a bimodule of the star
product algebra of regular functions.
\end{theorem}

\subsection{Consistency of the nonlinear equations}
\label{conscheck}

We want to show explicitly that the system of equations
(\ref{vas1})-(\ref{vas5}) is consistent with respect to both $x$ and
$Z$ variables.

We can start by acting on (\ref{vas1}) with the $x$-differential
$d$. Imposing $d^2=0$, one has $$ dW\*W-W\* dW=0 \ ,$$ which is
indeed satisfied by associativity, as can be checked by using
(\ref{vas1}) itself. So (\ref{vas1}) represents its own consistency
condition.

Differentiating (\ref{vas2}), one gets $$ dW\* B-W\*
dB-dB\*\widetilde{W}-B\* d\widetilde{W}=0 \ . $$ Using (\ref{vas2}),
one can substitute each $dB$ with $-W\* B+B\*\widetilde{W}$,
obtaining
$$ dW\* B+W\* W\*
B-B\*\widetilde{W}\*\widetilde{W}-B\*d\widetilde{W}=0 \ .$$ This is
identically satisfied by virtue of (\ref{vas1}), which is thus the
consistency condition of (\ref{vas2}).

The same procedure works in the case of (\ref{vas3}), taking into
account that, although $S$ is a space-time zero-form,  $dx^\mu
dZ^A_i=-dZ^A_i dx^\mu$.  Again one gets a condition which amounts to
an identity because of (\ref{vas1}).

Hitting (\ref{vas4})  with $d$ and using (\ref{vas2}) and
(\ref{vas3}), one obtains $$ -W\* S\* B-S\* B\* \widetilde{W}+W\*
B\*\widetilde{S}+B\*\widetilde{S}\*\widetilde{W}=0 \ ,$$ which is
identically solved taking into account (\ref{vas4}).

Finally, (\ref{vas3}) turns the differentiated l.h.s. of
(\ref{vas5}) into $-W\* S\* S+S\* S\* W$, while using (\ref{vas2})
the differentiated r.h.s. becomes $-2\Lambda^{-1}dz_i
dz^i(-W\*B\*\ck+B\*\widetilde{W}\*\ck)$. Using (\ref{prop1}), one is
then able to show that the two sides of the equation obtained are
indeed equal if (\ref{vas5}) holds.

$S$ being an exterior derivative in the noncommutative directions,
in the $Z$ sector the consistency check is more easily carried on by
making sure that a covariant derivative of each equation does not
lead to any new condition, {\it i.e.} leads to identities. This
amounts to implementing $d^2_Z=0$.

So commuting $S$ with (\ref{vas1}) gives identically 0 by virtue of
(\ref{vas1}) itself. The same is true for (\ref{vas3}), (\ref{vas2})
and (\ref{vas4}), with the proviso that in these latter two cases
one has to take an anticommutator of the equations with $S$ because
one is dealing with odd-degree-forms (one-forms). The only
nontrivial case then turns out to be (\ref{vas5}), which is treated
in Section \ref{eqns}.

\pagebreak

\end{document}